\documentclass[11pt,notitlepage,aps,prd,preprintnumbers,superscriptaddress,nofootinbib,floatfix]{revtex4-2}

\pdfoutput=1
\usepackage{graphicx}
\usepackage{dcolumn}
\usepackage{bm}
\usepackage{extarrows}
\usepackage{multirow,array}
\usepackage{amsmath,amssymb,physics}
\usepackage{subfigure}
\usepackage{enumitem}
\setlist{nolistsep}
\usepackage{url}
\usepackage{slashed}

\usepackage{orcidlink}

\usepackage{xcolor}

\definecolor{nicered}{rgb}{0.5,0.,0.}
\definecolor{nicegreen}{rgb}{0.,0.5,0.}
\definecolor{niceblue}{rgb}{0.,0.,0.5}

\usepackage{braket}
\usepackage{colortbl}
\usepackage{gensymb}
\usepackage{comment}
\usepackage{setspace}
\usepackage[normalem]{ulem}

\usepackage{hyperref}
\usepackage[capitalise]{cleveref}
\hypersetup{colorlinks,citecolor=nicegreen,linkcolor=nicered,urlcolor=niceblue}

\def\mm{\mu\bar{\mu}}

\renewenvironment{eqnarray}{\begin{equation}\begin{aligned}}{\end{aligned}\end{equation}}

\newcommand{\TeV}{\textrm{TeV}}

\newcommand{\calL}{\mathcal{L}}

\begin{document}

\title{A High-Energy  Neutrino Collider within a 
Muon Collider}
\author{\vspace*{0.3cm}Innes Bigaran\orcidlink{0000-0001-6883-953X}}
\email{ibigaran@fnal.gov}
\affiliation{Center for Neutrino Physics, Virginia Polytechnic Institute and State University,
Blacksburg, VA 24061, USA}

\affiliation{Department of Physics and Astronomy, University of Tennessee, Knoxville, TN 37996, USA\looseness=-1}

\author{Andr\'e de Gouv\^ea\orcidlink{0000-0002-9835-7731}}
\email{degouvea@northwestern.edu}
\affiliation{Department of Physics and Astronomy, Northwestern University, Evanston, IL 60208, USA\looseness=-1}

\author{Tao Han\orcidlink{0000-0002-5543-0716}}
\email{than@pitt.edu}
\affiliation{Pittsburgh Particle Physics, Astrophysics, and Cosmology Center, Department of Physics and Astronomy, University of Pittsburgh, Pittsburgh, PA 15260\looseness=-1}

\author{\mbox{Ian Low}\orcidlink{0000-0002-7570-9597}}
\email{ilow@northwestern.edu}
\affiliation{Department of Physics and Astronomy, Northwestern University, Evanston, IL 60208, USA\looseness=-1}
\affiliation{High Energy Physics Division, Argonne National Laboratory, Argonne, IL 60439, USA}

\author{Yang Ma\orcidlink{0000-0002-9419-6598}}
\email{yang.ma@uclouvain.be}
\affiliation{Centre for Cosmology, Particle Physics and Phenomenology (CP3),
UCLouvain, 1348 Louvain-la-Neuve, Belgium}

\author{Zahra Tabrizi\orcidlink{0000-0002-0592-7425}}
\email{z\_tabrizi@pitt.edu}
\affiliation{Pittsburgh Particle Physics, Astrophysics, and Cosmology Center, Department of Physics and Astronomy, University of Pittsburgh, Pittsburgh, PA 15260\looseness=-1}

\author{Keping Xie\orcidlink{0000-0003-4261-3393}\vspace*{0.3cm}}
\email{kepingxie@sjtu.edu.cn}
\affiliation{State Key Laboratory of Dark Matter Physics \& Key Laboratory for Particle Astrophysics and Cosmology (MOE) \& Shanghai Key Laboratory for Particle Physics and Cosmology, Tsung-Dao Lee Institute \& School of Physics and Astronomy, Shanghai Jiao Tong University, Shanghai 201210, China\looseness=-1}
\affiliation{Department of Physics and Astronomy, Michigan State University, East Lansing, MI 48824, USA}

\preprint{
COMETA-2026-44,~~IRMP-CP3-26-33,~~PITT-PACC-2615}
	
\singlespacing

\begin{abstract}
\vspace{0.5cm}
A multi-TeV muon collider is simultaneously a collider for high-energy neutrinos: The radiation of soft $W^\pm$ gauge bosons from the muon beams gives rise to collisions of $\mu^+\nu$, $\mu^-\bar{\nu}$ and $\nu\bar{\nu}$ at center-of-mass energies close to that of the $\mu^+\mu^-$ collisions in a controlled and theoretically predictable setting. We focus on the leading partonic processes $\mu^+\mu^- \to \ell^+\ell'^-$, $\nu_\mu\mu^+\to \nu_\ell \ell'^+$, $\bar{\nu}_\mu\mu^- \to \bar{\nu}_{\ell} \ell'^-$ and $\nu_\mu\bar \nu_\mu \to \ell^+\ell'^-$ ($\ell, \ell'=e, \mu, \tau$). These are related to one another at high energies by the electroweak $SU(2)$ gauge symmetry, and therefore probe different components of the same underlying electroweak dynamics. These processes open qualitatively new windows to neutrino interactions at unprecedented energies, and open new windows on physics beyond the Standard Model. We demonstrate the combined sensitivity and the complementarity of the different purely-leptonic collisions using both effective field theory and simplified models of $W'^\pm$ and $Z'$ vector bosons. The radiative return mechanism leads to resonant production of the $W^{\prime\pm}$ and $Z'$ vector bosons when their masses are kinematically accessible, while their off-shell effects are captured by the corresponding effective operators. In particular, we show that supplementing the dominant charged-lepton collisions with neutrino-initiated collisions significantly improves the ability to resolve degeneracies among different effective operators, even for operators to which the neutrino-initiated processes are not directly sensitive. Including these channels is therefore essential for properly characterizing the structure of new physics.

\end{abstract}

\maketitle 

\newpage

\tableofcontents{}

\section{Introduction}
Colliders with $\cal{O}$(10 TeV) partonic center-of-mass (c.m.) energy have been identified as future-generation facilities for exploring the physics of electroweak symmetry breaking and potential new physics beyond the Standard Model (SM)~\cite{FCC:2018byv,AlAli:2021let}. Multi-TeV circular muon colliders are especially promising and have been the subject of intense experimental and theoretical investigations owing to the large muon mass relative to that of the electron and subsequent reduction in synchrotron energy loss. Muon colliders may be ideally suited for realizing a high-luminosity collider with a large partonic c.m.~energy~\cite{Aime:2022flm,Accettura:2023ked,InternationalMuonCollider:2025sys}.

It is well known that TeV-scale lepton colliders, including linear $e^+e^-$ colliders~\cite{LinearColliderVision:2025hlt,CLIC:2018fvx,Gessner:2025acq}, multi-TeV $\mu^+\mu^-$ colliders~\cite{Aime:2022flm,Accettura:2023ked,InternationalMuonCollider:2025sys}, and $\mu^+\mu^+$ colliders \cite{Hamada:2022mua,Kitano:2025xaj}, can also serve as effective vector-boson colliders owing to the substantial radiation of $\gamma$, $Z$, and $W^\pm$ gauge bosons~\cite{Chen:2016wkt,Bauer:2017isx}. This makes them powerful tools for exploring high-energy electroweak gauge-boson interactions through processes known as \emph{vector boson fusion (VBF)} and \emph{vector boson scattering (VBS)}~\cite{Costantini:2020stv,Han:2020uid}. A much less explored possibility is to exploit the neutrino content of the muon beams for collider phenomenology. The emission of $W^\pm$ bosons from high-energy $\ell^\pm$ beams generates neutrino partons, enabling high-energy $\ell^\pm\nu$ and $\nu\bar{\ell}$ collisions~\cite{Han:2020uid,Capdevilla:2024ydp}. This opens up a new and largely unexplored avenue for studying neutrino interactions in a controlled and theoretically predictable environment. In this work, we investigate the phenomenological potential of these neutrino-initiated collisions at an opposite-sign muon collider, which we refer to as the $\nu$MuC physics program.

Neutrinos are special. Almost 30 years after the discovery of nonzero neutrino masses, the dynamical mechanism responsible for them remains unknown. There is a distinct possibility that nonzero neutrino masses are a direct consequence of the existence of a mass scale in physics that is independent from that of electroweak symmetry breaking. We are currently unable to identify this hypothetical mass scale and need more experimental input to make meaningful progress. Furthermore, given the enigmatic nature of neutrino masses, it is possible that there is new physics that couples (almost) exclusively to neutrinos.

Neutrino scattering on different target nuclei and at different c.m.~energies are among the observables that can help resolve the neutrino-mass puzzle.  The NuTeV experiment reached laboratory neutrino energies of a few hundred GeV, with neutrino-nucleon interaction c.m.~energies of tens of GeV~\cite{NuTeV:2005wsg}. The FASER experiments have observed neutrinos with up to order $10^3$ GeV of energy, associated to neutrino-nucleon c.m.~energies of  tens of GeV~\cite{FASER:2019dxq}. Neutrino telescopes have directly observed neutrinos with laboratory energies up to $10^8$~GeV, colliding with nuclei in and around the detectors at c.m.~energies up to $10^4$~GeV, and with electrons at up to hundreds of GeV~\cite{IceCube:2013low}.  However, high-energy neutrino scattering on charged leptons and other neutrinos remains far less experimentally explored than neutrino–nucleon scattering, particularly at c.m.~energies of hundreds of GeV and above. A muon collider operating at $\mathcal{O}(10~\mathrm{TeV})$ c.m.~energy could open a new avenue for studying such interactions by serving simultaneously as a high-energy $\nu$MuC.

\begin{figure}
    \centering
    \includegraphics[width=0.29\linewidth]{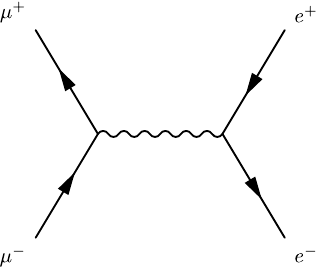}\hspace{0.8cm}
    \includegraphics[width=0.29\linewidth]{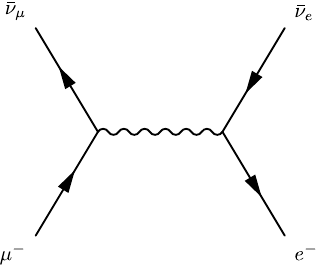}\hspace{0.8cm}
    \includegraphics[width=0.29\linewidth]{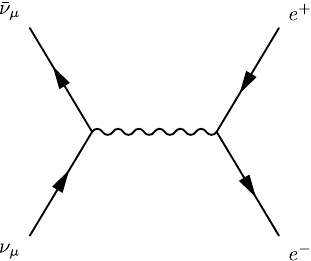}    
    \caption{Representative diagrams for the $\mu^+\mu^-(\mu\bar{\mu}),~\mu^-\bar{\nu}_\mu=\mu^+\nu_\mu(\mu\nu_\mu)$, and $\nu_\mu\bar{\nu}_\mu$ as partonic level collisions. }
    \label{fig:feynmupmum}
\end{figure}

Neutrinos belong to the weak $SU(2)$ lepton doublet together with the left-handed charged-leptons. Therefore, at energies much higher than the weak scale, when the masses of the $W^\pm$ and $Z$ bosons are much smaller than the c.m.~energy, scattering processes involving neutrino and charged-lepton initial states, such as the $\mm$, $\nu_{\mu}\bar{\nu}_{\mu}$, and $\mu{\nu}_{\mu}$ collisions depicted in~\cref{fig:feynmupmum}, are expected to be closely related by $SU(2)$ gauge symmetry \cite{Ciafaloni:2000df,Ciafaloni:2001mu,Chen:2016wkt,Bauer:2017isx,Capdevilla:2024amg}, therefore enriching the physics reach at $\nu$MuC. Here and henceforth we use the notation $\mu=\mu^-$, ${\bar \mu}=\mu^+$, and $\mu{\nu}_{\mu}$ includes both $\mu\bar{\nu}_\mu$ and its  conjugate initial states $\bar{\mu}\nu_\mu$.

The complementarity of $\mm$, $\nu_{\mu}\bar{\nu}_{\mu}$, and $\mu{\nu}_{\mu}$ collisions can be studied in a systematic, model-independent framework using effective field theory (EFT), particularly when the new physics lies well above the electroweak scale. In this regime, the Standard Model Effective Field Theory (SMEFT) provides a consistent description of possible deviations from the SM. The SMEFT extends the SM Lagrangian with higher-dimensional operators constructed from SM fields and respecting its gauge symmetries. In this case, searches for new physics can be interpreted as constraints on the Wilson Coefficients of these operators, allowing results from different experiments to be compared under a common framework. 
Hence, SMEFT is a useful benchmark for assessing the sensitivity of muon collider, including the neutrino collisions considered here, alongside current and future low-energy neutrino and other precision experiments. 

The large number of independent Wilson coefficients in the SMEFT can lead to flat directions in global fits, where different combinations of coefficients produce similar effects in observables. To break these degeneracies, one generally needs to combine a diverse set of observables, from different experiments and energy scales. In this work we will demonstrate that by combining $\mm$, $\nu_{\mu}\bar{\nu}_{\mu}$, and $\mu{\nu}_{\mu}$ collisions at  $\nu$MuC can lift flat directions among several four-lepton SMEFT operators within a single experimental facility. Furthermore, we show that the $\nu$MuC can probe new-physics scales of hundreds of TeV for most of the operators considered, surpassing the reach of many existing and projected precision experiments.

Beyond effective field theories, realizations of BSM physics often connect neutrinos and charged-leptons  systematically. Useful examples include BSM interactions characterized by the exchange of $SU(2)$ triplet or singlet bosons, including heavy Higgs-like scalars ($H^{0,\pm}$) or new heavy vector bosons ($Z',W'^{\pm}$). In these cases, the combined exploitation of $\nu_{\mu}\bar{\nu}_{\mu}$, and $\mu{\nu}_{\mu}$ collisions at high energies is necessary in order to properly characterize the new physics. For illustrative purposes, in this work we concentrate on a simplified $U(1)$ $Z'$ model and $SU(2)$ $W'$ model. The radiative return mechanism leads to resonant production  $\mu\bar \mu\to  Z'$ and $\mu\nu_\mu \to W^{\prime\pm}$ for their direct discovery at a high energy $\nu$MuC, if kinematically accessible to their masses. We also discuss how the results obtained using SMEFT can be translated to the $Z'/W'$ parameter space, if their masses lie far above the collider c.m.~energy.

The remainder of the manuscript is organized as follows. In \cref{sec:neutrino_pdf}, we introduce the treatment of muons and neutrinos as partons at a high-energy muon collider and discuss the corresponding parton distribution functions and parton luminosities. In \cref{sec:SMEFT}, we present the projected sensitivities of the $\nu$MuC to four-lepton SMEFT interactions, including the complementarity of the different muon- and neutrino-initiated collision channels. In \cref{sec:V'}, we investigate the discovery and characterization of heavy vector bosons at a muon collider, considering simplified $W'$ and $Z'$ scenarios and their relation to the SMEFT description. We summarize our results and conclude in \cref{sec:conclusion}. In \cref{App:SMEFTmatchbases}, we explain the mapping between the SMEFT convention used in this work and alternatives in the literature. Finally, in \cref{App:LEconstraints}, we collect the low-energy constraints on the four-lepton SMEFT coefficients relevant to our analysis and provide the basis for comparison with the projected $\nu$MuC sensitivities.

\section{Muon and neutrino as partons at high energies}
\label{sec:neutrino_pdf}
Following the factorization formalism in Refs.~\cite{Han:2020uid,Han:2021kes}, the semi-inclusive production cross section at a high-energy muon collider for a final state $\mathcal{F}$ can be written as a convolution of the parton luminosity $\calL_{ij}$ and the hard-scattering cross section for the partonic subprocess $\hat{\sigma}_{ij\to \mathcal{F}}(\hat{s})$:
\begin{align}
\label{eq:xsec}
&\sigma_{\mm \to \mathcal{F} +X}(s) = \sum_{ij}\int_{\tau_0}^1\dd\tau\frac{\dd\calL_{ij}}{\dd\tau}\hat{\sigma}_{ij\to \mathcal{F}}(\tau s)\,,  \\ 
&\frac{\dd\calL_{ij}}{\dd\tau} = \frac{1}{1+\delta_{ij}}\int_{\tau}^1\frac{\dd x}{x}f_{i/\mu}(x,Q^2)f_{j/\bar\mu}(\tau/x,Q^2)+(i\leftrightarrow j)\,, 
\end{align}
where $s=(p_\mu+p_{\bar{\mu}})^2$ is the c.m.~energy and $f_{i/\mu}(x,Q^2)$ or $f_{j/\bar\mu}(x,Q^2)$ denotes the parton distribution function (PDF) for finding a parton $i$ or $j$ carrying a fraction $x$ of the parent muon's energy at the factorization scale $Q$;
$\hat{s} = (p_i+p_j)^2$ is the partonic c.m.~energy squared  and $\tau = \hat s / s$. 
The integration lower limit $\tau_0$ is set by the production threshold, $\tau_0=m_{\mathcal F}^2/s$, or by any additional kinematical cuts, whichever is most restrictive. $X$ denotes the beam remnants accompanying the hard sub-process. The factorization scale $Q$ is an auxiliary scale introduced by the factorization procedure and is conventionally chosen to be of the order of the typical momentum transfer in the hard-scattering process $ij\to \mathcal{F}$. Its residual dependence provides an estimate of the theoretical uncertainty associated with the factorization procedure. 

The muon is the beam particle and thus serves as the ``valence'' parton, carrying the majority of the beam energy. The initial condition for the muon PDF evolution at leading order (LO) is
\begin{equation}
\label{eq:mumupdf}
f^{\rm LO}_{\mu/\mu}(x,Q^2=m_\mu^2)=   f^{\rm LO}_{\bar{\mu}/\bar{\mu}}(x,Q^2=m_\mu^2)=\delta(1-x) \,. 
\end{equation}
The muon PDF receives corrections from subsequent parton splittings and can be evolved to high scales following the DGLAP equations~\cite{Gribov:1972ri,Lipatov:1974qm,Altarelli:1977zs,Dokshitzer:1977sg}.\footnote{In the context of a 100 TeV hadron collider, the possibility of treating the top quark as a parton inside the hadron has been studied in Ref.~\cite{Dawson:2014pea,Han:2014nja}.} The evolution resums these corrections to all orders in perturbation theory at a given logarithmic accuracy. The corrections are dominated by photon radiation, commonly referred to as QED initial-state radiation (ISR).
As demonstrated in detail in Ref.~\cite{Han:2021kes}, in the vicinity of $x\to 1$ close to the full beam energy, the muon PDF can be expressed analytically when converted in the Mellin-$N$ space at large $N$ 
\begin{align}
    &f_{\mu/\mu}^{\textrm{large-}N}(x,Q^2)=\frac{e^{\beta\lambda}}{\Gamma(\beta)}(1-x)^{-1+\beta}\,, \\
    & \beta=\frac{\alpha}{\pi}\log\frac{Q^2}{m_\mu^2},~\lambda=\frac{3}{4}-\gamma_E\,,
\end{align}
where $\alpha$ is the fine structure constant, and $\gamma_E=0.577$ is the Euler–Mascheroni constant.
The large-$N$ treatment is valid only in the $x\to1$ region, and resums soft-photon contributions to all orders. Hard-photon contributions can be included by supplementing the QED ISR structure function with additional terms up to $\mathcal{O}(\alpha^3)$~\cite{Skrzypek:1990qs,Skrzypek:1992vk,Cacciari:1992pz}.\footnote{
For more details, see {\it e.g.} Sec.~2 of Ref.~\cite{Denner:2003iy}, Appendix A 
of~Ref.~\cite{Bertone:2022ktl}, and Appendix A 
of~Ref.~\cite{Ma:2024ayr}.} This QED ISR treatment is implemented in both \textsc{MadGraph\_aMC@NLO}~\cite{Alwall:2014hca,Frederix:2018nkq} and \textsc{Whizard}~\cite{Kilian:2007gr,Moretti:2001zz,Brass:2018xbv}.

Analogously, the neutrinos emerge as partons via weak ISR, i.e., $W$ boson radiation. At the LO in $\alpha_W=g_W^2/(4\pi)$, the muon-neutrino PDF takes the familiar form \cite{Chen:2016wkt} 
\begin{equation}
\label{eq:numuPDF}
   f_{\nu_\mu/\mu}^{\rm LO}(x,Q^2)=\frac{1}{2}\frac{\alpha_W}{2\pi}\frac{1+x^2}{1-x}\log\frac{Q^2}{M_W^2}\,,
\end{equation}
where the prefactor $1/2$ arises from the assumption of an unpolarized muon beam, for which 50\% of the muons are left-handed. In the electroweak PDF formalism~\cite{Han:2020uid,Han:2021kes}, all SM particles can be treated as effectively massless and can appear as partons and participate in the DGLAP evolution. In particular, this formalism allows for the evolution of both muon and neutrino PDFs at leading-log accuracy.

The necessary condition for the validity of the partonic picture is that the LO logarithmic terms are significantly  larger than the power corrections, that is, 
\begin{equation}
\alpha \log{\frac{Q^2}{m^2}} \gg {\frac{m^2}{Q^2}}\,. 
\end{equation}
This condition is easy to satisfy for the muon PDF, since the dominant contribution is from photon radiation off a light fermion ($m=m_\mu$). A much higher factorization scale $Q$ is required for the neutrino PDF to serve as a useful approximation due to the much larger $W$-boson mass. 

\begin{figure}[t]
    \centering
    \includegraphics[width=0.49\linewidth]{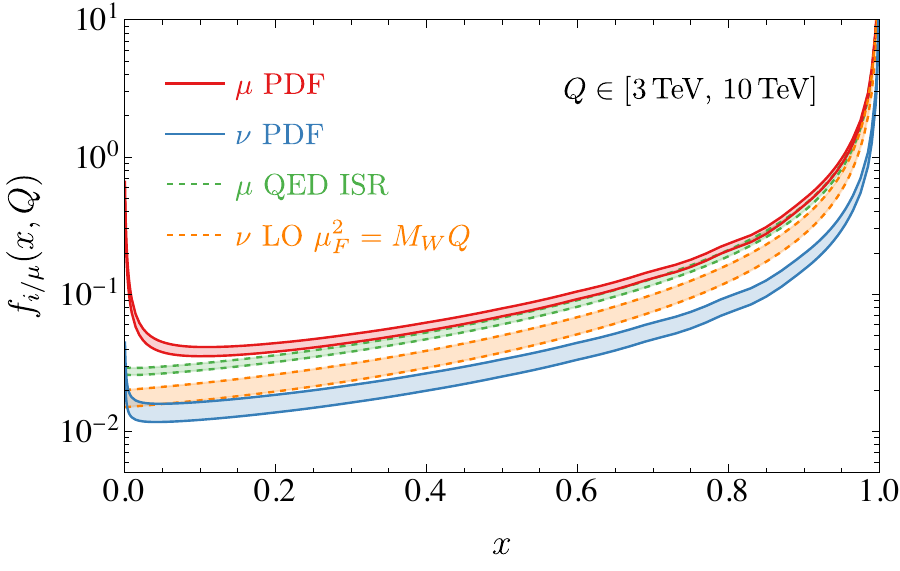}
    \includegraphics[width=0.49\textwidth]{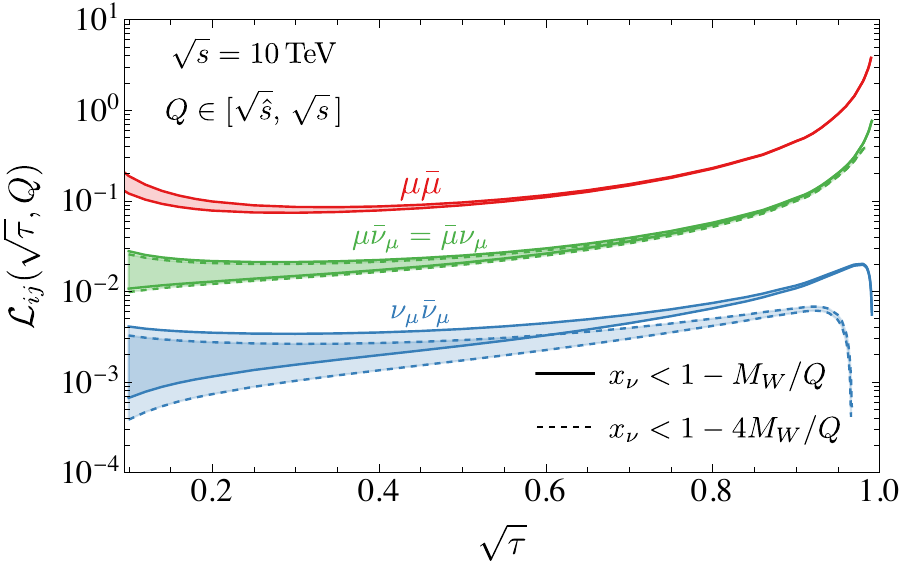}
    \caption{Left: DGLAP-evolved muon and neutrino PDFs, with the bands corresponding to the factorization scale $Q=3-10$ TeV, compared with the QED ISR muon PDF and the leading-order
(LO) neutrino PDF evaluated at the intermediate scale
$\mu_F^2=M_W Q$. Right: Parton luminosities for
$\mu\bar{\mu}$, $\mu\bar{\nu}_\mu=\bar{\mu}\nu_\mu$, and $\nu_\mu\bar{\nu}_\mu$ at a 10 TeV muon collider, with the bands corresponding to the scale variation $Q\in[1/2,1]\sqrt{s}$. For
$\mu\bar{\nu}_\mu=\bar{\mu}\nu_\mu$ and $\nu_\mu\bar{\nu}_\mu$, the solid (dashed) curves correspond to the cut $x_{\rm cut}=1- \xi x_W$, 
where $x_W=M_W/Q$ and $\xi=1$ or 4.} 
\label{fig:PDFlumi}
\end{figure}

In~\cref{fig:PDFlumi} (Left), we present the DGLAP-resummed PDFs for the muon and the $\nu_\mu$, shown as red and blue solid curves, respectively, together with the QED ISR muon structure function and the LO $\nu_\mu$ PDF with an intermediate scale $\sqrt{M_WQ}$, shown as green and orange dashed curves, respectively. Here, the intermediate scale $\sqrt{M_WQ}$ choice is motivated by the one-step solution to the DGLAP equation~\cite{Han:2020uid}.  Both the QED ISR muon PDF and the LO neutrino PDF capture the large-$x$ behavior of the fully resummed PDFs reasonably well but deviate substantially in the small-$x$ region due to the radiative-return effect. The bands correspond to the variation of the factorization scale $Q\in[3,{\rm TeV},\,10\,{\rm TeV}]$ and for the $\mu$ ($\nu_\mu$) PDF, the effect is smaller than  20\% (40\%).

In~\cref{fig:PDFlumi} (Right), we present the parton luminosities $\mathcal{L}_{ij}$ (see~\cref{eq:xsec}) for the combination of valence-like leptons $ij=\mu\bar \mu$ (red curves), $\mu\bar{\nu}_\mu=\bar{\mu}\nu_\mu$ (green curves), and $\nu_\mu\bar \nu_\mu$ (blue curves), respectively, as a function of $\sqrt{\tau}$, for $\sqrt s = 10$ TeV. A few important features are worth highlighting.
First, we illustrate the scale uncertainty by varying $Q$ between the collider energy, $Q=\sqrt{s}$, and the partonic center-of-mass energy, $Q=\sqrt{\hat{s}}$. The scale uncertainty is sizable at small $\sqrt{\tau} = \sqrt{\hat s/s}$,  especially for neutrino-initiated processes, due to electroweak gauge boson radiation, while it becomes much smaller at larger $\sqrt{\tau}$. Second, by construction, the valence-parton luminosities peak sharply toward the full collider energy, $\sqrt{\tau}\to1$.
For the neutrino PDF, however, the $W$-boson mass introduces a natural scale associated with $W$ radiation, leading to a cutoff of order $x_W\sim M_W/Q$. We therefore impose a hard cut 
\begin{equation}
    x<x_{\rm cut}=1-\xi x_W\,,
\end{equation} 
where $x_W=M_W/Q$ and $\xi=1$ or 4. This restricts the analysis to the collinear region. The results for both values of $\xi$ are shown as solid and dashed curves, respectively, in~\cref{fig:PDFlumi} (Right). The comparison of these two families of curves provides an estimate of the associated threshold uncertainty. Third, at small $\tau$, the effects from soft radiation become more and more significant, and we see an increase of the $\mm$ luminosities as $\sqrt{\tau}$ decreases. This is a consequence of the small-$x$ behavior of the muon PDF. On the other hand,  the decrease of $\mu\bar{\nu}_\mu=\bar{\mu}\nu_\mu$ and $\nu_\mu\bar{\nu}_\mu$ as $\sqrt{\tau}$ decreases comes from potentially large corrections of order $M_W^2/\hat s$.

Bearing in mind these features and uncertainties, the cross sections for $\mu \bar\mu,\ \mu\bar{\nu}_\mu$ and $\nu_\mu\bar{\nu}_\mu$ collisions can be computed using~\cref{eq:xsec}. The partonic cross sections $\hat{\sigma}_{ij\to \mathcal{F}}(x_i,x_j)$ as a function of the initial parton momentum fractions $x_{i(j)}$ can be easily obtained in a variety of ways for the different assumptions about the underlying physics (SM, SMEFT, UV-complete BSM models). More specifically, for the dimensional-six SMEFT operators, we use \textsc{FeynArts}~\cite{Hahn:2000kx} and \textsc{FeynCalc}~\cite{Shtabovenko:2023idz} to perform the semi-analytical calculations of the partonic cross sections. For comparison, one can perform fixed-order (FO) diagrammatic calculations\footnote{These may also be referred to as \emph{full matrix element} calculations.} of $\mu \bar\mu$ collisions. The FO calculations serve as reference points and can guide the scale choice for the resummed PDF treatment. Therefore, whenever necessary, our numerical results are cross checked with \textsc{MadGraph\_aMC@NLO}~\cite{Alwall:2014hca,Frederix:2018nkq} and/or \textsc{Whizard}~\cite{Kilian:2007gr,Moretti:2001zz,Brass:2018xbv,Christensen:2010wz} with the Universal \textsc{FeynRules} Output (UFO)~\cite{Degrande:2011ua} model files generated with \textsc{FeynRules}~\cite{Alloul:2013bka} or directly taken from \textsc{SMEFTsim}~\cite{Brivio:2020onw}. For the UV models for heavy vector bosons, we use \textsc{MadGraph\_aMC@NLO}~\cite{Alwall:2014hca,Frederix:2018nkq} and \textsc{Whizard}~\cite{Kilian:2007gr,Moretti:2001zz,Brass:2018xbv}.

\section{Projected sensitivities to SMEFT}
\label{sec:SMEFT}

In this Section we consider the projected sensitivities of $\nu$MuC to a subset of the Standard Model Effective Field Theory (SMEFT), focusing on dimension-six, exclusively leptonic SMEFT operators. Working with four-component spinors, these are, in the Warsaw basis~\cite{Grzadkowski:2010es}
\begin{align}
\label{eq:SMEFT}
    \mathcal{L} &=\frac{1}{2}C_{ ijkl}^{\ell\ell} (\overline{\ell_i} \gamma^\mu  \ell_j) (\overline{\ell_k} \gamma_\mu  \ell_l)+ \frac{1}{2}C^{ee}_{ijkl} (\overline{e_i} \gamma^\mu e_j) (\overline{e_k} \gamma_\mu  e_l) + C^{\ell e}_{ ijkl}(\overline{\ell_i} \gamma^\mu \ell_j) (\overline{e_k} \gamma_\mu  e_l) \,.
\end{align}    
Here $\ell_i$ is the $SU(2)_L$ lepton doublet, $e_i$ is the $SU(2)_L$ lepton singlet, and the summation is taken over all Latin indices, $i,j,k,l\in\{1,\,2,\,3\}$, where the indices $1$, $2$, and $3$ denote the electron, muon, and tau flavors, respectively. 
The Wilson Coefficients~(WCs) $C^X_{ijkl}$, with $X=\ell\ell,\ell e,ee$, have mass-dimension $-2$. We can also define dimensionless WCs $c^X_{ijkl}$, related to the dimensionful ones via a BSM scale $\Lambda$,
\begin{equation}
C^X_{ijkl}\equiv \frac{c^X_{ijkl}}{\Lambda^2}\,,
\label{eq:dim}
\end{equation}
For the remainder of this work, we will make the simplifying assumption that the WCs are real-valued. At the appropriate scale $\Lambda$, the dimensionless coupling of the underlying  theory should be naturally the order of unity \cite{Manohar:1983md,Cohen:1997rt}. To illustrate the physics reach of $\nu$MuC, we set $c^X_{ijkl}=1$ and present the corresponding constraints in $\Lambda=1/\sqrt{C^X_{ijkl}}$ in ~\cref{sec:SMEFT_single}.  For the EFT description to be valid, the experimentally probed energy scale should be below the cutoff $\Lambda$.

Not all of the four-fermion coefficients in a general summation over flavor indices are independent. For example, Hermiticity of the SMEFT Lagrangian implies
\begin{align}
    C^X_{ijkl}= C^{X,*}_{jilk}\,,
    \label{eq:hermitian}
\end{align}
for $X\in \{\ell\ell, ee, \ell e\}$. Furthermore, the symmetry properties of the effective interactions reduce the number of independent operators through flavor-index redundancies. For  operators  containing two identical currents,  $X\in \{\ell\ell, ee\}$, there is a further exchange symmetry
\begin{align}
C^X_{ijkl}= C^X_{klij}\,, 
\label{eq:samecurrent}
\end{align}
and, as the $e_i$ field is a SM singlet, the following relation holds by Fierz rearrangement: 
\begin{align}
    C^{ee}_{ijkl}= C^{ee}_{ilkj}\,. \label{eq:eebonus}
\end{align}
The application of these relations to the flavor indices of the four-lepton operators in~\cref{eq:SMEFT}, and taking the WCs to be real, leaves thirty independent operators entering the interactions studied in this work, which are listed in~\cref{tab:WCDEF}.\footnote{The treatment of flavor redundancies in SMEFT WCs differs slightly among works in the literature. For convenience, the relevant mapping between the \textsc{SMEFTsim}~\cite{Brivio:2020onw} convention  (which was utilized as a cross-check) and ours can be found in~\cref{App:SMEFTmatchbases}. In addition, for flavor-violating coefficients (right sub-table of~\cref{tab:WCDEF}), our work includes one additional independent coefficient $C^{\ell \ell }_{1223}$ omitted in Ref.~\cite{Glioti:2025zpn}.} Note that for this work we restrict our attention to operators for which at least two of the flavor indices correspond to muon flavor.

\begin{table}[t!]
\resizebox{0.4\textwidth}{!}{\def\arraystretch{1.5}
\centering
\begin{tabular}{|c|c|}
\hline 
WC & Partonic processes\\
\hline 
$C^{ee}_{1122}$ & $\mu^+ \mu^- \to e^+ e^-$\\
$C^{\ell e}_{1122}$ &  $\mu^+ \mu^- \to e^+ e^-$\\
$C^{\ell e}_{2211}$ & $\mu^+ \mu^- \to e^+ e^-, \nu_\mu \bar{\nu}_\mu \to e^+ e^-$\\
$C^{\ell e}_{1221}$ & $\mu^+ \mu^- \to e^+ e^-, \mu^+ \nu_\mu \to e^+ \nu_e$\\
$C^{\ell \ell}_{1122}$ & $\mu^+ \mu^- \to e^+ e^-, \nu_\mu \bar{\nu}_\mu \to e^+ e^-$\\
$C^{\ell \ell}_{1221}$ & $\mu^+ \mu^- \to e^+ e^-, \mu^+ \nu_\mu \to e^+ \nu_e$\\
\hline     
$C^{ee}_{2222}$ &  $\mu^+ \mu^- \to \mu^+ \mu^-$\\
$C^{\ell e}_{2222}$ &  $\mu^+ \mu^- \to \mu^+ \mu^-, \mu^+ \nu_\mu \to \mu^+ \nu_\mu, \nu_\mu \bar{\nu}_\mu \to \mu^+ \mu^-$\\
$C^{\ell \ell}_{2222}$ & $\mu^+ \mu^- \to \mu^+ \mu^-, \mu^+ \nu_\mu \to \mu^+ \nu_\mu, \nu_\mu \bar{\nu}_\mu \to \mu^+ \mu^-$ \\
\hline
$C^{ee}_{2233}$ & $\mu^+ \mu^- \to \tau^+ \tau^-$\\
$C^{\ell e}_{3322}$ &$\mu^+ \mu^- \to \tau^+ \tau^-$ \\
$C^{\ell e}_{2233}$ & $\mu^+ \mu^- \to \tau^+ \tau^-,  \nu_\mu \bar{\nu}_\mu \to \tau^+ \tau^-$\\
$C^{\ell e}_{2332}$ & $\mu^+ \mu^- \to \tau^+ \tau^-, \mu^+ \nu_\mu \to \tau^+ \nu_\tau$\\
$C^{\ell \ell}_{2233}$ & $\mu^+ \mu^- \to \tau^+ \tau^-, \nu_\mu \bar{\nu}_\mu \to \tau^+ \tau^-$\\
$C^{\ell \ell}_{2332}$ &$\mu^+ \mu^- \to \tau^+ \tau^-, \mu^+ \nu_\mu \to \tau^+ \nu_\tau$ \\
\hline 
\end{tabular}}
\hspace{0.5cm}
\resizebox{0.5\textwidth}{!}
{\def\arraystretch{1.5}
\centering
\begin{tabular}{|c|c|}
\hline 
    WC & Partonic processes\\
    \hline 
    $C^{ee}_{1222}$ &  $\mu^+ \mu^- \to \mu^+ e^-$\\
    $C^{\ell e}_{1222}$ &  $\mu^+ \mu^- \to e^+ \mu^-, \mu^+ \nu_\mu \to \mu^+ \nu_e$\\
    $C^{\ell e}_{2212}$ &$\mu^+ \mu^- \to e^+ \mu^-, \nu_\mu \bar{\nu}_\mu \to e^+ \mu^-$ \\
    $C^{\ell \ell}_{1222}$ & $\mu^+ \mu^- \to e^+ \mu^-,
   \mu^+ \nu_\mu \to e^+ \nu_\mu,
   \mu^+ \nu_\mu \to \mu^+ \nu_e,
   \nu_\mu \bar{\nu}_\mu \to e^+ \mu^-$\\
    \hline 
    $C^{ee}_{2223}$ &  $\mu^+ \mu^- \to \mu^+ \tau^-$\\
    $C^{\ell e}_{2322}$ & $\mu^+ \mu^- \to \tau^+ \mu^-, \mu^+ \nu_\mu \to \mu^+ \nu_\tau$\\
    $C^{\ell e}_{2223}$ & $\mu^+ \mu^- \to \tau^+ \mu^-, \nu_\mu \bar{\nu}_\mu \to \tau^+ \mu^-$\\
    $C^{\ell \ell}_{2223}$ &$\mu^+ \mu^- \to \tau^+ \mu^-, \mu^+ \nu_\mu \to \tau^+ \nu_\mu,\mu^+ \nu_\mu \to \mu^+ \nu_\tau, \nu_\mu \bar{\nu}_\mu \to \tau^+ \mu^-$\\
    \hline 
    $C^{ee}_{1322}$
& $\mu^+ \mu^- \to \tau^+ e^-$\\

$C^{\ell e}_{1322}$
& $\mu^+ \mu^- \to \tau^+ e^-$\\

$C^{\ell e}_{2213}$
& $\mu^+ \mu^- \to \tau^+ e^-,
   \nu_\mu \bar{\nu}_\mu \to \tau^+ e^-$\\

$C^{\ell e}_{2312}$
& $\mu^+ \mu^- \to \tau^+ e^-,
   \mu^+ \nu_\mu \to e^+{\nu}_\tau$\\

$C^{\ell e}_{1223}$
& $\mu^+ \mu^- \to \tau^+ e^-,
   \mu^+ \nu_\mu \to \tau^+ {\nu}_e$\\

$C^{\ell \ell}_{1322}$
& $\mu^+ \mu^- \to \tau^+ e^-,
   \nu_\mu \bar{\nu}_\mu \to \tau^+ e^-$\\
$C^{\ell \ell }_{1223}$ & $\mu^+ \mu^- \to \tau^+ e^-,
   \mu^+ \nu_\mu \to \tau^+ {\nu}_e ,  \mu^+ \nu_\mu \to e^+ {\nu}_\tau$ \\
\hline  
\end{tabular}}
\caption{
The set of independent Wilson Coefficients~(WCs) considered in this work, highlighting the partonic processes that each will contribute to in muon collision within the electroweak PDF framework. We highlight only processes which could give a visible final state. For brevity, the charge-conjugate processes are omitted from this table but are implied.}
\label{tab:WCDEF}
\end{table}

For all $\nu$MuC constraints presented, we assume a benchmark muon collider  with a c.m.~energy $\sqrt{s}=10$ TeV and 10 ab$^{-1}$ integrated luminosity. 
As already mentioned, at the $\nu$MuC we  consider not only ``on-shell" muon--muon collisions but also  collisions with electroweak partonic neutrinos, namely $\mm$, $\mu\nu_\mu$, and $\nu_\mu\bar{\nu}_\mu$. \cref{tab:WCDEF} lists the partonic subprocesses considered in this analysis and the dimension-six SMEFT operators that directly contribute to each process. The corresponding total cross sections are obtained using the electroweak PDFs formalism in~\cref{eq:xsec}. 
As discussed in more detail in the previous section, the partonic cross sections $\hat{\sigma}_{ij\to\mathcal{F}}$ are calculated analytically using \textsc{FeynArts}~\cite{Hahn:2000kx} and \textsc{FeynCalc}~\cite{Shtabovenko:2023idz} together with a dimension-six SMEFT model file. The analytical results are cross-checked against numerical results obtained with \textsc{MadGraph5\_aMC@NLO} and \textsc{Whizard} using the \textsc{SMEFTsim} model~\cite{Brivio:2020onw}. Throughout this study, we use $Q=\sqrt{{s}}/2$ as the benchmark choice for illustrating the sensitivity to the SMEFT operators.

To reduce the beam-induced backgrounds (BIB) from muon decay~\cite{Ally:2022rgk,MuonCollider:2022glg,MuonCollider:2022ded,Accettura:2023ked},  muon-collider detectors are expected to contain nozzles around the beam pipe, which limit the angular coverage to typically $10^\circ<\theta<170^\circ$, corresponding to $|\eta|<2.44$~\cite{MAIA:2025hzm,Andreetto:2025mrd}.
Motivated by this angular acceptance, we impose a minimal transverse momentum cut $p_T^{{\text {FS-lep.}}}>p_T^{\min}$ on the final-state charged leptons (denoted `FS-lep.'). For the $2\to 2$ partonic subprocesses considered here, this requirement provides a direct proxy for the detector angular acceptance. For a massless final-state lepton, its energy and transverse momentum satisfy $E_{{\text {FS-lep.}}}=p_T^{{\text{FS-lep.}}}\cosh\eta^{{\text{FS-lep.}}}$. Since the final-state energy cannot exceed the beam energy, $E_{{\text{FS-lep.}}}\leq \sqrt{s}/2$ and one obtains
\begin{align}
    |\eta^{{\text{FS-lep.}}}|< \text{arccosh} \left(\frac{\sqrt{s}}{2p_T^{{\text{FS-lep.}}}}\right)\,.
\end{align}
For $\sqrt{s}=10$~TeV,  $p_T^{{\text {FS-lep.}}} \geq p_T^{\min}=850$~GeV implies $|\eta^{{\text {FS-lep.}}}|\lesssim2.46$, very close to the detector acceptance. This $p_T$ cut also avoids large logarithmic radiative corrections~\cite{Glioti:2025zpn}. For flavor-violating processes, we assume that two independent measurements, corresponding to charge flips in the final-state charged leptons, can be made, so we present projections for the reach of both measurements combined.

Throughout, we neglect any tagging or reconstruction efficiencies for final-state particles, as these are highly dependent on the collider design and the detector performance. For this reason, the same results are quoted for either taus or electrons in the final state. We acknowledge that due to detector reconstruction systematics the sensitivity to processes with taus in the final state will be relatively reduced, but assessing the magnitude of this distinction is beyond the scope of this work. A comprehensive assessment of $\nu$MuC's reach is left for future work, once a projected detector design is accessible. 

In restricting our attention to the Lagrangian in~\cref{eq:SMEFT} (mass-dimension six), we have assumed that we are working in a regime where the contributions of higher mass-dimension operators are consistently relatively smaller. For example, mass-dimension eight operators can be phrased akin to~\cref{eq:dim} but instead with four powers of a scale $\Lambda$ in the denominator, \emph{i.e.} $C^{(8)} = c_8/\Lambda^4$, and we are interested in $\Lambda$ values that are larger than the c.m.~energy. 

In more detail, the cross sections in SMEFT at tree-level can be  expanded in powers of the new-physics scale $\Lambda$:
\begin{equation}
\sigma_{\rm SMEFT} = \sigma_{\rm SM} +\sigma_{\rm int}^{(6)} +\sigma_{\rm quad}^{(6)} +\sigma_{\rm int}^{(8)} +\cdots\,,
\end{equation}
where $\sigma_{\rm int}^{(6)}$ denotes the interference between the SM and dimension-six amplitudes, $\sigma_{\rm quad}^{(6)}$ the quadratic dimension-six contribution, and $\sigma_{\rm int}^{(8)}$ the interference between the SM and dimension-eight amplitudes. 
Assuming a common scale $\Lambda$, these terms parametrically scale as 
\begin{equation}
\sigma_{\rm SM}\propto \frac{g^4}{s}\,, \qquad \sigma_{\rm int}^{(6)} \propto \frac{c_6 g^2}{\Lambda^2}\,,\qquad \sigma_{\rm quad}^{(6)}\propto \frac{c_6^2 {\hat s}}{\Lambda^4}\,, \qquad \sigma_{\rm int}^{(8)} \propto \frac{c_8 g^2 {\hat s}}{\Lambda^4}\,,
\end{equation}
where $c_6$ ($c_8$) is the corresponding dimensionless dimension-six (dimension-eight) WC as in~\cref{eq:dim}, governed by the unspecified  underlying theory, and $g$ is the SM gauge coupling. As such
\begin{equation}
    \frac{\sigma_{\rm quad}^{(6)}}{\sigma_{\rm int}^{(6)}} \propto \frac{c_6 {\hat s}} {g^2\Lambda^2}\,, \qquad \frac{\sigma_{\rm lin}^{(8)}}{\sigma_{\rm quad}^{(6)}} \propto \frac{g^2 c_8}{c_6^2}\,.
\end{equation}
For some of our analyses, we present results for two cases: (i) only the SM-interference dimension-six contributions, which are linear with respect to the corresponding WCs, are included,\footnote{For the process where the SM-interference term vanishes, the leading contribution will be $\sigma_{\rm quad}^{(6)}$, which we keep as the leading term.} (ii) both the linear and quadratic dimension-six contributions are included. 
The latter should be regarded as a phenomenological dimension-six benchmark rather than a complete $\mathcal{O}(\Lambda^{-4})$ prediction. If the quadratic dimension-six contribution were found to have a substantial impact on the phenomenological results, it would indicate that $\mathcal{O}(\Lambda^{-4})$ effects are important and that the corresponding dimension-eight--SM-interference terms should be considered in a more complete analysis.

When estimating the sensitivity to the different SMEFT operators, the presence of both $\sigma_{\rm int}^{(6)}$ and $\sigma_{\rm quad}^{(6)}$ contributions lead to degenerate best-fit solutions, only one of which is centered around a vanishing WC (corresponding to the SM).  When quoting single-coefficient constraints, where we turn on only one effective operator at a time,  we concentrate on the region around the vanishing WC to define the sensitivity.

To estimate the sensitivity to the SMEFT parameters, we compute \cite{ParticleDataGroup:2026mpi} 
\begin{eqnarray}
\mathcal{S} = \sqrt{2(S+B)\log\left(1+\frac{S}{B}\right) - 2S} \,, 
\label{eq:signal_strength}
\end{eqnarray}
where $S$ is the expected number of signal events as a function of the SMEFT parameters, while $B$ is the expected number of events for background (i.e., SM) hypothesis. 
In the limit $S \ll B$, $\mathcal{S}$ can be approximated as  
\begin{eqnarray}
\label{eq:calS2}
\mathcal{S} \simeq \frac{S}{\sqrt{B}}\,.
\end{eqnarray}
$\mathcal{S}=1(2)$ corresponds to the 68\% (95\%) confidence level (C.L.) sensitivity. 
\cref{eq:signal_strength} is not valid in the limit of zero background events. In this work, we encounter processes for which we have zero SM contributions (e.g., for flavor-violating processes), and therefore we need to also consider this limit. In these cases, using Poisson statistics, we adopt the following test statistic~\cite{ALRead_2002}
\begin{align}
\mathcal{C}_{S}= e^{-S}\,.
\label{CLeq}
\end{align}
The signal can be excluded with a confidence of $\alpha$ if $\mathcal{C}_{S}< 1-\alpha$. For example, the exclusion limit is set for $\alpha= 0.95(0.68)$ (corresponding to a 95\% (68\%) C.L.) with $S=3.00~(1.14)$ signal events.

\subsection{Projected single-operator constraints }
\label{sec:SMEFT_single}
We begin by turning on one operator at a time, recalling that we are assuming that the WCs are real-valued. This allows the projected muon-collider constraints to be compared directly, operator by operator, with existing low-energy bounds in the literature. We refer to the resulting sensitivity throughout as the \emph{single-operator sensitivity}.

\begin{figure}[t!]
    \centering
    \includegraphics[width=\linewidth]{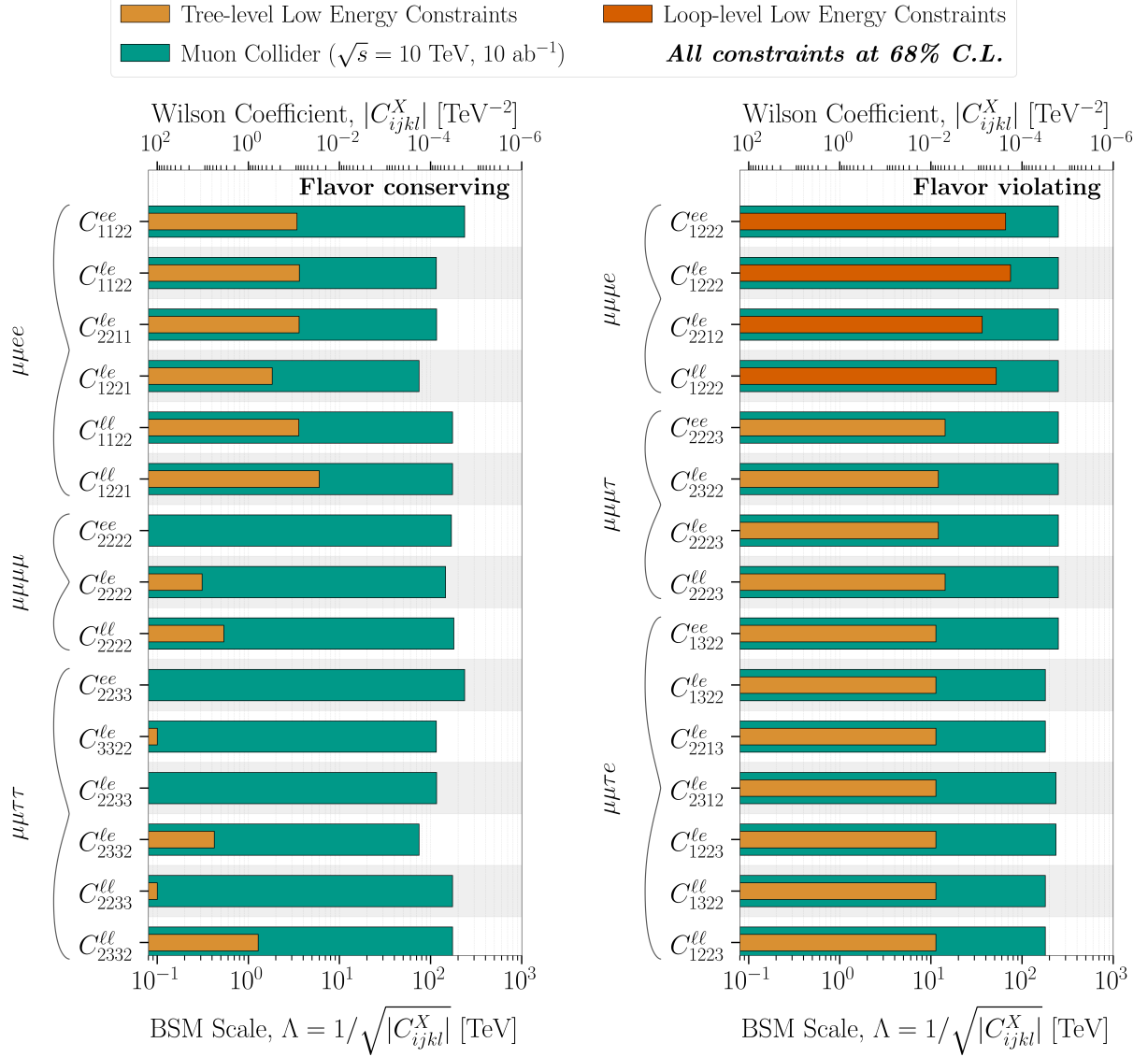}
    \caption{
    Projected single-operator reach of a 10 TeV c.m. $\nu$MuC with 10\ ab$^{-1}$ at 68\% C.L., contrasted with existing (relatively) low-energy constraints. Electroweak PDF implementation is utilized for the muon collider projections, with $Q=\sqrt{\hat{s}}$ and $\xi=1$. Low energy constraints are enumerated with appropriate references in~\cref{tab:SmeftvLowE}. Low energy constraints which are derived at tree-level are shown in light orange, and those which only appear if loop-level corrections are considered appear in dark orange. The grouping of operators with adjacent braces indicates the flavor content of each respective set of coefficients.  
    All the relevant constraints are tabulated in~\cref{tab:SmeftvLowE}. 
    }
    \label{fig:FigSMEFTLow}
\end{figure}

In this single-operator analysis, the strongest muon-collider constraints arise from dilepton final states containing two charged leptons. Channels involving neutrinos are comparatively suppressed by the relevant parton luminosities. Moreover, because only one operator is considered, some of the potential complementarity between the different final states is necessarily absent.
The PDF treatment does lead to dependence on the scale choice $Q$ and the momentum fraction cut $1-\xi \,x_W$ (see~\cref{sec:neutrino_pdf}). We find that this leads to, at most, a 4\% variation in the extracted sensitivity to the WCs considered. This value was obtained by considering benchmarks for all possible combinations of $Q\in \{\sqrt{s}, \sqrt{s}/2, \sqrt{\hat{s}}, \sqrt{\hat{s}}/2\}$ and $\xi\in \{1, 4\}$ and examining the spread of each coefficient's single-operator sensitivity.

\cref{fig:FigSMEFTLow} contains the projected 68\% C.L. single-operator sensitivities of the $\nu$MuC using the processes listed in~\cref{tab:WCDEF}.
For comparison, we also show the existing constraints from low-energy flavor measurements (orange and dark orange bars),  where available (see~\cref{App:LEconstraints} for details). Low-energy interactions are typically constrained using a Weak Effective Field Theory~(WEFT) framework, with a basis of fields and associated WCs that are dynamical below the electroweak symmetry breaking scale. In WEFT, all the SM heavy particles (electroweak gauge bosons, Higgs, and the top quark) are integrated out. These constraints are translated to constraints on SMEFT coefficients through a procedure of running and matching to the high-energy scale (See e.g.~\cite{Falkowski:2021bkq} and other references in~\cref{App:LEconstraints}). These currently represent some of the strongest constraints on a subset of the leptonic WC's of interest. For most low energy constraints summarized in this section, a tree-level matching was assumed to translate the WEFT constraint into the bounds on the SMEFT WCs. However, as it will be described below, 
there is a subset of charge-lepton-flavor violating interactions which does not mediate kinematically allowed process at tree-level, and the relevant observables are first sensitive to these operators at one loop. 
We indicate these loop-induced constraints using darker-orange bars in~\cref{fig:FigSMEFTLow}. In~\cref{tab:SmeftvLowE} we tabulate all the single-operator sensitivities of the $\nu$MuC and the relevant low-energy constraints, along with the related references.

For all four--lepton operators considered here, no existing experiment can match the projected $\nu$MuC reach; for all cases, the 10~TeV $\nu$MuC is substantially more sensitive than the best available low-energy bounds.  For some of the flavor conserving operators (left panel of~\cref{fig:FigSMEFTLow}), $\nu$MuC can constrain $\Lambda$ (\emph{i.e.}~\cref{eq:dim} with $c^X_{ijkl}=1$) values up to $235~$TeV (for both $C^{e e}_{1122}$ and $C^{ee}_{2233}$). By contrast, the strongest low-energy constraint is only $6~$TeV for the $C^{\ell \ell}_{1221}$ operator, in which case $\nu$MuC can probe it up to $174~$TeV.

For lepton-flavor-violating (LFV) new physics, $\nu$MuC can probe $\Lambda$ values up to $\sim180-250$~TeV. For interactions that can mediate tree-level LFV processes, one can find sensitivity around $11-14$~TeV from existing low-energy experiments (light orange bars in the right panel of~\cref{fig:FigSMEFTLow}), among which the best upper bound comes from the lepton flavor violating tau decay probed by the Belle-II experiment~\cite{Belle-II:2025urb}. Belle-II expects to improve the upper bounds on the branching fractions of the LFV tau decays by a factor of 70-80 after collecting $50~$ab$^{-1}$ of data~\cite{Belle-II:2018jsg}. This translates into an upper bound of approximately $40~$TeV for the relevant scale of new physics, which will still be five times smaller than the reach of $\nu$MuC.

The discussion in the last paragraph is restricted to tree-level comparisons, where the low-energy constraints and the $\nu$MuC projections are unambiguous and directly comparable. The flavor-violating operators with WCs $C^{ee}_{1222}$, $C^{\ell e}_{1222}$, $C^{\ell e}_{2212}$ and $C^{\ell\ell}_{1222}$, however, are only constrained by low-energy experiments at the one-loop level (dark orange bars in the right panel of~\cref{fig:FigSMEFTLow}). From $\mu\to e\gamma$, $\mu\to e$ conversion on atomic nuclei, and $\mu\to 3e$, one can currently extract $16-75~$TeV bounds on the effective new physics scale, compared to the (approximate) $250$~TeV reach of $\nu$MuC. Future low-energy experiments (Mu3e~\cite{Mu3e:2020gyw}, Mu2e~\cite{Mu2e:2014fns} and COMET~\cite{COMET:2018auw} ) are expected to be sensitive to new-physics scales around $100-600~$TeV, comparable to or exceeding the $\nu$MuC reach computed here (see~\cref{App:LEconstraints} and especially Ref.~\cite{Asadi:2025dii} for more details). 

There are effective operators not currently constrained by low-energy experiments. For flavor-conserving operators, they are $C^{ee}_{2222}$, $C^{ee}_{2233}$ and $C^{\ell e}_{2233}$. The operator $C^{ee}_{2222}$ can be constrained at the LHC through measurements of $Z\to 4\mu$, with a lower bound on $\Lambda$ of $60~$GeV~\cite{Boughezal:2020klp}. This is several orders of magnitude weaker than the reach of $\nu$MuC, $\sim170~$TeV. On the other hand, $C^{ee}_{2233}$ could also be similarly probed through measurements of $Z\to 2\mu2\tau$; however there is currently no existing constraint in the literature of which we are aware, and thus dedicated searches would be motivated.

\subsection{The Phenomenological impact of neutrino-initiated channels}
\label{sec:SMEFT_multi}
At the $\nu$MuC, hard collisions can occur between the parton pairs $\mu^+\mu^-$, $\mu\nu_\mu$, and $\nu_\mu\bar{\nu}_\mu$. The electroweak PDF approach accounts for electroweak radiative corrections to the conventional $\mu^+\mu^-$ channel, and incorporates the additional $\mu\nu_\mu$ and $\nu_\mu\bar{\nu}_\mu$ channels. Since the neutrino PDF is much smaller than the muon PDF, \emph{i.e.}, $f_{\nu_\mu/\mu}(x,Q^2) \ll f_{\mu/\mu}(x,Q^2)$, the $\nu_\mu\bar{\nu}_\mu$ channel provides only a small correction to a conventional analysis based solely on $\mu^+\mu^-$ collisions. On the other hand, the $\mu\nu_\mu$ channel provides complementary information that can lift degeneracies in parameter space. In this subsection, we present a detailed analysis of two-dimensional parameter spaces to demonstrate the impact of neutrino-initiated collisions at the $\nu$MuC.

\begin{figure}[t]
    \centering
    \includegraphics[width=0.48\textwidth]{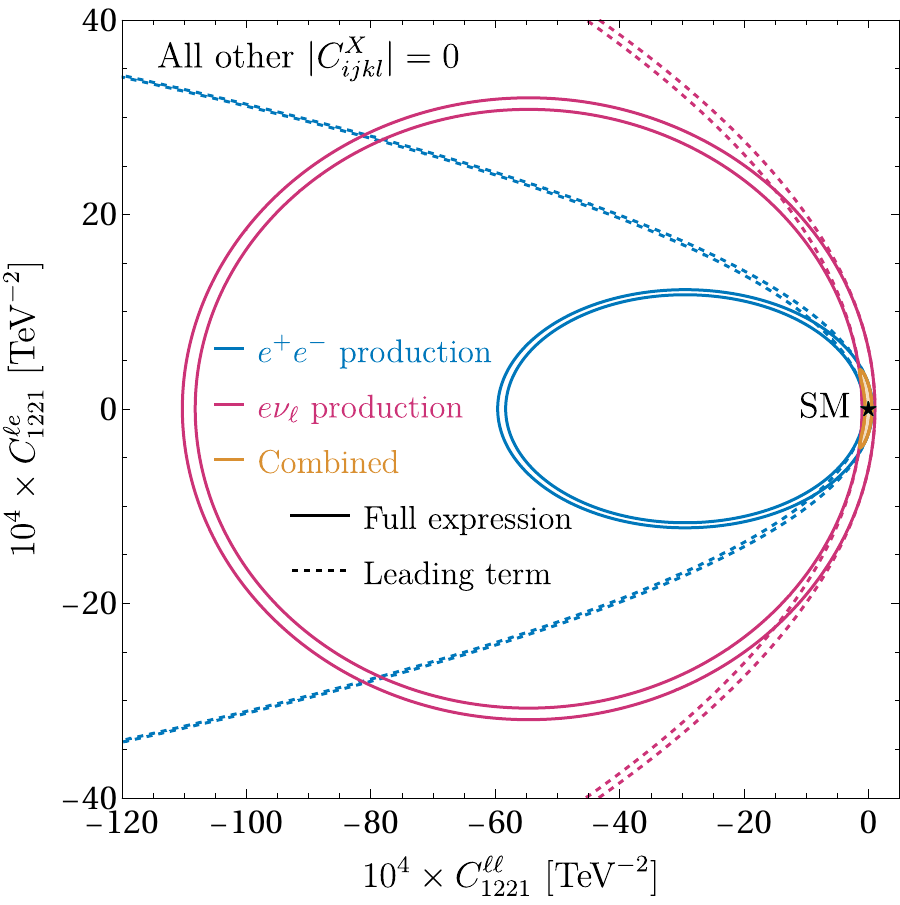}
    \includegraphics[width=0.485\textwidth]{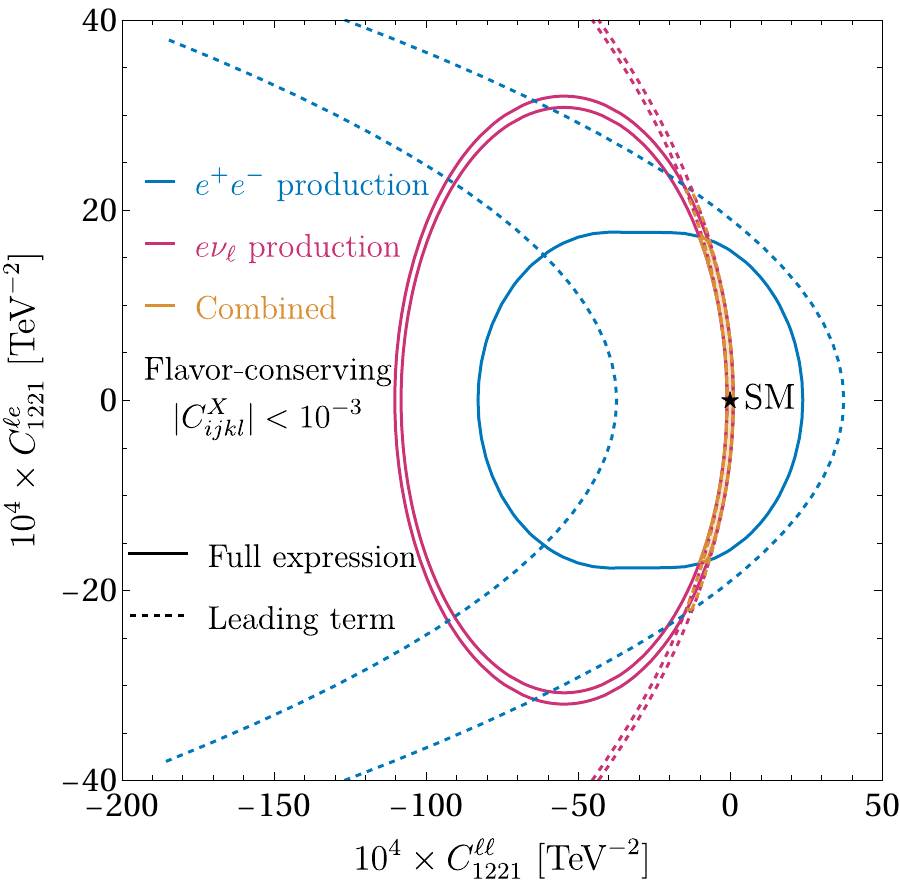}
    \includegraphics[width=0.48\textwidth]{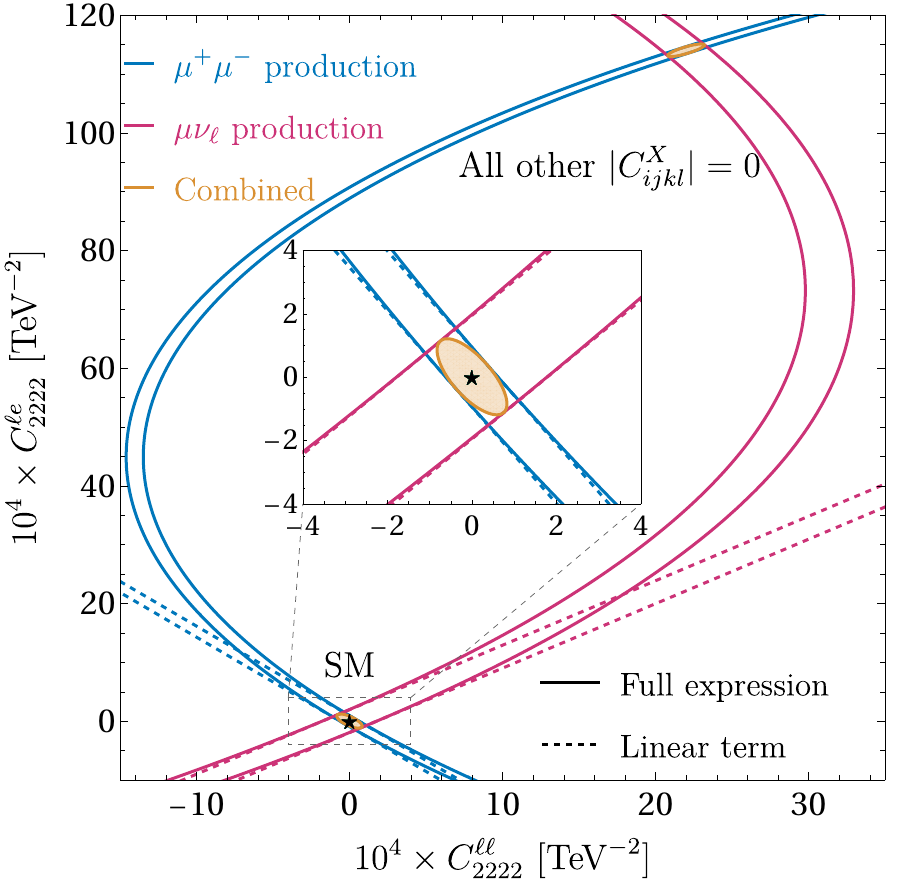}
    \includegraphics[width=0.48\textwidth]{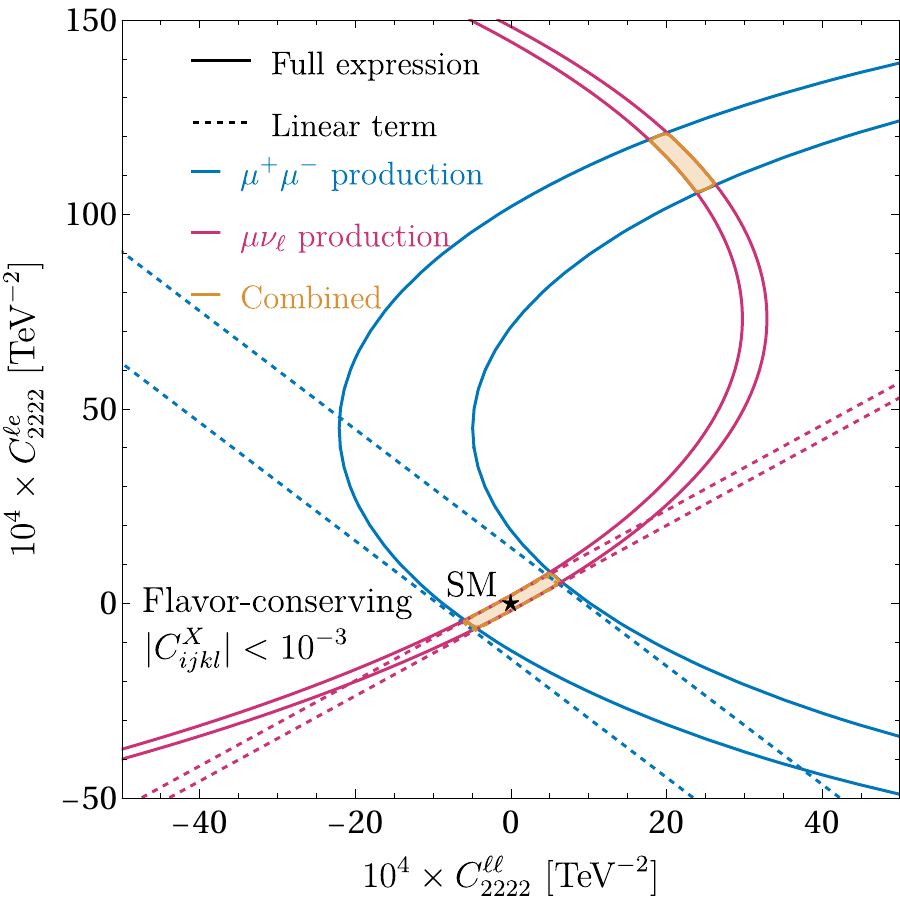}
    \caption{The 95\% C.L. sensitivity contours for electron-flavor production (upper panels) and muon-flavor production (lower panels) in the $(C_{1221}^{\ell\ell},\,C_{1221}^{\ell e})$ and $(C_{2222}^{\ell\ell},\,C_{2222}^{\ell e})$ planes, respectively. The left panels assume that all WCs not illustrated vanish, while the right panels allow flavor-conserving WCs not illustrated to vary within the region $|C_{ijkl}^X|<10^{-3}$~TeV$^{-2}$, while fixing the flavor-violating WCs to zero. The solid contours correspond to the full expression, while the dashed contours correspond to the linear term only for the muon-flavor and  leading term only for electron-flavor productions. The blue, pink, and yellow contours correspond to $\ell^+\ell^-$ pair production, single-lepton production, and the combined sensitivity, respectively. The yellow shaded regions indicate the combined sensitivity of both channels obtained from their full expression. The SM point is marked by a black star.}
    \label{fig:newchannel1}
\end{figure}

In the following analyses, we consider two benchmark prescriptions for the BSM signal in the dimension-six SMEFT. 
\begin{itemize}
    \item {\bf Linear term:} This prescription includes only the SM--BSM interference contribution, $\sigma_{\rm int}^{(6)}$, which is linear in the dimension-six WCs and represents the leading BSM contribution at $\mathcal{O}(\Lambda^{-2})$ in the SMEFT power counting. For operators whose linear interference contribution vanishes, we instead retain the quadratic contribution, which then provides the leading nonzero BSM effect. We refer to this prescription as the ``leading term'' prescription. 
    \item {\bf Full expression:} This prescription includes both $\sigma_{\rm int}^{(6)}$ and the pure BSM contribution, $\sigma_{\rm quad}^{(6)}$, which is quadratic in the dimension-six WCs.
\end{itemize}
\begin{figure}[t]
    \centering
    \includegraphics[width=0.46\textwidth]{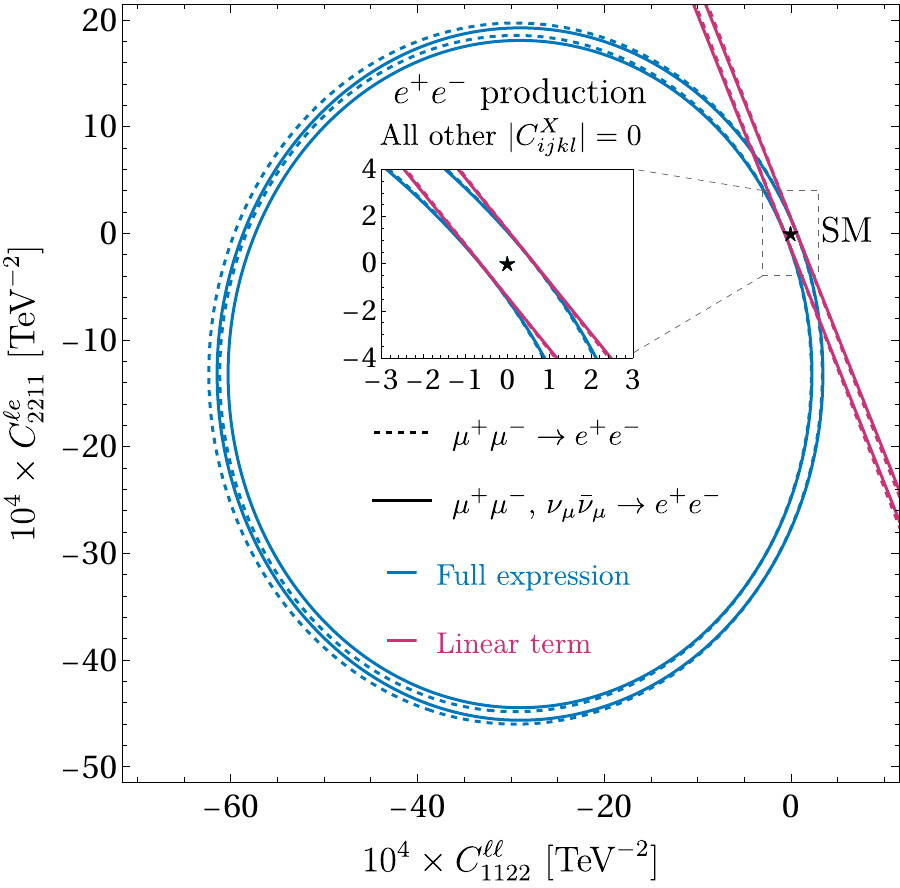}
    \includegraphics[width=0.48\textwidth]{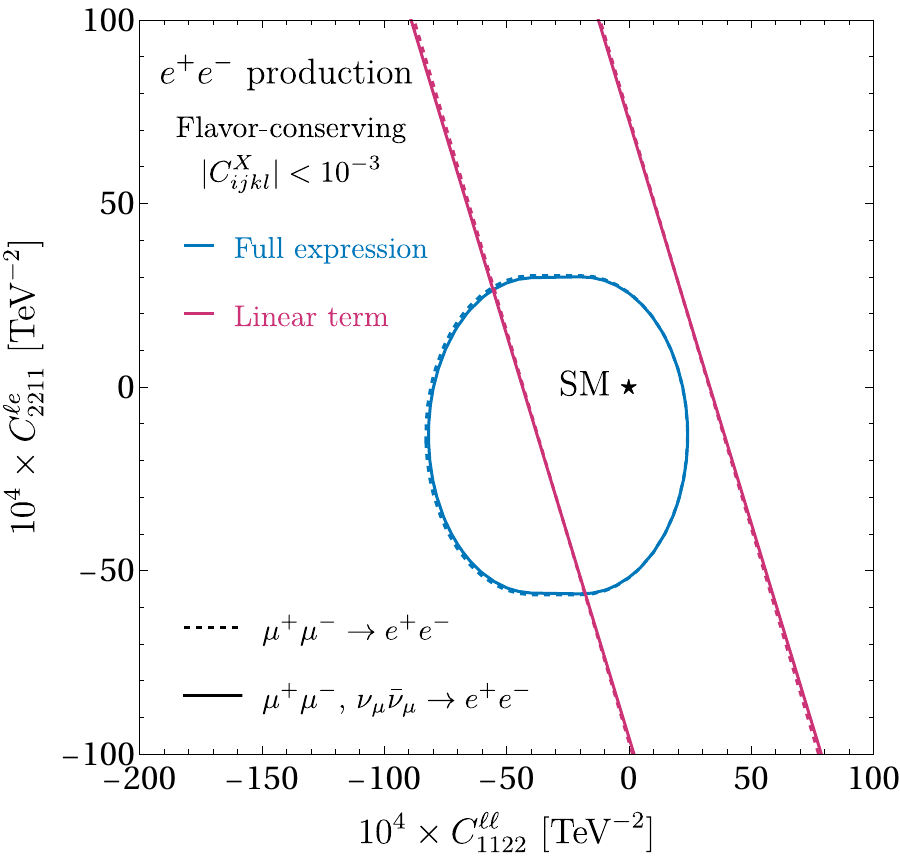}
    \includegraphics[width=0.465\textwidth]{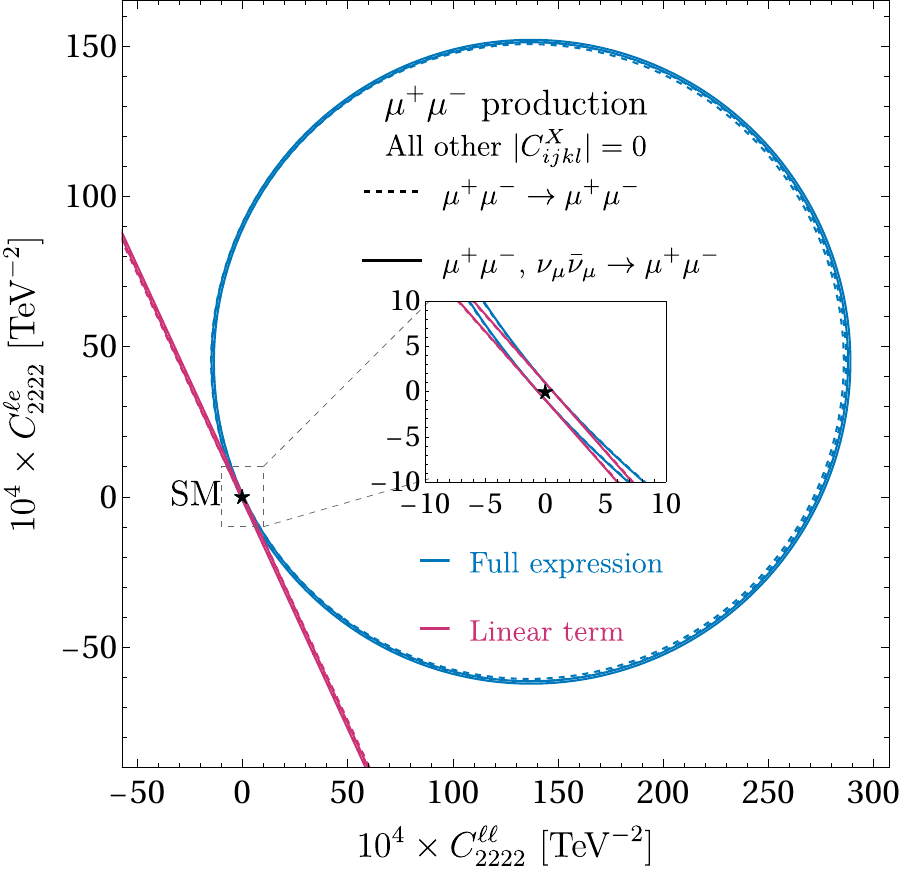}
    \includegraphics[width=0.48\textwidth]{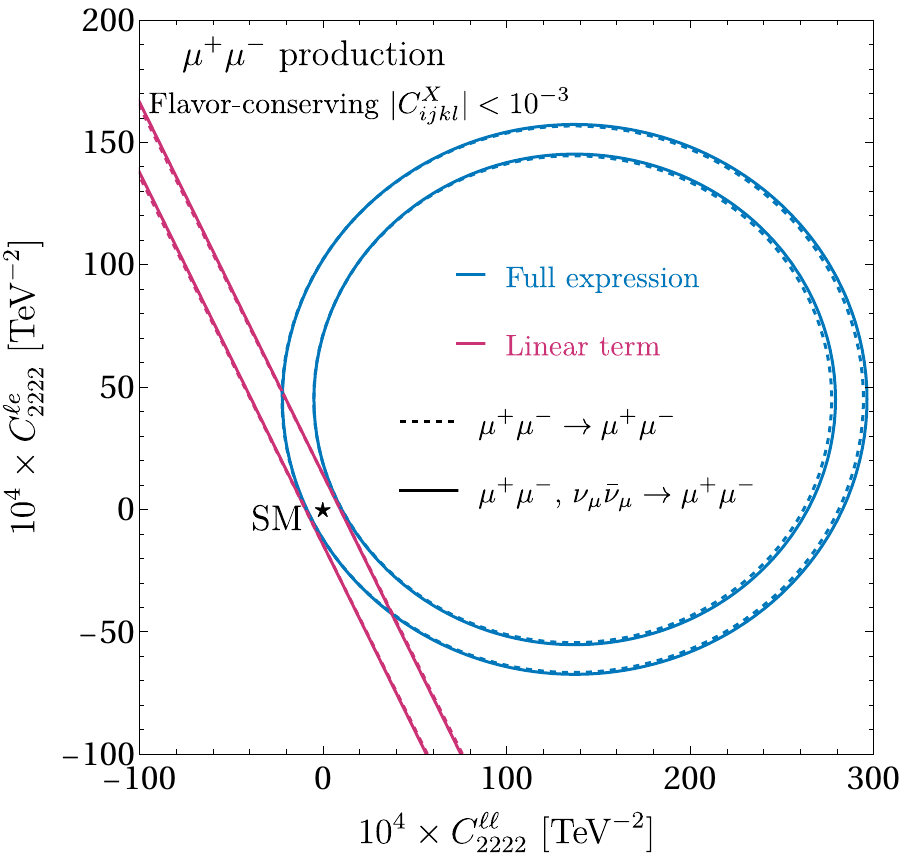}
    \caption{The 95\% C.L. sensitivities of $e^+e^-$ production (top panels) and $\mu^+\mu^-$ production (bottom panels). The left panels assume $C_{ijkl}^X=0$ for all WC not illustrated on the axes, whereas the right panels adopt a prior on all not-illustrated flavor-conserving WCs of $|C_{ijkl}^X|<10^{-3}$ [TeV$^{-2}$] with flavor-violating WCs fixed to zero. The blue (pink) contours are for the full expression (linear term only). The solid contours include both the $\mu^+\mu^-$ collision channel and the $\nu_\mu\bar{\nu}_\mu$ collision channel, while the dashed contours correspond to the conventional $\mu^+\mu^-$ -- only analysis. The SM point is marked by a black star.}
    \label{fig:nunu}
\end{figure}
These define the conventions used in~\cref{fig:nunu,fig:newchannel1,fig:newchannel2}. Significant differences between the constraints obtained using these two prescriptions indicate that the $\mathcal{O}(\Lambda^{-4})$ contributions may be sizable. In such cases, a consistent analysis at this order should also account for the SM--BSM interference involving dimension-eight operators $\sigma_{\rm int}^{(8)}$.

The cross sections considered in this work are sensitive to more than two WCs. When presenting sensitivities in the $(C_{abcd}^X,\,C_{a'b'c'd'}^{X'})$ plane, with $X, X'\in \{ee, \ell e, \ell\ell\}$ we set \emph{all other} WCs to zero, i.e.
\begin{align}C_{ijkl}^Y=0\;\;\text{for}\;\; C_{ijkl}^Y \notin \{C_{abcd}^X,\,C_{a'b'c'd'}^{X'}\}, \end{align}
with $Y\in \{ee, \ell e, \ell\ell\}$. To represent more conservative constraints, we also consider a marginalization scheme in which flavor-conserving WCs are allowed to vary within the prior $|C_{ijkl}^Y|<10^{-3}$ TeV$^{-2}$, while all flavor-violating WCs are fixed to zero. The $|C_{ijkl}^Y|<10^{-3}$ TeV$^{-2}$ prior is chosen to be one order of magnitude larger than the weakest 95\% C.L. single-operator bound presented in~\cref{tab:SmeftvLowE}.\footnote{Because of the unresolved flat directions, the marginalization depends strongly on the choice of prior. Our results should therefore be taken as a benchmark for illustrative purposes only.}

\subsubsection{Neutrino--neutrino collision}
We first take $e^+e^-$ production and $\mu^+\mu^-$ production as examples to show the numerical impact of including the $\nu_\mu\bar{\nu}_\mu$ collision channels. The 95\% C.L. constraints are shown in~\cref{fig:nunu} as contours in the $(C_{1122}^{\ell\ell},\, C_{2211}^{\ell e})$ (for the  $e^+e^-$ production, top panels) and $(C_{2222}^{\ell\ell},\, C_{2222}^{\ell e})$ (for the $\mu^+\mu^-$ production, bottom panels) planes for these two processes. These cross sections are functions of subsets of the WCs as follows:
\begin{align}
    &\sigma_{e^+e^-}=\sigma_{\mu^+\mu^-\to e^+e^-}(C_{1122}^{\ell\ell},C_{1221}^{\ell\ell},C_{1122}^{\ell e},C_{1221}^{\ell e},C_{2211}^{\ell e},C_{1122}^{e e})+\sigma_{\nu_\mu \bar{\nu}_\mu \to e^+e^-} (C_{1122}^{\ell\ell},C_{2211}^{\ell e})\,, \\
    &\sigma_{\mu^+\mu^-}=\sigma_{\mu^+\mu^-\to \mu^+\mu^-}(C_{2222}^{\ell\ell},C_{2222}^{\ell e},C_{2222}^{e e})+\sigma_{\nu_\mu \bar{\nu}_\mu \to \mu^+\mu^-} (C_{2222}^{\ell\ell},C_{2222}^{\ell e})\,.  \label{eq:SigmasXsec}
\end{align}
By replacing the electron-flavor index $1$ with the $\tau$-flavor index $3$ in the $e^+e^-$ production results, one obtains the corresponding constraints for $\tau^+\tau^-$ production, assuming a 100\% $\tau$-tagging efficiency. The blue (pink) contours are for the full expression (linear term only). The solid contours correspond to the analyses that include both the $\mu^+\mu^-$ collision channel and the $\nu_\mu\bar{\nu}_\mu$ collision channel, while the dashed contours are for the conventional $\mu^+\mu^-$-- only analysis, in which the muon PDF is also applied to account for the radiative corrections. 

We begin by drawing the reader's attention to the top and bottom left panels in~\cref{fig:nunu}, which follow the assumption that all WCs not illustrated are fixed to zero. In these figures, the full-expression contours form ring-shaped allowed regions, with the allowed region lying between the inner and outer circles. This behavior arises from the interplay between the linear and quadratic contributions in the dimension-six WCs, which can cancel each other and lead to a degeneracy along the ring. The ring is narrower for $\mu^+\mu^-$ production because its SM--BSM interference is larger than that in $e^+e^-$ production. Near the SM point at the origin, which is marked by a black star, the full-expression contours overlap with those for the linear-term-only case, showing the validity of the SMEFT power counting in this region. We observe that, around the SM point, the addition of the $\nu_\mu\bar{\nu}_\mu$ collision channels has limited numerical impact: the solid and dashed curves overlap in the linear-term-only case and for the full expression in the linear-term-dominant region. Farther from the SM point, along the ring-shaped allowed region, the differences are more pronounced, reflecting the change in the interference contribution induced by the $\nu_\mu\bar{\nu}_\mu$ channel. In this region, the dimension-eight interference contribution, $\sigma_{\rm int}^{(8)}$, should also be considered for a consistent EFT expansion at $\mathcal{O}(\Lambda^{-4})$. However, as we will show below, this region can be excluded by combining information from different processes. 

In our flavor-conserving $C_{ijkl}^Y<10^{-3}$ TeV$^{-2}$ marginalization scheme (the right panels in~\cref{fig:nunu}), the conclusion regarding the $\nu_\mu\bar{\nu}_\mu$ channel remains unchanged; although this channel should be included in principle, it has little numerical impact in practice.
Marginalization weakens the constraints and its effects differ between the two processes. Due to the strong dominance of the linear contribution in $\mu^+\mu^-$ production, the full-expression contours remain ring-shaped after the marginalization, and still overlap with the linear-term-only contours near the SM point. In contrast, the prior $|C_{ijkl}^Y|<10^{-3}$ TeV$^{-2}$ pushes the allowed region for $e^+e^-$ production to where quadratic contributions become important: the full-expression contour forms a closed region rather than a ring and no longer fully overlaps with the linear-term-only contour. This indicates that, when the other WCs are only weakly constrained, the analysis of $\ell^+\ell^-$ production can extend into a region where dimension-eight contributions may be relevant and should be considered for a consistent EFT expansion at $\mathcal{O}(\Lambda^{-4})$. As will be demonstrated below, combining the $\ell^+\ell^-$ production processes with the $\mu\nu_\mu$ collision channel restores the constraints to a region dominated by the linear dimension-six contributions.

\subsubsection{Muon--neutrino collision}
Including processes initiated by $\mu\nu_\mu$ collisions introduces new observables for probing the lepton sector at the $\nu$MuC. These processes include both $\mu^+\nu_\mu$ and $\mu^-\bar{\nu}_\mu$ collisions, which give rise to single-charged-lepton production, $\mu^+\nu_\mu\to\ell^+\nu_\ell$ and $\mu^-\bar{\nu}_\mu\to\ell^-\bar{\nu}_\ell$. These channels probe combinations of WCs that are complementary to those probed by $\ell^+\ell^-$ pair production, thereby helping to lift flat directions in parameter space. We take electron-flavor production ($e^+e^-$ and $e\nu_\ell$ production) and muon-flavor production  ($\mu^+\mu^-$ and $\mu\nu_\ell$ production) as examples to illustrate how this combination can improve the sensitivity in the $(C_{abcd}^{\ell\ell},\,C_{a'b'c'd'}^{\ell e})$ and $(C_{abcd}^{\ell\ell},\,C_{a'b'c'd'}^{ee})$ planes and to demonstrate the relevant features of these results. Again, replacing the electron-flavor index $1$ with the $\tau$-flavor index $3$ yields the corresponding results for $\tau^+\tau^-$ production, assuming a 100\% $\tau$-tagging efficiency.

For both electron-flavor and muon-flavor production, the pair-production and single-lepton channels depend on the corresponding two WCs, and the constraints from the former are stronger than those from the latter because the muon PDF is larger than the neutrino PDF. These cross sections are functions of a subset of the WCs as follows, where we have separated out the flavor-violating and flavor-conserving pieces. 
\begin{align}
    &\sigma_{e \nu_\ell }=\sigma_{\mu \nu_\mu\to e \nu_e}(C_{1221}^{\ell\ell},C_{1221}^{\ell e})+\sigma_{\mu \nu_\mu \to e \nu_{\ell\neq e}} (C_{1222}^{\ell\ell},C_{1223}^{\ell \ell},C_{2212}^{\ell e},C_{2312}^{\ell e})\,, \\
    &\sigma_{\mu\nu_\ell}=\sigma_{\mu^+\nu \to \mu \nu_\mu}(C_{2222}^{\ell\ell},C_{2222}^{\ell e})+\sigma_{\mu \nu_\mu \to \mu \nu_{\ell \neq \mu}} (C_{1222}^{\ell\ell},C_{2223}^{\ell \ell},C_{1222}^{\ell e},C_{2322}^{\ell e})\,.  \label{eq:SigmasXsec2}
\end{align}
In this subsection, for all results illustrated, we do not consider nonzero flavor-violating coefficients. 
In~\cref{fig:newchannel1}, we present the 95\% C.L. sensitivities for electron-flavor production (upper panels) and muon-flavor production (lower panels) as contours in the $(C_{1221}^{\ell\ell},\, C_{1221}^{\ell e})$ and $(C_{2222}^{\ell\ell},\, C_{2222}^{\ell e})$ planes, where the blue, pink, and yellow contours correspond to $\ell^+\ell^-$ pair production, single-$\ell$ producion, and the combined sensitivity, respectively. For illustration, we show the sensitivities using both the full expression (solid curves) and the linear term  (dashed curves) prescriptions. The linear term of $C_{1221}^{\ell e}$ vanishes in $\sigma_{e \nu_\ell }$, so that we retain its quadratic term while keeping only the linear term of $C_{1221}^{\ell\ell}$ for the leading BSM contribution. 
Consequently, the ``leading term'' only contours are curved bands with flat directions rather than straight bands. In contrast, the leading BSM contributions to muon-flavor production arise from the interference terms linear in the dimension-six WCs.

Under the assumption that all other WCs vanish (left panels in~\cref{fig:newchannel1}), the full-expression contours for both pair production and single-lepton production exhibit ring-shaped allowed regions in both flavor cases, with the allowed regions lying between the inner and outer circles. These rings arise from cancellations between the linear and quadratic contributions in the dimension-six WCs. By combining the two channels, the degeneracy is broken, and the combined sensitivities lie around the SM point, where the linear term dominates. For muon-flavor production, a small additional allowed region remains far from the SM point, where the quadratic dimension-six contributions become important and dimension-eight effects may also be relevant. We also Near the SM point the bounds on the two WCs from the combined sensitivities are very close to those from the one-operator-at-a-time single-operator bound, as expected.

When the flavor-conserving WCs are allowed to vary within the prior $|C_{ijkl}^Y|<10^{-3}\,{\rm TeV}^{-2}$ (right panels in~\cref{fig:newchannel1}), the resulting constraints on the WCs become weaker. For electron-flavor production, $e^+e^-$ is pushed into the quadratic-term-dominant region, where the solid pink contour is not a ring but a closed circle and is not fully covered by the leading-term-only band near the SM point. However, after combining with $e\nu_\ell$ production, the combined sensitivity returns to the linear-term-dominant region near the SM point. For muon-flavor production, the marginalization broadens the ring-shaped allowed region and enlarges the additional allowed region far from the SM point. Nevertheless, the combined constraint near the SM point from the full expression prescription still overlaps with the linear term only approach.

\begin{figure}[t]
    \centering
    \includegraphics[width=0.48\textwidth]{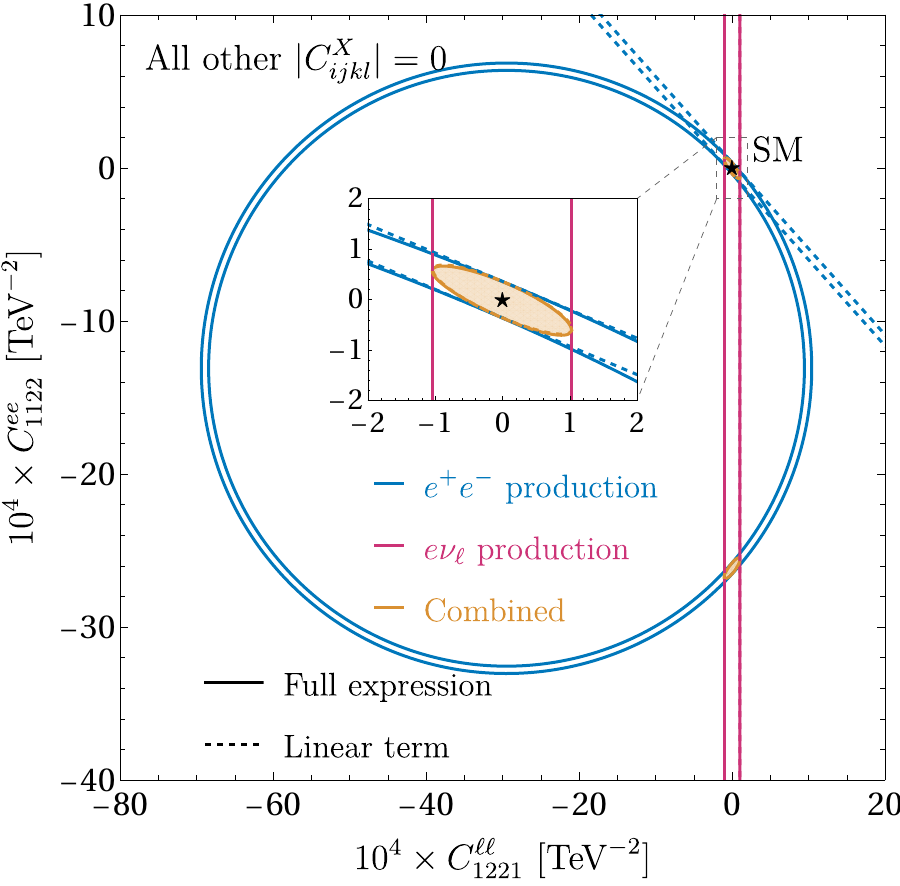}
    \includegraphics[width=0.475\textwidth]{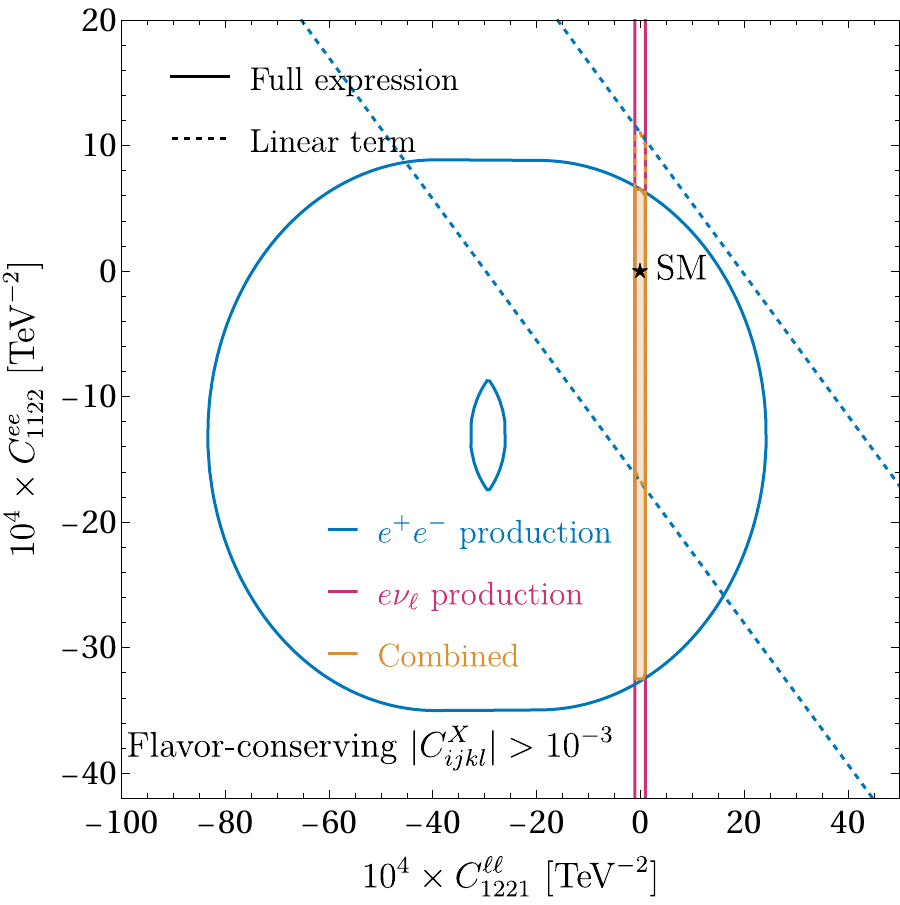}
    \includegraphics[width=0.48\textwidth]{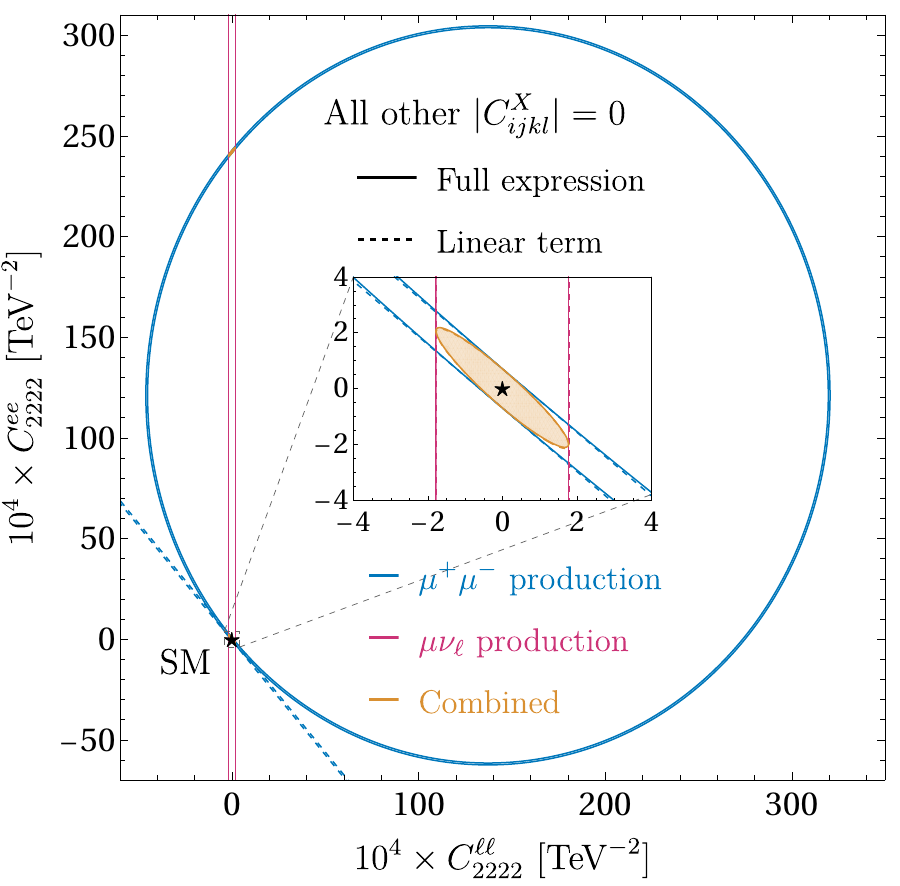}
    \includegraphics[width=0.48\textwidth]{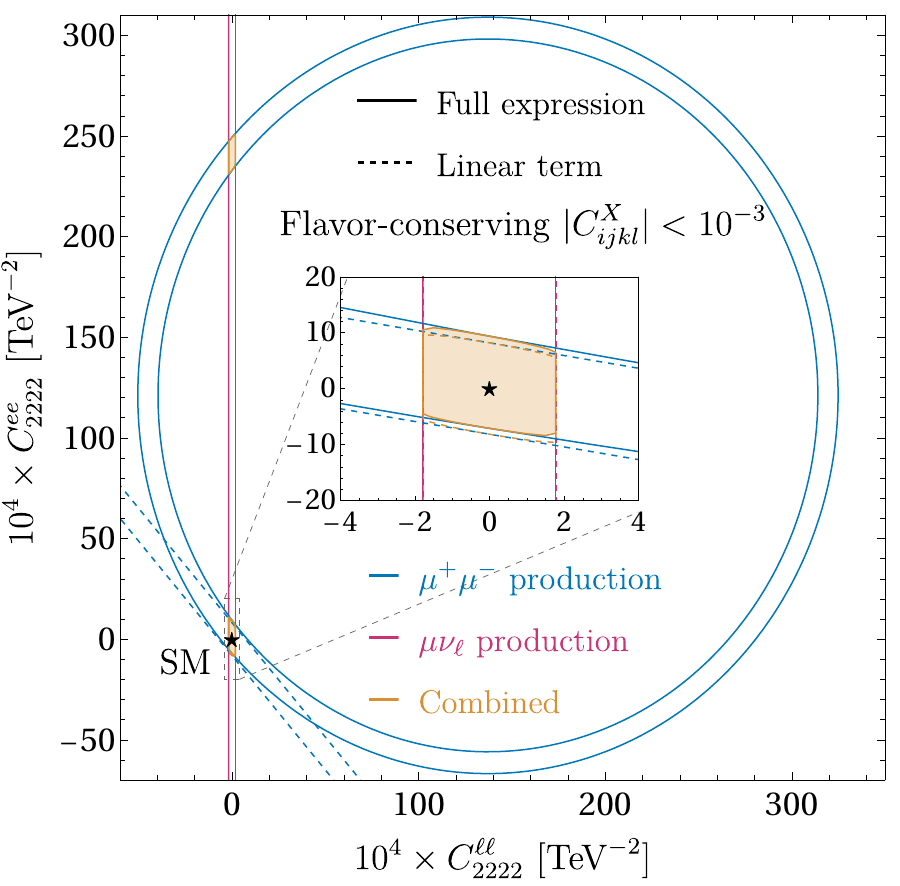}
    \caption{Same as~\cref{fig:newchannel1} but in the $(C_{1221}^{\ell\ell},\,C_{1122}^{e e})$ and $(C_{2222}^{\ell\ell},\,C_{2222}^{e e})$ planes.}
    \label{fig:newchannel2}
\end{figure}

In~\cref{fig:newchannel2}, we present the 95\% C.L. sensitivity contours for electron-flavor production (top panels) and muon-flavor production (bottom panels) in the $(C_{1221}^{\ell\ell},\,C_{1122}^{ee})$ and $(C_{2222}^{\ell\ell},\,C_{2222}^{ee})$ planes, respectively. The $\ell^+\ell^-$ pair production is shown in blue and single-lepton production in pink; solid contours correspond to the full expression and dashed contours to the linear term only. As discussed above, the full-expression contours for pair production exhibit ring-shaped allowed regions due to cancellations between the linear and quadratic contributions in the dimension-six WCs, and the muon-flavor case has thinner rings since it is more linear-term dominant. Since single-lepton production is independent of the $C^{ee}_{ijkl}$ WCs, it exhibits vertical straight bands in these planes.
Combining pair production with single-lepton production breaks the degeneracies and substantially improves the constraints in both flavor cases, even though single-lepton production does not directly constrain the $C^{ee}_{ijkl}$ WCs. This improvement arises because the single-lepton channel constrains the $C^{\ell\ell}_{ijkl}$ direction, thereby eliminating much of the ring-shaped degeneracy left by pair production. A small allowed region still remains far from the SM point, where the quadratic dimension-six contributions become important and dimension-eight effects may also be relevant. For the full-expression prescription, single-lepton production yields two separate allowed bands in each plane. We show only the band near the SM point, since the other band lies farther away along the $C^{\ell\ell}_{ijkl}$ axis and does not intersect the allowed region from pair production. It is therefore excluded when the two channels are combined.

Marginalization over the other flavor-conserving WCs weakens the constraints in both flavor cases, but their effects differ. For muon-flavor production, marginalization broadens the ring-shaped pair-production contours and the bands from single-lepton production, while the combined constraint still provides a substantial improvement. For electron-flavor production, marginalization pushes the pair-production constraint into a region where the quadratic contributions become important, resulting in a much weaker combined constraint. Nevertheless, the newly introduced $\mu\nu_\mu$ channel still substantially improves the sensitivity to BSM effects by lifting the flat direction, even though it is insensitive to one of the two WCs being constrained. This demonstrates that a new channel can provide valuable complementary information and break degeneracies without having direct sensitivity to every WC in the parameter space.

\section{Discovery of heavy vector bosons at a muon collider }
\label{sec:V'}

Marching to the new energy frontier, the hope for a muon collider is to open a new physics threshold. Going beyond the model-independent SMEFT approach, we next consider the resonant production of the new heavy vector bosons. We note that the radiative return mechanism \cite{Denig:2006kj,Chakrabarty:2014pja,Dasgupta:2023zrh} via the initial state radiation (ISR)   will lead to a resonant signal as long as kinematically accessible. 
As in the SMEFT case, we are especially interested in the impact of the different partonic-level leptoninc initial states available at the $\nu$MuC on these complementary  searches. 

Several BSM models predict the existence of, for example, charged ($W_{R}^\pm$) and neutral ($Z_R$) gauge bosons associated to left-right symmetric models \cite{Mohapatra:1974hk}, $Z'$ bosons from U(1)$_{B-L}$ gauge models \cite{Davidson:1978pm}, and Kaluza--Klein excitations of the electroweak gauge bosons from extra-dimensional models \cite{Cheng:2001vd}. Here we use $W'^\pm$ and $Z'$ to generically denote heavy vector bosons mediating new charged- and neutral-current interactions, respectively. As representative benchmarks, we focus on the sequential $W'$ and $Z'$ scenarios \cite{Altarelli:1989ff,Barger:1980dx}. 

\subsection{Sequential \texorpdfstring{$W'$}{W'} boson}
\label{sec:W'}
In the sequential model, the $W'$ boson is a heavy analogue of the SM $W$ boson, with the same  $V-A$ interaction structure to that of the SM interactions but suppressed couplings  in order to avoid constraints from precision electroweak measurements \cite{ParticleDataGroup:2026mpi}.  The relevant interaction can be written as 
\begin{equation}
\label{eq:WprimeLagrangian}
\calL_{W'}\supset-\frac{(g_{W'})_i}{\sqrt{2}}W'_\mu\Big(\bar{e}_{L,i}\gamma^\mu\nu_{L,i}+\bar{d}_{L,i}\gamma^\mu u_{L,i}\Big)+{\rm{h.c.}}\,,
\end{equation}
where $i$ denotes to flavor indices. For simplicity, we assume flavor-universal couplings, $(g_{W'})_i=g_{W'}$, $\forall~i$. We define the ratio of the new coupling to that of the SM as 
\begin{equation}
\zeta_{\rm SSM}\equiv \frac{g_{W'}}{g_{W}}, 
\end{equation}
where $\zeta_{\rm SSM}=1$ corresponds to the original sequential model in Ref.~\cite{Altarelli:1989ff}. The decay width of $W'$ to fermions can be approximated as
\begin{equation}
\label{eq:width}
\Gamma(W')=\Gamma_0(N_\ell+N_qN_c),~~~~~
\Gamma_0=\frac{g^2_{W'}M_{W'}}{48\pi},
\end{equation}
where $N_\ell=N_q=3$ is the number of lepton and quark generations, respectively, and $N_c=3$ is the number of colors. Here we focus on $W'$ masses $M_{W'}\gtrsim 1~\TeV$,  significantly heavier than the masses of all SM fermions, including the top quark. We would also like to keep the $W'$ boson relatively narrow and perturbatively well-defined and hence only consider $W'$ decay widths such that $\Gamma(W')/M_{W'}\lesssim50\%$. This translates into an upper bound on the effective coupling $\zeta_{\rm SSM}\lesssim3.8$.  

\begin{figure}
    \centering
    \includegraphics[width=0.35\linewidth]{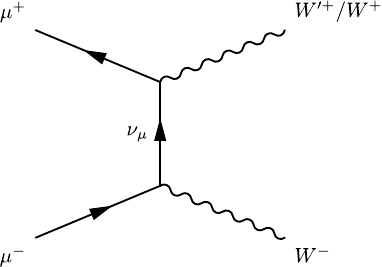} \hspace{1.5cm}
    \includegraphics[width=0.3\linewidth]{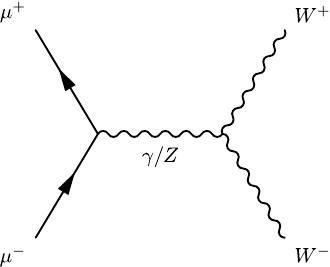}
    \caption{Fixed-order Feynman diagrams, left for $W'^+W^-$ and $W^+W^-$ production; and right for the SM $s$-channel $\gamma/Z$ mediated $W^-W^+$ production. The corresponding $s$-channel $W'^+W^-$ is absent in the SSM, due to the null triple gauge coupling $g_{VW'W}=0$
    ($V=\gamma/Z$). 
    }
    \label{fig:feynWp}
\end{figure}

\begin{figure}[tb]
\centering
\includegraphics[width=0.49\textwidth]{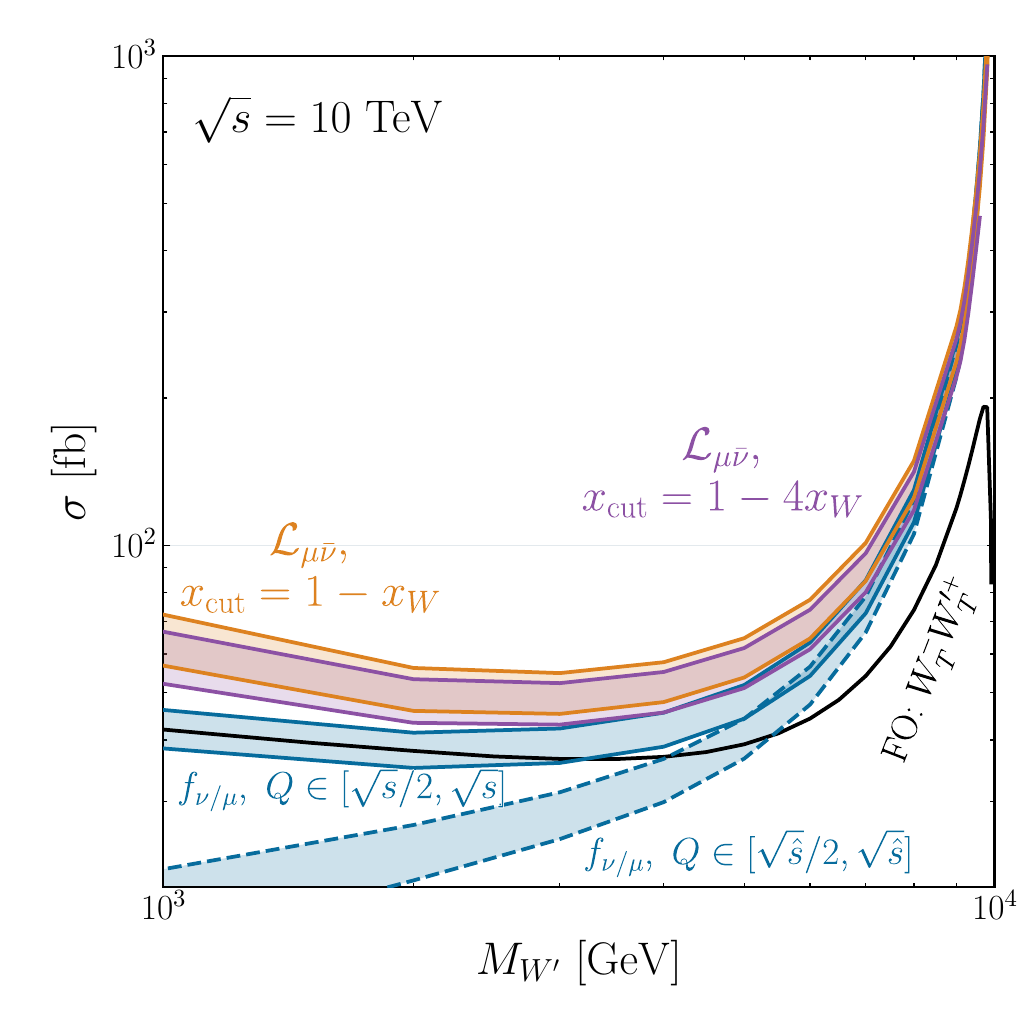}
\includegraphics[width=0.49\textwidth]{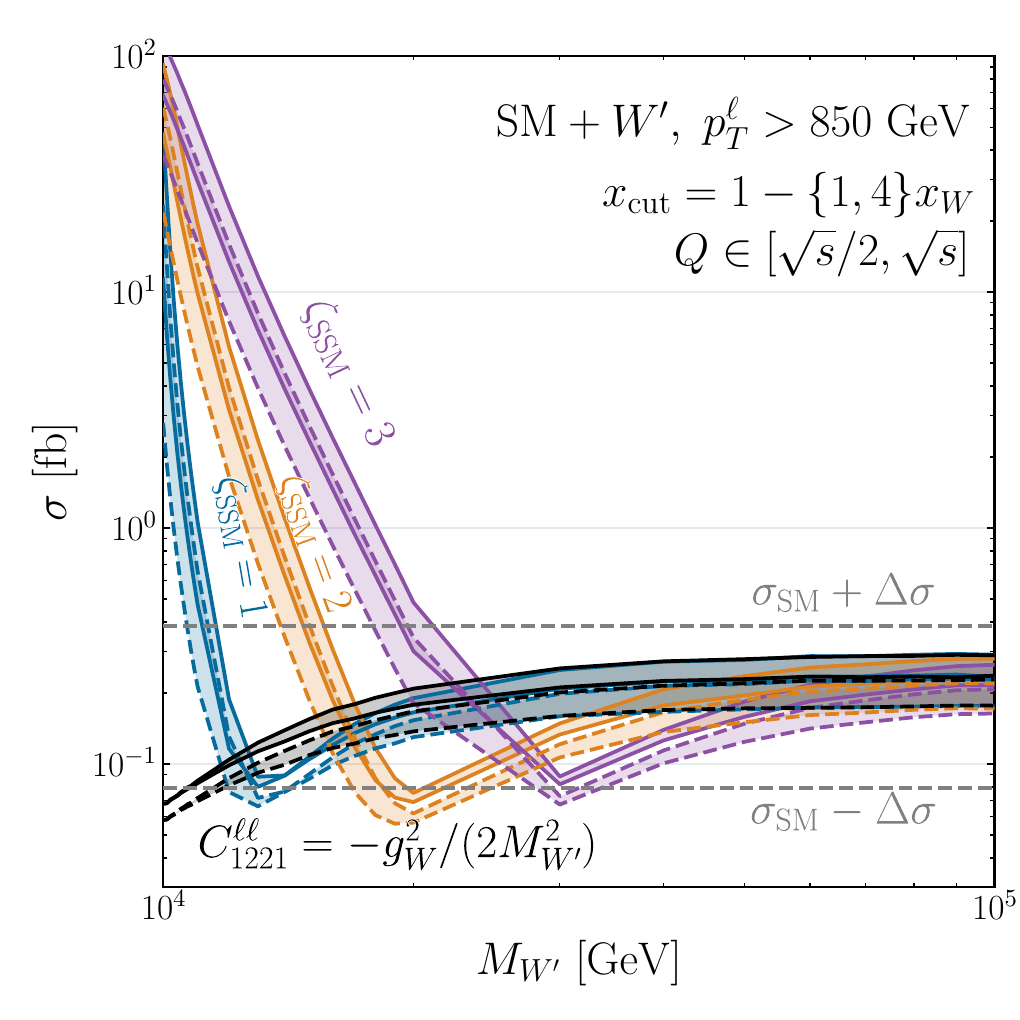}
\caption{For a 10 TeV muon collider, \textbf{Left:} Production cross section of the sequential $W'$ boson versus its mass ($\zeta_{\rm SSM}=1$) in the transverse mode. The FO result is black; the one-sided neutrino-PDF result in~\cref{eq:oneside} is blue. The luminosity $\calL_{\mu\bar{\nu}}$ with $x_{\rm cut}=1-x_W$ and $1-4x_W$ is shown in orange and purple, respectively. The blue dashed band corresponds to $Q\in[\sqrt{\hat{s}}/2,\sqrt{\hat{s}}]$; the solid bands use $Q\in[\sqrt{s}/2,\sqrt{s}]$.
\textbf{Right:} Production cross sections for off-shell $W'$ plus full SM contribution versus the $W'$ mass with $\zeta_{\rm SSM}=1,2,3$ (blue, orange, purple) and the matched SMEFT result (black), with $C^{\ell\ell}_{1221}=-g_{W}^2/(2M_{W'}^2)$. Solid and dashed curves distinguish the two factorization-scale settings in $Q\in[\sqrt{s}/2,\sqrt{s}]$; each shaded band spans $x_{\rm cut}=1-\xi x_W$ ($\xi=1$ or 4) at fixed scale.
The gray dashed lines denote the allowed $2\sigma$ range of $\sigma_{\rm SM}\pm\Delta\sigma$, translated from~\cref{eq:2sigma}. 
}
\label{fig:Wp}
\end{figure}

We first compute $W'^+$ production using the full matrix element for $\mu^+\mu^-\to W_T^-W_T'^{+}$ as depicted in~\cref{fig:feynWp} (left), \emph{i.e.}, the fixed-order (FO) calculation, using \textsc{MadGraph\_aMC@NLO}~\cite{Alwall:2014hca,Frederix:2018nkq}. 
At this point, some care needs to be taken. To be conservative on the signal prediction, we neglect the production of longitudinal $W'$ in the FO calculation. In the absence of a full UV-complete model and the symmetry-breaking dynamics for $W'$-mass generation, the production of the longitudinal gauge boson $W'_L$ grows as $g_{W'}^2 s/M_{W'}^2$, thus leading to unreliable results when $\sqrt s \gg M_{W'}/g_{W'}$. While this energy behavior is expected inside a specific energy range and can be used to test the physics of $W'$ mass-generation, it is beyond the scope of this exercise. The resulting $W_T^-W_T'^+$ cross section as a function of the $W'$ mass is shown in~\cref{fig:Wp} (left panel, black curve). This result will serve as a baseline for the collinear factorization treatment in the following. 

In the collinear limit of $\mu^-\to W_T^-\nu_\mu$ splitting, the $W_T'^+$ production cross section can be factorized as~\cref{eq:xsec} with convolution with the partonic cross section\footnote{There is no spin-average for a neutrino since there is only one contributing helicity state.}
\begin{equation}
\label{eq:Wp}
\hat{\sigma}(\mu^+\nu_\mu\to W'^+)=12\pi g_{W'}^2\delta(\hat{s}-M_{W'}^2)\,.
\end{equation}
Here, we have employed the narrow width approximation (NWA) 
$\Gamma(W')\ll M_{W'}$. 
Using~\cref{eq:mumupdf} as the LO  muon PDF, the cross section can be written as 
\begin{equation}\label{eq:oneside}
\sigma(\mu^+\mu^-\to W'^+) =\int d xf_{\nu_\mu/\mu}(x)\hat{\sigma}(xs)
\approx f_{\nu_\mu/\mu}(\tau)\frac{12\pi g_{W'}^2}{s}\,,
\end{equation}
where $\tau=M_{W'}^2/s$. We consider two scale choices, $Q=\sqrt{s}$ and $Q=\sqrt{\hat{s}}$, and vary the scales by a factor of 1/2 between  $(\sqrt s/2,\sqrt{s})$ and $(\sqrt{\hat s}/2,\sqrt{\hat s})$, respectively. The  corresponding cross sections are represented by  the blue bands bounded by solid and dashed curves  in the left panel of~\cref{fig:Wp}. 

In comparison, we see that the $Q\in (\sqrt s/2,\sqrt{s})$ choice gives a cross section closer to the \textsc{MadGraph\_aMC@NLO} FO prediction, especially in the low $M_{W'}$ region. It can be understood as the $t$-channel integration ends up with a logarithm $\log(s/M_W^2)$, which implies choosing  $Q\sim\sqrt{s}$ would minimize the logarithm in the factorization formalism. For this reason, we will take $Q\sim\sqrt{s}$ in computing the $W'$ production, and use the variation in the range $(Q/2,Q)$ to quantify the corresponding scale uncertainty. In the left panel of~\cref{fig:Wp}  we also include the scale uncertainty from enforcing a cut on  the neutrino momentum fraction cut as $x<x_{\rm cut}=1-\xi x_W$, with $\xi=1$ shown as the orange curves and $\xi=4$ as the purple curves. We see that the one-side neutrino PDF in~\cref{eq:oneside} gives a good approximation already, while the additional resummed muon PDF gives a small enhancement, in particular in the small $M_{W'}$ region.

\begin{figure}[tb]
\centering
\includegraphics[width=0.49\textwidth]{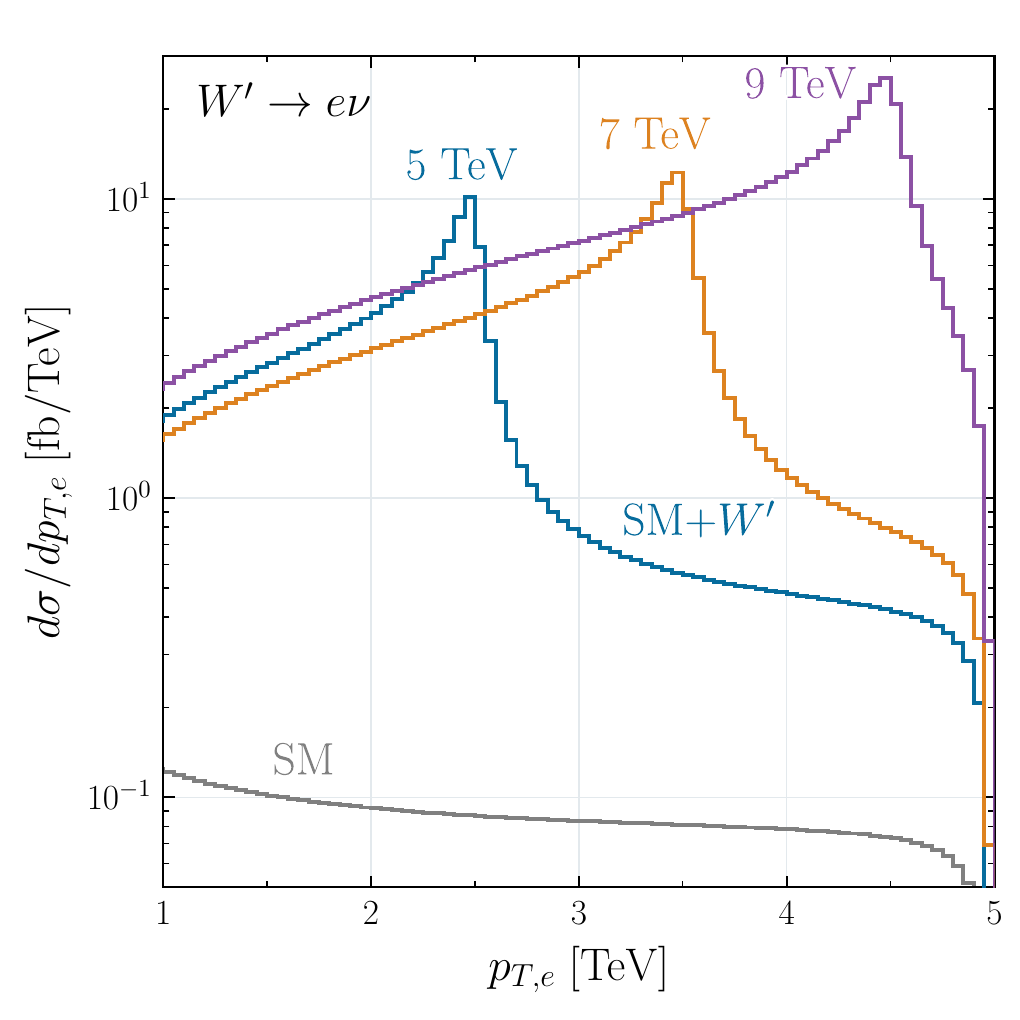}
\includegraphics[width=0.49\textwidth]{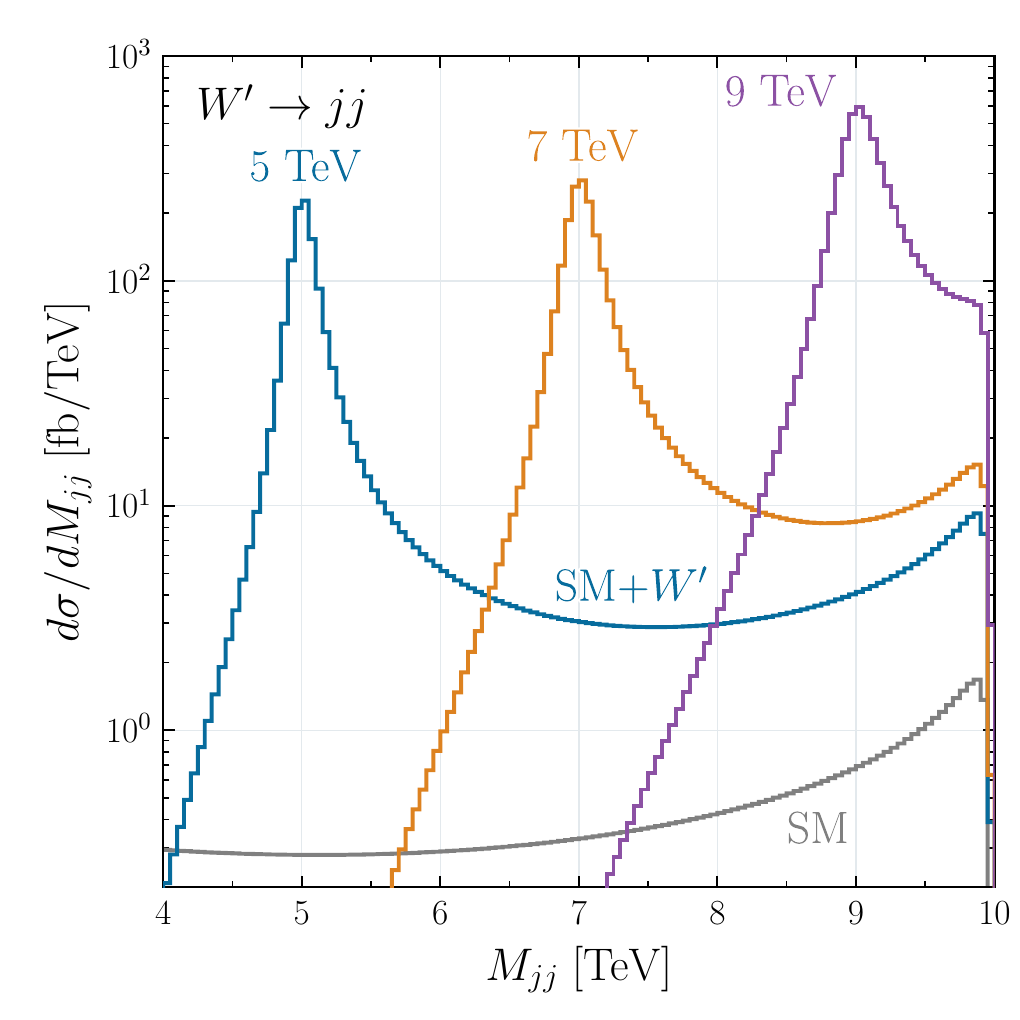}
\caption{Electron transverse momentum distribution for $W'\to e\nu$ (left) and dijet invariant mass distribution for $W'\to jj'$ (right) at a 10 TeV muon collider. Solid blue, orange, and purple curves show the coherent SM+$W'$ predictions, including interference, for $M_{W'}=5,7,9$ TeV, respectively. Gray curves show the SM background in the corresponding decay channel.
}
\label{fig:W}
\end{figure}

It is important to appreciate the key feature of a resonance significantly below the collider energy. In~\cref{fig:W} we show the expectation for the transverse momentum distribution $p_T(e)$ for $W'\to e\bar \nu$ and the dijet invariant mass distribution for $W'\to jj$ for $M_{W'}=5,~7,~9$~TeV at $\sqrt s=10$~TeV. We see the distinct kinematical peak features for the resonant production, standing out above the gray SM background curves. The closer the $W'$ mass is to the c.m.~energy, the larger the production rate is.

If the $W'$ mass lies slightly above the collider center-of-mass energy, \emph{i.e.}, $M_{W'}>\sqrt{s}$, the resonant kinematics above is absent. Its effects can still be probed through virtual production. As a demonstration, we consider only the electronic decay channel, $W'\to e\bar{\nu}$, shown in~\cref{fig:feynmupmum} (middle), while the analysis can be straightforwardly recast for other decay channels. The corresponding cross sections for several representative values of $\zeta_{\rm SSM}$ are shown by the blue, orange, and purple bands for $\zeta_{\rm SSM}=1,2,3$, respectively, as well as a comparison with the SMEFT matching $C^{\ell\ell}_{1221}=-g_{W}^2/(2M_{W'}^2)$ by the black band (\emph{i.e.}, $\zeta_{\rm SSM}=1$) in~\cref{fig:Wp} (right).
As before, we vary the factorization scale over $Q\in[\sqrt{s}/2,\sqrt{s}]$, with solid and dashed curves distinguishing the two scale settings. Each shaded band spans the neutrino momentum-fraction cuts $x_{\rm cut}=1-\{1,4\}x_W$ at fixed scale. In addition, we impose the cut $p_T^\ell>850$~GeV, following the setup in~\cref{sec:SMEFT}.
As $M_{W'}$ increases beyond $\sqrt{s}$, the cross section decreases rapidly, owing to the off-shell $W'$ production. When both the $W'$ and SM $W$ contributions are off shell, their interference becomes non-negligible and is fully included in our calculation. The interference is destructive, further suppressing the cross section and giving rise to the characteristic dip shown in~\cref{fig:Wp}. At still larger $M_{W'}$, the $W'$ contribution gradually decouples, and the cross section asymptotically approaches the SM prediction, $\sigma_{\rm SM}$, as a consequence of the dying-out $W'$ contribution. 

When $M_{W'}\gg\sqrt{s}$, one can integrate out the $W^\prime$ and match the resulting effective theory to SMEFT. The tree-level matching relation between SMEFT and the parameters of the sequential model is
\begin{equation}
\label{eq:WprimeMatching}
C_{ijji}^{\ell\ell}\Big|_{W'}=-\frac{(g_{W'})_{i}(g_{W'})_{j}}{2M_{W'}^2}\,.
\end{equation}
Therefore, integrating out a heavy sequential $W'$ only simultaneously generates the operators
\begin{equation}
    C_{1221}^{\ell\ell}\,,\quad C_{2332}^{\ell\ell}\quad {\rm and}\quad C_{ 2222}^{\ell\ell}\, . 
\end{equation} 
This therefore strongly motivates the complementary searches of the three lepton flavors in the channel of $\mu^\pm \nu$ annihilation at $\nu$MuC.
We note that, since the charged $W'$ field by itself is not an $SU(2)$ gauge-invariant object, in the unbroken SMEFT we can only compare the coefficient with the $(\bar\nu \nu) (\bar e e)$ part of the $O_{\ell\ell}$ operator. This matching allows for a direct comparison of the production rates in $\mu^-\bar{\nu}_\mu\to e^-\bar{\nu}_e$ at $\nu$MuC, calculated from both the $W^\prime$ mediation and the $C^{\ell\ell}_{1221}$ operator with the matching in~\cref{eq:WprimeMatching} and $\zeta_{\rm SSM}=1$. The result is shown in the right panel in~\cref{fig:Wp}. We see that the cross section of $C_{1221}^{\ell\ell}$ operators matches to the off-shell $W'$ one very well, except a small region around $M_{W'}\sim\sqrt{s}$, due to the residual resonance effect.

To estimate the projected sensitivity, we take the background-free assumption for the on-shell $W'$ production since any potential background events  can be effectively suppressed, \emph{e.g.}, by examining the kinematic distributions and via a definite mass window cut.
The signal event number is calculated as  $S=2\mathcal{L}_{\rm int}\sigma$, where the factor of $2$ comes from the sum of positive and negative charges of the $W'$ boson, and the integrated luminosity is taken as $\mathcal{L}_{\rm int}=10~\textrm{ab}^{-1}$~\cite{MuCoL:2025quu}.
As discussed in~\cref{sec:SMEFT}, we take the signal event number $S=14.4$ to define the $5\sigma$ discovery, and a 95\% confidence level (C.L.) exclusion limit corresponding to $S=3.09$, in a background-free environment~\cite{ALRead_2002,ParticleDataGroup:2026mpi}. 

For the off-shell production, the SM background event number is about $B=2\mathcal{L}_{\rm int}\sigma_{\rm SM}=5.2\times 10^{3}$, where $\sigma_{\rm SM}$ is shown in~\cref{fig:Wp}  in the right panel (the horizontal asymptote), and again the factor of 2 indicates the sum of the positive and negative charges. The signal event is determined by $S=2\mathcal{L}_{\rm int}|\sigma_{\rm BSM}-\sigma_{\rm SM}|$. 
Slightly different from the measure in~\cref{eq:signal_strength} or~\cref{eq:calS2}, we adjust the 95\% C.L. sensitivity as
\begin{equation}
\label{eq:2sigma}
\mathcal{S}\simeq\frac{S}{\sqrt{B+(\Delta B)^2}}=2,
\end{equation}
to include the statistical error $\sqrt B$ and the systematic errors added in quadrature $\Delta B=\sqrt{\Delta B^2(x_{\rm cut})+\Delta B^2(Q)}$. Here, $\Delta B(x_{\rm cut})$ and $\Delta B(Q)$ correspond to the estimated background event number variation from the neutrino momentum fraction cut $x_{\rm cut}=1- \xi x_W$ and the scale choice $Q\in[1/2,1]\sqrt{s}$.
With such a criterion, we can derive the envelope of the BSM cross sections as $\sigma_{\rm BSM}=\sigma_{\rm SM}\pm\Delta\sigma$ (where $\Delta\sigma=S/2\mathcal{L}_{\rm int}$), 
shown as the gray dashed lines in~\cref{fig:Wp} (right). The exclusion limit on the $M_{W'}$ with different $\zeta_{\rm SSM}$ as well as SMEFT operator $C_{1221}^{\ell\ell}$ can be directly read out from the intersection points with the gray dashed lines. More explicitly, we obtain the $|C_{1221}^{\ell\ell}|<9.7\times10^{-4}~\TeV^{-2}$ at 95\% C.L.\footnote{The two-side bound is asymmetric as $-9.7\times10^{-4}<C_{1221}^{\ell\ell}<7.7\times10^{-4}~\textrm{TeV}^{-2}$. Here, the absolute one corresponds to the larger value and negative direction, matching to the $W'$ model.} If we assume $\Delta B=0$, we obtain the 95\% C.L. bound as $|C_{1221}^{\ell\ell}|<1.1\times10^{-4}~\TeV^{-2}$, consistent with the $e\nu_\ell$ bound in~\cref{fig:newchannel1} and~\cref{fig:newchannel2}. In comparison, the bound displayed in~\cref{tab:SmeftvLowE} in the appendix comes from a stronger bound from the $e^+e^-$ channel. 

\begin{figure}
    \centering
    \includegraphics[width=0.6\textwidth]{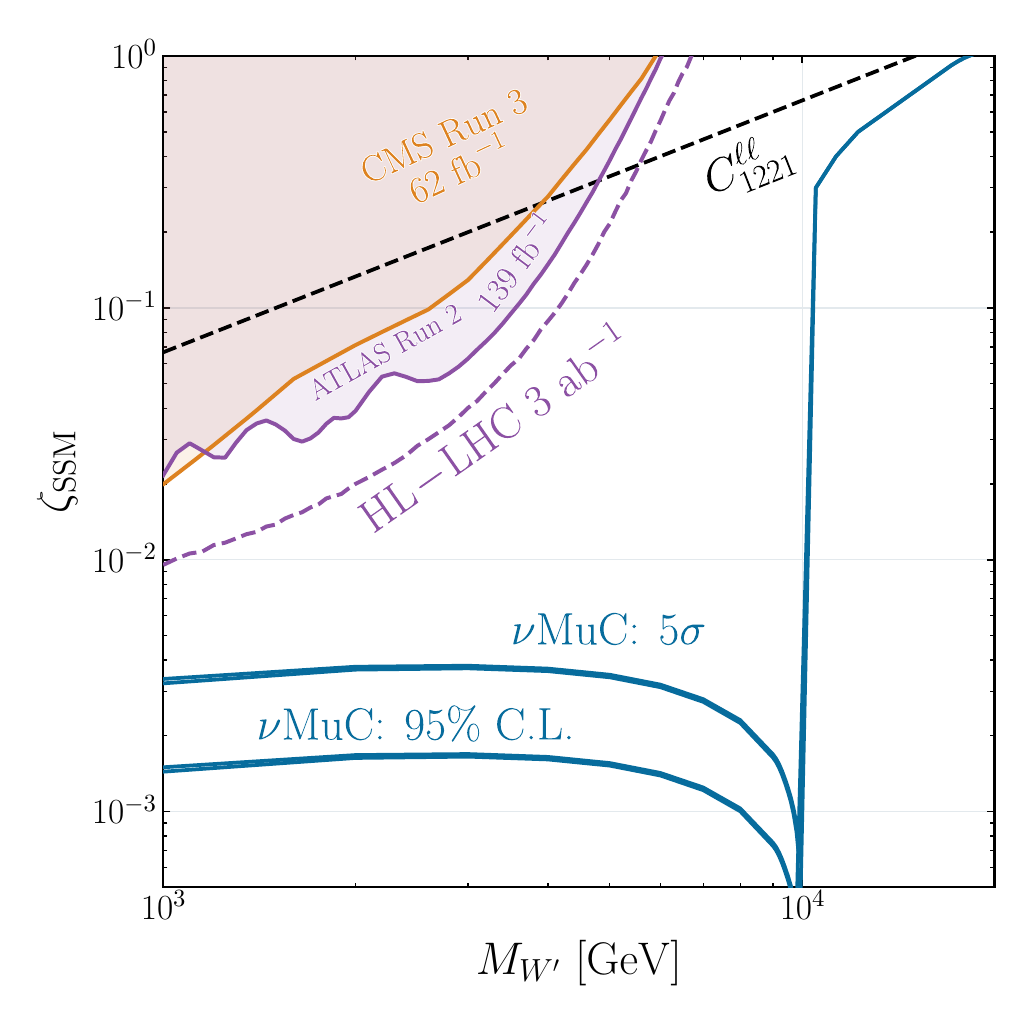}
    \caption{Sensitivity to the sequential $W'$ in the $(M_{W'},\zeta_{\rm SSM})$ plane. Blue curves show the neutrino-PDF $95\%$ C.L. and $5\sigma$ reaches; black dashed     the SMEFT reference for $C_{1221}^{\ell\ell}=-9.7\times10^{-4}~\mathrm{TeV}^{-2}$. The high-mass $W'$ branch retains its first archived asymptotic completion, independently of the updated SMEFT reference. The observed CMS Run 3 bound at 13.6 TeV with $62~\mathrm{fb}^{-1}$~\cite{CMS_Wprime_Run3_62fb} is orange and the ATLAS Run 2 bound at 13 TeV with $139~\mathrm{fb}^{-1}$~\cite{ATLAS_Wprime_Run2_139fb} is purple, with excluded regions shaded upward. The purple dashed curve is the luminosity-scaled HL-LHC projection at $3~\mathrm{ab}^{-1}$.}
    \label{fig:Wpsens}
\end{figure}

We summarize the combined on-shell and off-shell sensitivities of a 10-TeV $\nu$MuC to the $W'$ boson in the $(M_{W'},\zeta_{\rm SSM})$ plane in~\cref{fig:Wpsens}. Through the on-shell $W'$ production channel, the $5\sigma$ discovery could be made all the way to the kinematic limit of $M_{W'}\sim \sqrt s$ for the effective coupling $\zeta_{\rm SSM}\sim 3\times 10^{-3}$, and a $95\%$ C.L.~limit could be probed down to $\zeta_{\rm SSM}\sim10^{-3}$. One would reach the maximum sensitivity when the c.m.~energy approaches the mass peak $M_{W'}\approx \sqrt s$. In the off-shell region, $M_{W'}>\sqrt{s}$, we retain the first asymptotic continuation of the original $W'$ curve. The black dashed curve independently shows the SMEFT reference inferred from the precision measurement of the $e^-\bar{\nu}_e/e^+\nu_e$ production rate, using the negative coefficient endpoint quoted above.
For comparison, we include the CMS Run 3 bound at 13.6 TeV with $62~\mathrm{fb}^{-1}$~\cite{CMS_Wprime_Run3_62fb} (orange) and the ATLAS Run 2 bound at 13 TeV with $139~\mathrm{fb}^{-1}$~\cite{ATLAS_Wprime_Run2_139fb} (purple), with the excluded regions shaded upward. Only the most recent CMS result is displayed. The LHC coupling curves here and in~\cref{fig:Zpoff} are approximate reinterpretations of the published limits. The purple dashed HL-LHC curve is a luminosity-only extrapolation of the ATLAS expected limit to $3~\mathrm{ab}^{-1}$, with the coupling reach scaled as $\zeta_{\rm SSM}\propto\mathcal{L}_{\rm int}^{-1/4}$. We find that, through the on-shell $W'$ production mode, $\nu$MuC can improve the LHC sensitivity to $\zeta_{\rm SSM}$ by more than one order of magnitude in the region $M_{W'}<\sqrt{s}$. Meanwhile, in the off-shell regime, $\nu$MuC retains indirect sensitivity to $\zeta_{\rm SSM}$ for $W'$ masses beyond the direct kinematic reach, probing a region that is inaccessible to both the LHC and the HL-LHC.

\subsection{Sequential \texorpdfstring{$Z'$}{Z'} boson}
\label{sec:Z'}

The interaction of a  $Z'$ boson with fermions can be written as~\cite{Fuks:2017vtl}
\begin{equation}
\label{eq:ZprimeLagrangian}
\calL_{Z'}\supset -\frac{g_W}{c_W} Z'_\mu\sum\bar{f_i}\gamma^\mu\Big(\zeta^f_{L,i} P_L+\zeta^f_{R,i}P_R\Big)f_i
+\textrm{h.c.}\,,
\end{equation}
where $f_i$ includes both leptons and quarks, and $g_W$ and $c_W$ are the weak coupling constant and the cosine of the weak mixing angle, respectively. We assume there is no right handed neutrinos in the SM, hence $\zeta^{\nu_L}_{R,i}=0$. The sequential  limit is obtained when the $Z^\prime$ coupling to fermion is directly proportional to the SM $Z$:
\begin{equation}
\label{eq:zetaLR}
\zeta^f_{R,i/L,i}=(\zeta_{\rm SSM})_{i} (g_V^f\pm g_A^f),
\end{equation}
where
\begin{equation}
g_V^f=\frac{1}{2}T_3^f-Q_fs_W^2,\qquad g_A^f=-\frac{1}{2}T_3^f\,,
\end{equation}
and as before, $(\zeta_{\rm SSM})_i\equiv (g_{W'})_i/g_{W}$. 
We note that the sequential $Z'$ boson can be easily generalized to other models including a new neutral vector boson, \emph{e.g.}, a leptophilic $Z'$ gauged under the $U(1)_{L_\mu-L_\alpha}$ symmetry~\cite{Dasgupta:2023zrh}.

Similar to~\cref{eq:Wp}, the partonic cross section for the sequential $Z'$ production can be written as
\begin{align}
\label{eq:Zp}
\hat{\sigma}_{\mu\bar{\mu}\to Z'}&=\frac{12\pi^2\Gamma(Z'\to\mu\bar{\mu})}{M_{Z'}}\delta(\hat{s}-M_{Z'}^2)\,,\\
\hat{\sigma}_{\nu_\mu\bar{\nu}_\mu\to Z'}&=\frac{48\pi^2\Gamma(Z'\to\nu_\mu\bar{\nu}_\mu)}{M_{Z'}}\delta(\hat{s}-M_{Z'}^2)\,,\nonumber
\end{align}
where
\begin{align}
\Gamma(Z'\to\mu\bar{\mu})=\frac{M_{Z'}g_W^2}{24\pi c_W^2}\left[(\zeta_{L,2}^\ell)^2+(\zeta_{R,2}^\ell)^2\right]\,,~~~~~~
\Gamma(Z'\to\nu_\mu\bar{\nu}_\mu)=\frac{M_{Z'}g_W^2}{24\pi c_W^2}(\zeta_{L,2}^{\nu_L})^2,
\end{align}
and $\zeta_{L,R}^{\ell,\nu_L}$ relate to $\zeta_{\rm SSM}$ in terms of~\cref{eq:zetaLR}.
Here we have a factor of 2 with respect to the $W'$ case in~\cref{eq:Wp}, due to the initial-state neutrino polarization. The total width is a sum over the fermions including the leptons and quarks.

\begin{figure}
    \centering
    \includegraphics[width=0.49\linewidth]{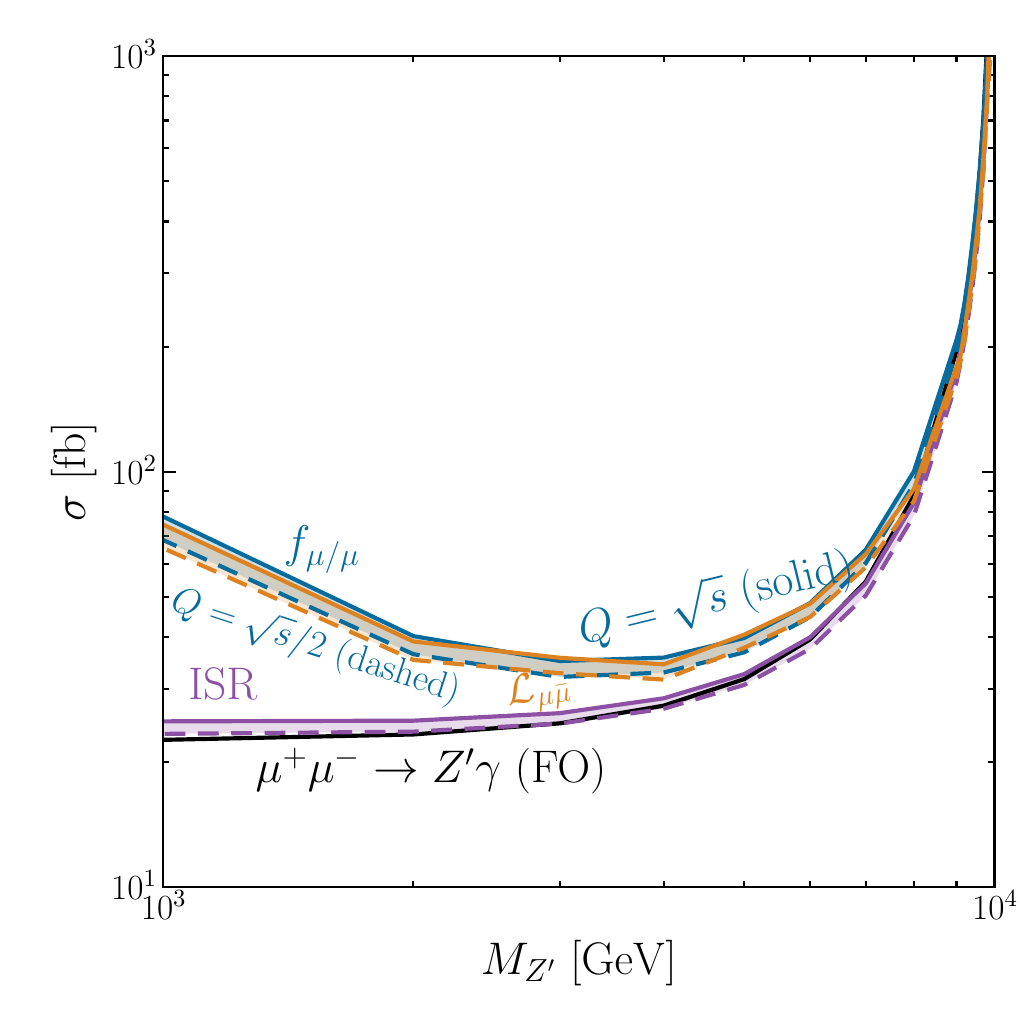}
    \includegraphics[width=0.49\linewidth]{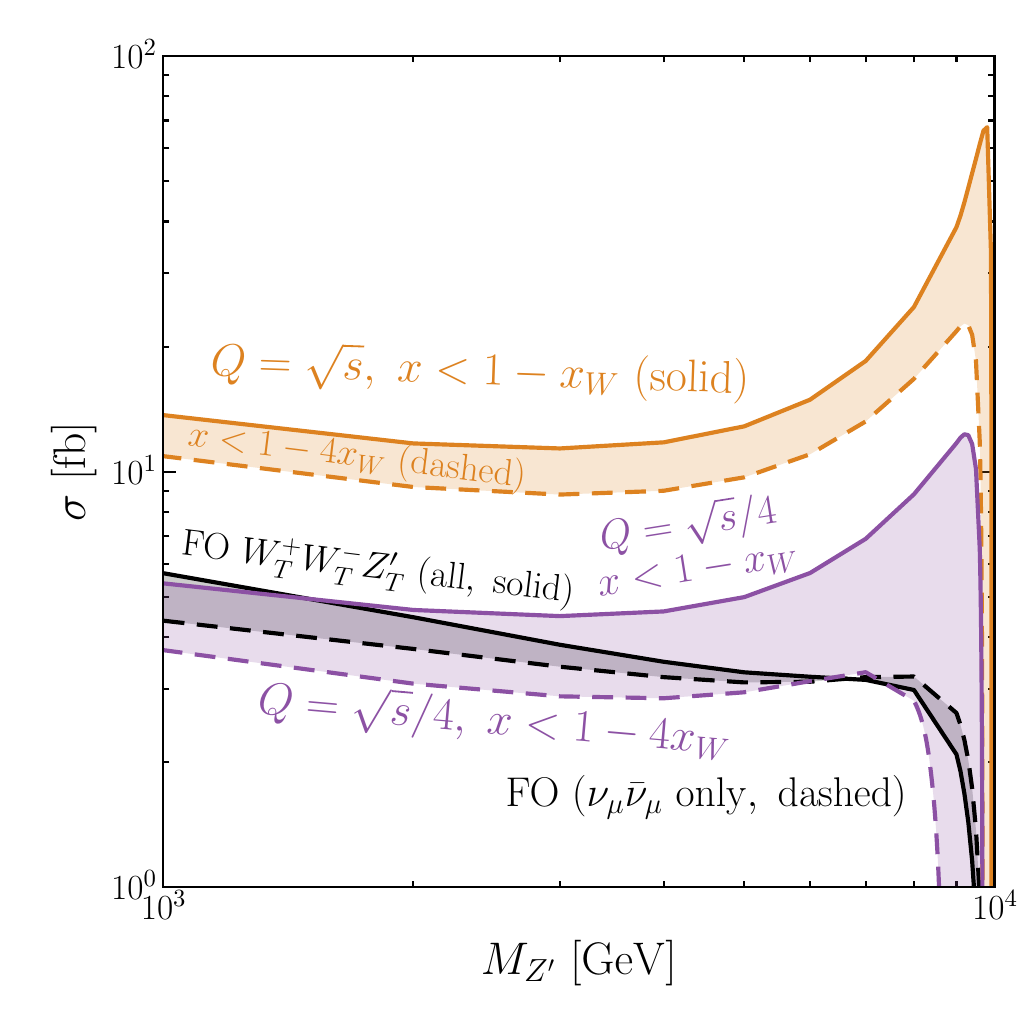}
    \caption{
    The $Z'$ production cross section through $\mu\bar{\mu}$ (left) and $\nu_\mu\bar{\nu}_\mu$ (right) annihilation. Left: black is the FO result, purple is QED ISR, blue is the one-sided muon-PDF approximation, and orange is the two-muon luminosity result. Solid and dashed curves correspond to $Q=\sqrt{s}$ and $Q=\sqrt{s}/2$, respectively; the bands span this scale variation. Right: black solid and dashed curves show the FO predictions with all $\mu^+\mu^-\to W_T^+W_T^-Z_T'$ diagrams and only the neutrino-fusion diagrams, respectively. Orange and purple bands show the neutrino-luminosity predictions at $Q=\sqrt{s}$ and $Q=\sqrt{s}/4$, respectively. Their solid and dashed boundaries correspond to $x_{\rm cut}=1-x_W$ and $1-4x_W$, with $x_W=M_W/Q$.
    }
    \label{fig:Zp}
\end{figure}

We start with the $Z'$ production with the FO calculation of $\mu\bar{\mu}\to Z'\gamma$ process, with the cross section shown as the black line in~\cref{fig:Zp} (left), which serves as a baseline. The soft-photon can be well approximated with the inital-state radiation (ISR), which is verified explicitly with the WHIZARD ISR option~\cite{Kilian:2007gr,Moretti:2001zz}. We find good agreement between the two as shown in the purple band, with the dashed and solid lines corresponding to scale choice $Q\in (\sqrt s/2,\sqrt{s})$, respectively. Alternatively, with the collinear PDF approach, the $Z'$ production cross section can be written in terms of the partonic cross sections in~\cref{eq:Zp} convoluted with the luminosity function $\mathcal{L}_{\mu\bar{\mu}}$. Using the LO muon PDF as $f_{\mu/\mu}(x)=\delta(1-x)$, the cross section for $\mu\bar{\mu}$ collisions can be simplified as 
\begin{equation}
\sigma(\mu\bar{\mu}\to Z')=2f_{\mu/\mu}(\tau)\frac{12\pi^2\Gamma(Z'\to\mu\bar{\mu})}{sM_{Z'}}\,,
\end{equation}
where the factor of 2 comes from two incoming  beams. The corresponding cross sections are shown as the blue band in~\cref{fig:Zp} (left).
Finally, we show the result with collinear muon PDFs for both beams, \emph{i.e.}, $\mathcal{L}_{\mu\bar\mu}$ as the orange band.
We see that the one-sided PDF approximation $f_{\mu/\mu}$ agrees very well with the two-parton luminosity $\mathcal{L}_{\mu\bar\mu}$ approach throughout the whole $M_{Z'}$ range. In comparison, they differ from the FO/ISR predictions, mainly in the low $M_{Z'}$ region, reflecting the PDF resummation effect. 

\begin{figure}[t]
    \centering
    \includegraphics[width=0.29\linewidth]{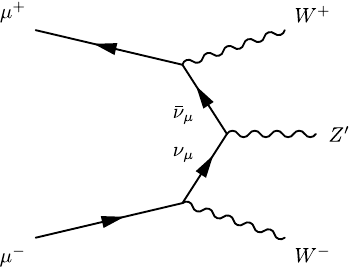} \hspace{.8cm}
    \includegraphics[width=0.29\linewidth]{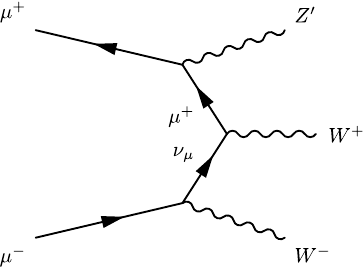}\hspace{.8cm}
    \includegraphics[width=0.29\linewidth]{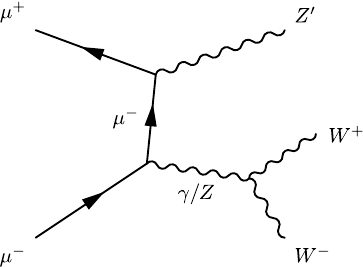}
    \caption{Fixed-order Feynman diagrams for $Z'$ production at the muon colliders. The left one with $\nu_\mu\bar{\nu}_\mu\to Z'$ can be factorized into annihilation of the neutrino PDF. }
    \label{feyn:Zp}
\end{figure}

We now illustrate the $Z'$ production through neutrino annihilation, $\nu_\mu\bar{\nu}_\mu\to Z'$, as indicated by the center sub-process in~\cref{feyn:Zp} with the $W$ boson radiation off the initial beams. We first calculate the FO prediction for $\mu^+\mu^-\to W_T^+W_T^-Z'_T$ with all diagrams in~\cref{feyn:Zp}. The result is shown by the black solid line in the right panel of~\cref{fig:Zp}. Since only the fusion diagrams associated with $\nu_\mu\bar{\nu}_\mu\to Z'$ as~\cref{feyn:Zp} (left) can be factorized in terms of collinear neutrino PDFs, we retain only this diagram in the FO calculation, with the corresponding prediction shown by the black dashed line. The two FO predictions agree remarkably well, validating the collinear approximation for $\nu_\mu\bar{\nu}_\mu\to Z'$ production.
As before, we focus only on the transverse components of the final-state gauge bosons $Z'_T$ to avoid unphysical growth over $s/M_{Z'}^2$.

In comparison, we now present the neutrino-PDF predictions obtained from the $\mathcal{L}_{\nu_\mu\bar{\nu}_\mu}$ luminosity convoluted with the partonic cross section in~\cref{eq:Zp}. On the right panel of~\cref{fig:Zp}, the orange and purple bands correspond to the factorization-scale choices $Q=\sqrt{s}$ and $\sqrt{s}/4$, respectively, while the solid and dashed curves denote the neutrino momentum-fraction cuts $x_{\rm cut}=1-x_W$ and $1-4x_W$, respectively. We find that only the lower scale choice $Q=\sqrt{s}/4$ leads to a good description of the FO prediction in this case, in contrast to the $W'$ scenario.
In addition, by comparing the left panel (with photon radiation for $\mu \bar \mu$) and the right panel (with $W$ radiation for $\nu \bar \nu$) of~\cref{fig:Zp}, we find that the neutrino-annihilation cross section is approximately one order of magnitude smaller than that of the muon-annihilation channel. Consequently, the contribution to the $Z'$ sensitivity obtained from neutrino annihilation is expected to be weaker than that from muon annihilation.

\begin{figure}[tb]
\centering
\includegraphics[width=0.49\textwidth]{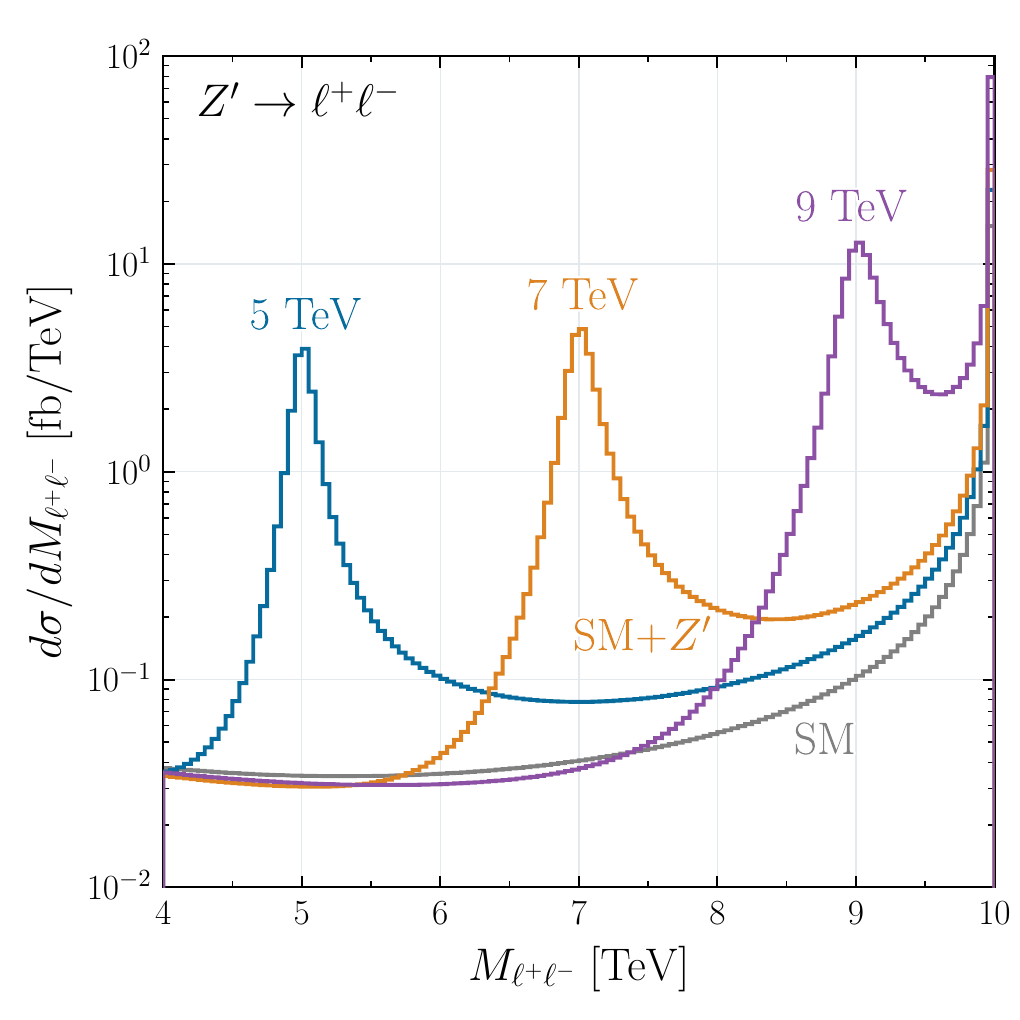}
\includegraphics[width=0.49\textwidth]{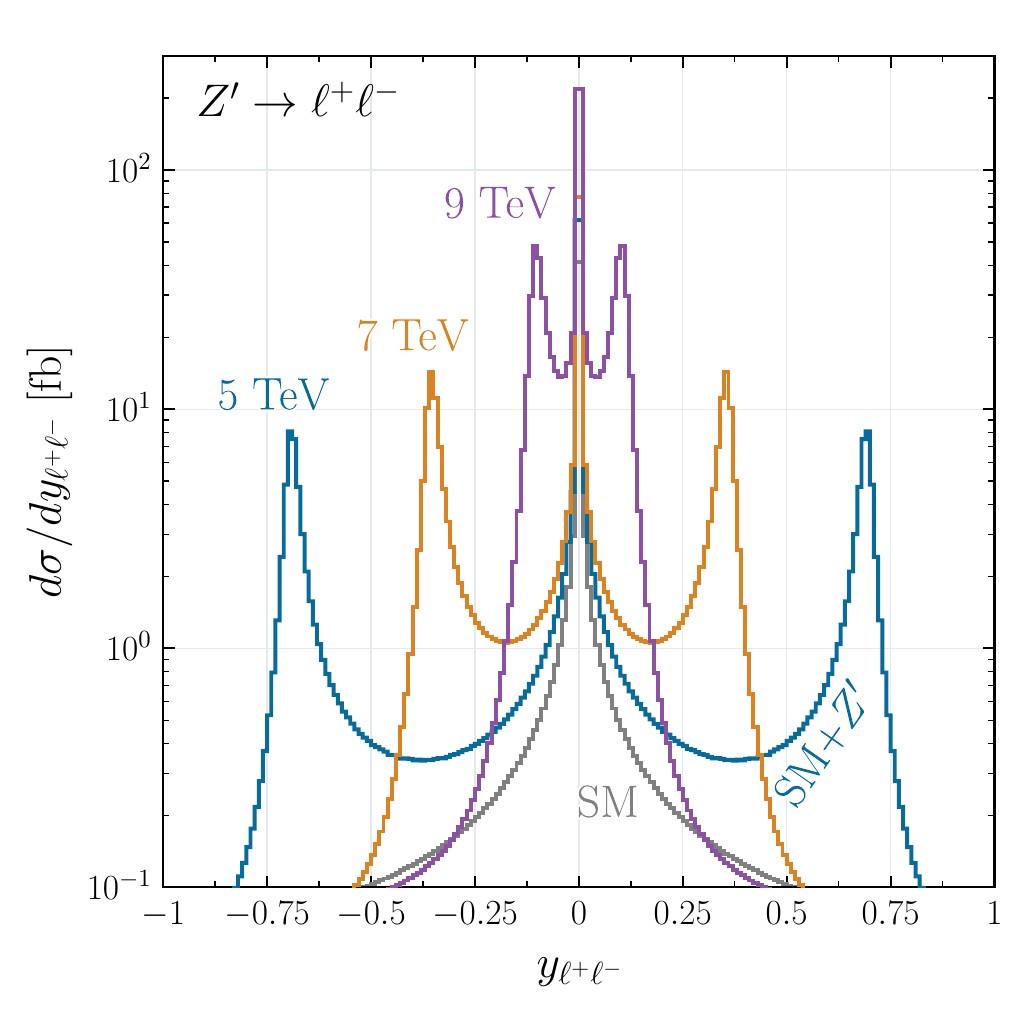}
\caption{Signal for $Z'\to\ell^+\ell^-$ at a 10 TeV muon collider: (left) dilepton invariant mass and (right) rapidity distributions. Solid blue, orange, and purple curves show the coherent SM+$Z'$ predictions, including interference, for $M_{Z'}=5,7,9$ TeV, respectively. Gray curves show the SM $\gamma/Z\to\ell^+\ell^-$ background.
}
\label{fig:Z}
\end{figure}

Similar to the $W'$ signal, we show the expectation for the invariant mass distribution for the $Z'$ production in~\cref{fig:Z} (left) for $M_{Z'}=5,~7,~ 9$~TeV at $\sqrt s=10$~TeV. The resonant production is manifest. Again, the  closer the $Z'$ mass is to the c.m.~energy, the larger the production rate is. We point out that the radiative return kinematics is dominated by the two-body kinematics $Z'\gamma$, with the photon mostly along the beam direction. As such, $Z'$ has no transverse motion with a monochromatic energy $E_{Z'}=\frac{\sqrt s}{2}(1+M^2_{Z'}/s)$ and the rapidity distribution for the resonant $Z'$ develops a kinematic peak at~\cite{Dasgupta:2023zrh}
\begin{equation}
    y_{Z'} = \frac{1}{2} \ln{\frac{E_{Z'} +p_z}{E_{Z'}-p_z}  }= \ln\frac{\sqrt s}{M_{Z'}}\,,
\end{equation} 
as seen in~\cref{fig:Z} (right). Such a distinct feature not only stands out above the gray SM background curves, but also provides additional means for the $M_{Z'}$ determination.

For a $Z'$ mass slightly above  the collider energy ($M_{Z'}>\sqrt{s}$), we can again study the virtual effects from the off-shell production. Far above $\sqrt{s}$, it is appropriate to integrate out the $Z^\prime$ and matching to SMEFT can be performed. At the LO, the dim-6 operators generated include
\begin{align}
\label{eq:C4fZp}
C^{\ell\ell}_{iijj}\Big|_{Z'} &= -\frac{g^2_W}{c^2_W M_{Z'}^2}\zeta^f_{L,i}\zeta^f_{L,j}\,,  \\
C^{\ell e}_{iijj}\Big|_{Z'}   &= -\frac{g^2_W}{c^2_W M_{Z'}^2}\zeta^f_{L,i}\zeta^f_{R,j}\,, \\
C^{ee}_{iijj}\Big|_{Z'}       &= -\frac{g^2_W}{c^2_W M_{Z'}^2}\zeta^f_{R,i}\zeta^f_{R,j}\,, 
\end{align}
with $i,j=1,2,3$, the lepton generation indices. Again, the complete searches of the three lepton flavors in the complementary channels of $\mu\bar \mu,  \nu\bar \nu$ annihilation at $\nu$MuC would be beneficial.

\begin{figure}
    \centering
    \includegraphics[width=0.49\textwidth]{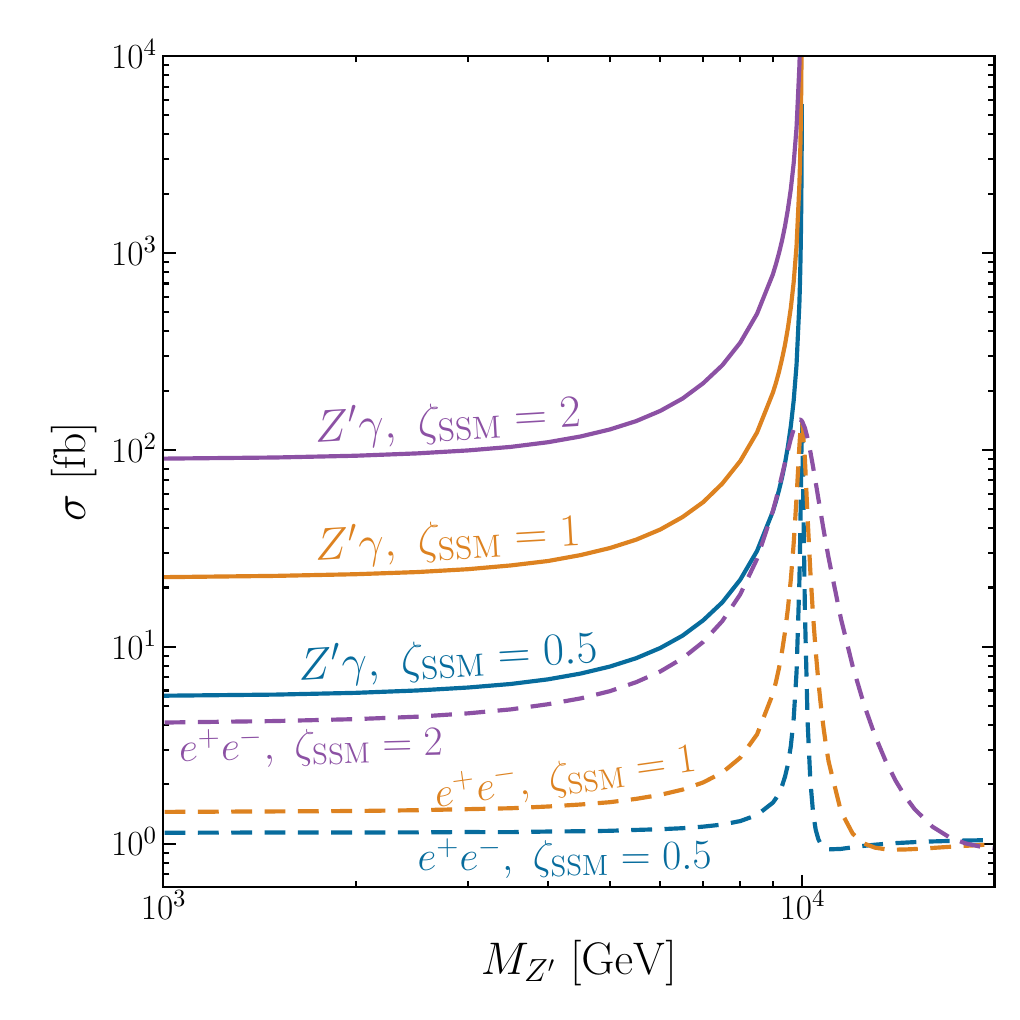}
    \includegraphics[width=0.49\textwidth]{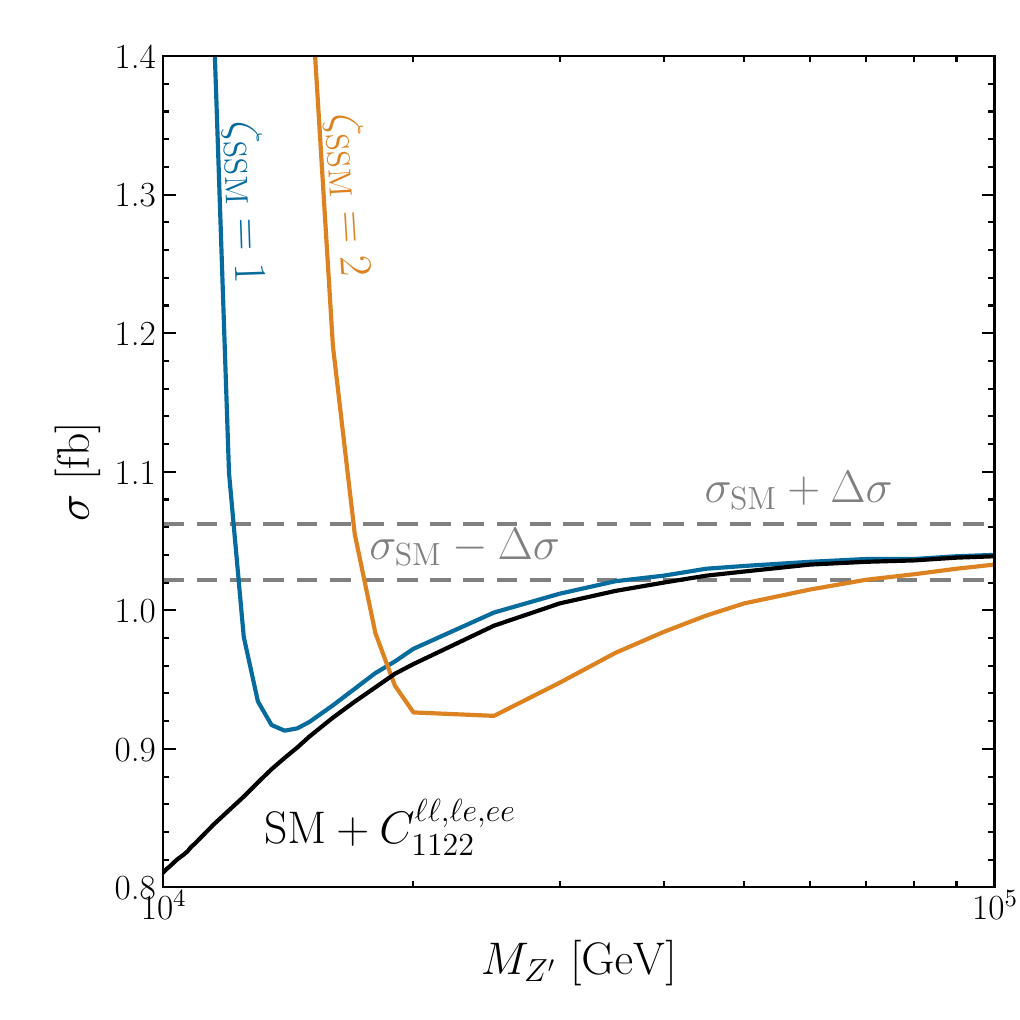}
    \caption{For a 10 TeV muon collider, \textbf{Left}:  
    Production cross sections of the sequential $Z'$ through $Z'_T\gamma$ associated production (solid) and $\mu^+\mu^-\to e^+e^-$ including SM+$Z'$ (dashed). Blue, orange, and purple correspond to $\zeta_{\rm SSM}=0.5,1,2$, respectively.
    \textbf{Right}: The off-shell $\mu^+\mu^-\to e^+e^-$ cross section for $\zeta_{\rm SSM}=1,2$ (blue, orange) and the matched SMEFT result from~\cref{eq:C4fZp} at $\zeta_{\rm SSM}=1$ (black). Gray dashed lines bound the $95\%$ C.L. allowed cross section region.
    }
    \label{fig:Zpoff}
\end{figure}

In the left panel of~\cref{fig:Zpoff}, we compare the on-shell $Z'$ production cross section with that of $\mu^+\mu^-\to Z/\gamma/Z'\to e^+e^-$ with a few representative $\zeta_{\rm SSM}$ values. We see that in the on-shell region when $M_{Z'}<\sqrt{s}$, the $e^+e^-$ cross section scales down consistently from the $Z'$ production by a branching fraction BF$(Z'\to e^+e^-)\sim3\%$. In comparison, when $M_{Z'}>\sqrt{s}$, we can have the off-$Z'$ non-resonance production, in which we only focus on the $e^+e^-$ channel for simplicity in this work. In this region, similar to the $W'$ situation, there is significant destructive interference between the $Z'$ and the SM $Z/\gamma$ contribution shown in the right panel of~\cref{fig:Zpoff}. 
We have also shown the cross section of the SMEFT operators matched with~\cref{eq:C4fZp} as the black curve in the right panel of~\cref{fig:Zpoff}.
We see that the $Z'$ off-shell cross section with $\zeta_{\rm SSM}=1$ largely reproduces the SMEFT operator results for $C_{1122}$, except around the $M_{Z'}\sim\sqrt{s}$ region when close to the resonance.

As for the sensitivity reach, we again take the background-free assumption for the on-shell production channel. We adopt the criteria with the signal events of $S=14.4$ for the $5\sigma$ discovery, and $S=3.09$ for $95\%$ C.L.~exclusion. For the off-shell channel, however, we take into account the SM backgrounds. Following~\cref{eq:2sigma}, we require $\mathcal{S}\simeq S/\sqrt{B+(\Delta B)^2}=2$. Unlike the $W'$ scenario, the prediction for the off-shell $Z'$ channel to final state leptons can be determined with much higher precision. Consequently, the corresponding systematic uncertainty is negligible, $\Delta B\simeq 0$, different from the off-shell $W'$ scenario. Following the same procedure as above, the 95\% C.L. allowed band for the cross section, $\sigma_{\rm SM}\pm\Delta\sigma$, is shown as the gray dashed envelope in the right panel of~\cref{fig:Zpoff}. The cross section asymptote for $M_{Z'}\to \infty$ leads to $\sigma_{\rm SM}$. By comparing the $Z'$ predictions with this band, we can directly extract the exclusion limits on the $Z'$ mass for different values of $\zeta_{\rm SSM}$, as well as the corresponding constraints on the SMEFT operators.

\begin{figure}
    \centering
    \includegraphics[width=0.6\textwidth]{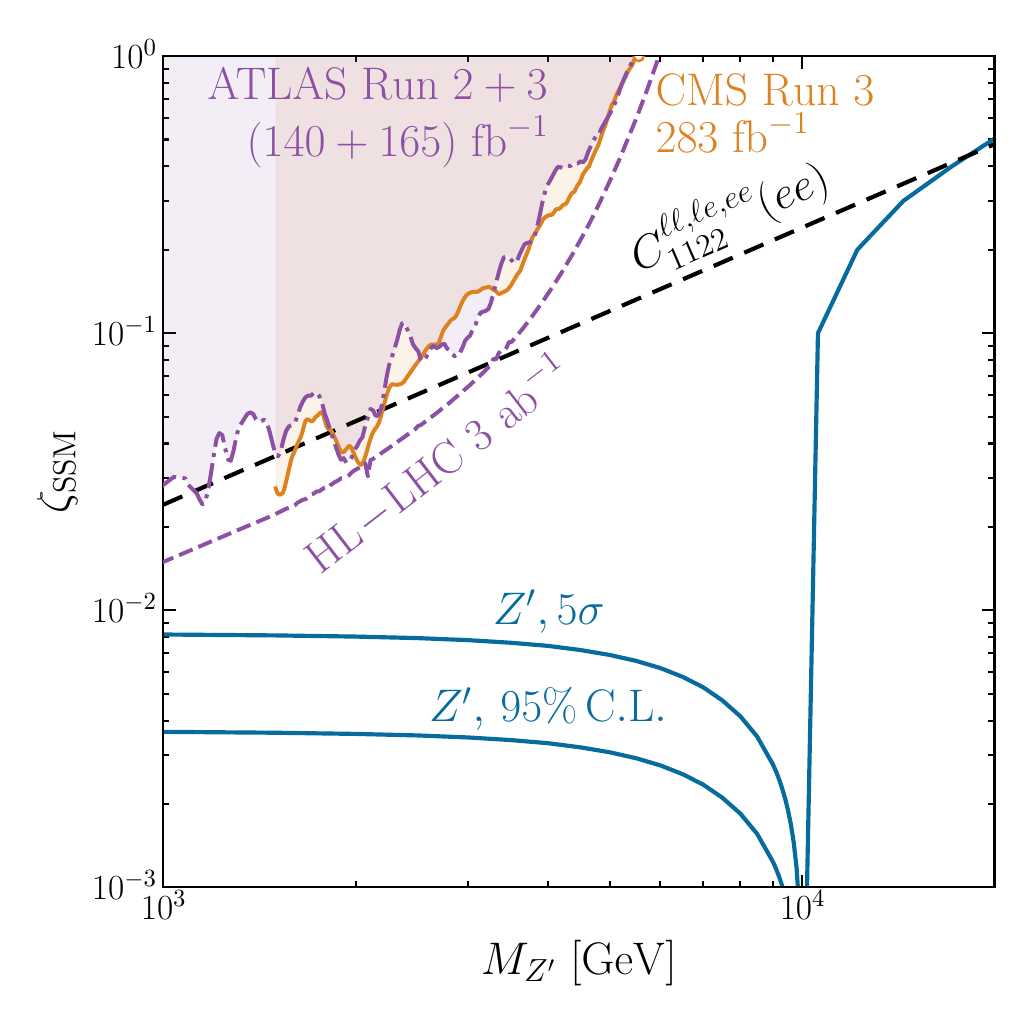}
    \caption{Sensitivity in the $(M_{Z'},\zeta_{\rm SSM})$ plane: blue curves show the $95\%$ C.L. and $5\sigma$ reaches, and black dashed shows the matched SMEFT constraint. The CMS Run 3 bound with $283~\mathrm{fb}^{-1}$ at 13.6 TeV~\cite{CMS_Zprime_Run3_283fb} is orange, and the ATLAS Run 2+3 bound~\cite{ATLAS_Zprime_Run23_305fb} is purple, with observed exclusions shaded upward. The purple dashed curve is the luminosity-scaled HL-LHC projection at $3~\mathrm{ab}^{-1}$.}
    \label{fig:Zpsens}
\end{figure}

\cref{fig:Zpsens} shows the combined on-shell and off-shell sensitivities to the sequential $Z'$ in the $(M_{Z'},\zeta_{\rm SSM})$ plane. We find that the on-shell sensitivity to the coupling strength $\zeta_{\rm SSM}$ can be two to three orders of magnitude stronger than that from the off-shell channel. This substantial improvement results from both the larger production cross section, without the branching-fraction suppression relevant to the off-shell analysis, and the essentially background-free nature of the on-shell search. Again, one would reach the maximum sensitivity when the c.m.~energy approaches the mass peak $M_{Z'}\approx \sqrt s$.
For comparison, we show the CMS Run 3 bound with $283~\mathrm{fb}^{-1}$ at 13.6 TeV~\cite{CMS_Zprime_Run3_283fb} (orange) and the ATLAS Run 2+3 bound combining $140~\mathrm{fb}^{-1}$ at 13 TeV and $165~\mathrm{fb}^{-1}$ at 13.6 TeV~\cite{ATLAS_Zprime_Run23_305fb} (purple). Only the most recent CMS result is displayed. The purple dashed HL-LHC curve extrapolates the ATLAS expected limit to $3~\mathrm{ab}^{-1}$ using the same luminosity-only scaling as above. The on-shell $Z'$ sensitivity is slightly weaker than that for the $W'$ by about a factor of 2, owing to its smaller production cross section. In contrast, the off-shell $Z'$ sensitivity is slightly stronger, since it is primarily driven by the more precise determination of the $e^+e^-$ final state together with the larger rate, compared with the $e^-\bar{\nu}_e/e^+\nu_e$ channels relevant to the $W'$ case.  Moreover, as in the $W'$ scenario, the off-shell $Z'$ bound asymptotically approaches the corresponding SMEFT constraint, demonstrating the consistency of the matching relation in~\cref{eq:C4fZp}. Finally, we note that the indirect constraint on the off-shell $Z'$ model inferred from the SMEFT analysis differs slightly from the results presented in~\cref{tab:SmeftvLowE}, where one operator is varied at a time. In the $Z'$ model, by contrast, the three operators in~\cref{eq:2sigma} appear in a fixed linear combination and therefore vary simultaneously. Consequently, the sensitivity is largely controlled by the weakest constrained
direction, namely $C_{1122}^{\ell e}$.

\section{Summary and Conclusion}
\label{sec:conclusion}

A high-energy muon collider is simultaneously a collider of high-energy neutrinos. Electroweak radiation of $W^\pm$ bosons from the energetic muon beams gives rise to $\mu\nu_\mu$ and $\nu_\mu\bar{\nu}_\mu$ collisions at partonic center-of-mass energies approaching that of the underlying $\mu^+\mu^-$ collisions. In this work, we have explored the physics opportunities associated with these neutrino-initiated processes and, in particular, their interplay with the conventional $\mu^+\mu^-$ collision channels. Together, these different initial states provide access to complementary components of the electroweak $SU(2)$ structure of interactions involving leptons and offer a qualitatively new environment in which to search for physics beyond the Standard Model. We dub this aspect of muon colliders the $\nu$MuC.

In~\cref{sec:neutrino_pdf} we established the partonic framework needed to describe these processes, using the muon and neutrino electroweak  parton distribution functions (PDFs) and their associated luminosities to characterize the different leptonic initial states available at a high-energy muon collider. Although the neutrino luminosities are suppressed relative to the dominant muon luminosity due to the $W^\pm$ radiation, the resulting neutrino-initiated collision rates are sufficiently large to provide useful probes of new interactions. More importantly, the different initial states probe different, complementary, interactions and therefore provide information that cannot, in general, be reproduced simply by accumulating more charged-lepton collision events.

In~\cref{sec:SMEFT} we demonstrated this complementarity explicitly using four-lepton operators in the Standard Model Effective Field Theory~(SMEFT). In~\cref{sec:SMEFT_single} we studied the conventional $\mu^+\mu^-$ channels at a $10~\TeV$ muon collider with $10~{\text{ab}}^{-1}$ of integrated luminosity, demonstrating that they provide extremely strong sensitivity to a broad range of four-lepton Wilson coefficients, with projected single-operator constraints typically reaching the $10^{-5} - 10^{-4}~\TeV^{-2}$ level (corresponding to an effective new physics scale, $\Lambda$, of $75-250$ TeV). For most all the four-lepton operators considered here, these sensitivities significantly improve upon existing constraints from electroweak precision measurements, low-energy neutrino scattering, and charged-lepton flavor observables.
For most operators, the reach of $\nu$MuC will also exceed the projections expected from future low energy experiments such as Belle-II. The only exceptions are the flavor-violating $C^{ee}_{1222}$ and $C^{\ell e}_{2212}$ WCs, for which the Mu2e~\cite{Mu2e:2014fns} and COMET~\cite{COMET:2018auw} experiments are expected to probe them up-to $600~$TeV~\cite{Asadi:2025dii}.     

The reach of $\nu$MuC therefore corresponds to sensitivity to effective BSM scales extending well beyond the collider center-of-mass energy, therefore beyond the accessible energies for BSM direct production.  The importance of the neutrino-initiated channels becomes especially clear once more than one Wilson coefficient is allowed to vary, as detailed in~\cref{sec:SMEFT_multi}. Since the neutrino PDFs are smaller than the corresponding muon PDFs, $\nu_\mu\bar{\nu}_\mu$ collisions generally provide only a modest improvement to constraints that are already dominated by $\mu^+\mu^-$ annihilation. Likewise, considered in isolation, the mixed $\mu\nu_\mu$ channels typically yield weaker constraints than the corresponding charged-lepton channels. Their value, however, lies in their different dependence on the underlying Wilson coefficients. On the other hand, combining $\mu\nu_\mu$ with $\mu^+\mu^-$ collisions can lift degeneracies and substantially reduce otherwise weakly constrained directions in the multidimensional SMEFT parameter space. This remains true even for coefficients to which the neutrino-initiated process has no direct sensitivity: by constraining correlated directions, the additional channel can improve the combined determination of those coefficients. Thus, the neutrino component of a muon collider is not simply an additional source of events, but an independent probe of the electroweak and chiral structure of possible new interactions.

We further illustrated this point using simplified models containing sequential $W'$ and $Z'$ vector bosons in~\cref{sec:V'}. The charged-current nature of $W'$ production makes the $\mu\nu_\mu$ initial state particularly important. For $M_{W'}<\sqrt{s}$, resonant production via the radiative return mechanism allows a $10~\TeV$ $\nu$~MuC to probe effective couplings down to approximately $\zeta_{\rm SSM}\sim 10^{-3}$ all the way to $M_{W'}\approx 10$ TeV to reach the maximum sensitivity, extending substantially beyond the corresponding LHC sensitivity. For masses above the direct kinematic reach, virtual $W'$ exchange remains accessible through precision measurements of charged-current final states. In the limit $M_{W'}\gg\sqrt{s}$, this sensitivity smoothly approaches that obtained from the corresponding four-lepton SMEFT operator, providing a direct connection between the resonance and effective-field-theory descriptions. For a sequential $Z'$, the dominant sensitivity instead arises from the much larger $\mu^+\mu^-$ luminosity, while $\nu_\mu\bar{\nu}_\mu$ annihilation provides a complementary neutral-current production mode. As in the $W'$ case, the collider is sensitive both to resonant production below the kinematic threshold and to virtual effects above it. In the heavy-mass limit, the off-shell $Z'$ results approach the correlated combination of SMEFT operators obtained by integrating-out the new vector boson. The $W'$ and $Z'$ examples therefore provide concrete illustrations of how the different leptonic initial states can distinguish the charged- and neutral-current structures of new physics and connect direct searches continuously to indirect EFT probes.

The central observation of this work is broader than the particular EFT operators or simplified benchmark new physics models considered here. Electroweak radiation transforms a multi-TeV muon collider into a collection of related leptonic colliders, with $\mu^+\mu^-$, $\mu\nu_\mu$, and $\nu_\mu\bar{\nu}_\mu$ initial states probing different components of the same underlying electroweak dynamics. While the charged-lepton collisions will often provide the greatest statistical sensitivity, the neutrino-initiated processes supply additional information that can be essential for identifying the structure of a deviation from the Standard Model. Fully exploiting the physics potential of a future high-energy muon collider therefore calls for treating its high-energy neutrino collisions not as a secondary effect, but as an integral part of its experimental program.
Although we restrict our discussion in this work to opposite-sign muon colliders, our results are also more broadly applicable to high-energy electron--positron collisions~\cite{LinearColliderVision:2025hlt,CLIC:2018fvx,Gessner:2025acq} and same-sign muon colliders \cite{Hamada:2022mua,Kitano:2025xaj}. 

\noindent 
\section*{Acknowledgements}
We would like to thank Dario Buttazzo and Florentin Jaffredo for participation in the initial stage of this project. We acknowledge Lukas Allwicher, Darius A. Faroughy, Florentin Jaffredo, Olycr Sumensari and Felix Wilsch, as the authors of one of the SMEFT UFO files utilized in this work. IB would like to thank Ilaria Brivio for useful discussions.
The work of IB was supported by the U.S. Department of Energy under award DE-SC0020262.
The work of AdG was supported in part by the US Department of Energy under award \#de-sc0010143 and in part by the National Science Foundation under award No.~PHY-1630782. 
The work of TH and ZT was supported in part by the US Department of Energy under grant No.~DE-SC0007914 and in part by Pitt PACC. He also thanks the Aspen Center for Physics for hospitality during the final stage of this project, which is supported by NSF Grant PHY-2210452. 
IL is supported in part by the U.S. Department of Energy under contracts No. DEAC02-06CH11357
at Argonne and by the U.S. Department of Energy under contract No. DE-SC0010143 at Northwestern.
The work of YM is supported as an F.R.S.-FNRS Postdoctoral Researcher (Belgium) and he also acknowledges support from the COMETA COST Action CA22130.
The work of KX at MSU is supported in part by the U.S. National Science Foundation under Grants No. PHY-2112829, PHY-2013791, and PHY-2310497.
The work of KX at TDLI/SJTU is also supported by the National Natural Science Foundation of China and by the State Key Laboratory of Dark Matter Physics.

\appendix
\section{Mapping between SMEFTsim and our SMEFT convention}
\label{App:SMEFTmatchbases}
The definition of the independent set of WCs utilized in SMEFTsim~\cite{Brivio:2020onw} follows from the below Lagrangian, 
\begin{align}
    \hat{\mathcal{L}} &=\hat{C}_{ ijkl}^{\ell\ell} (\overline{\ell_i} \gamma^\mu  \ell_j) (\overline{\ell_k} \gamma_\mu  \ell_l)+ \hat{C}^{ee}_{ijkl} (\overline{e_i} \gamma^\mu e_j) (\overline{e_k} \gamma_\mu  e_l) + \hat{C}^{\ell e}_{ ijkl}(\overline{\ell_i} \gamma^\mu \ell_j) (\overline{e_k} \gamma_\mu  e_l)+\text{h.c.}.
\end{align}
Writing out the explicit hermitian conjugate terms, this reads
\begin{align}
    \hat{\mathcal{L}} &=\hat{C}_{ ijkl}^{\ell\ell} (\overline{\ell_i} \gamma^\mu  \ell_j) (\overline{\ell_k} \gamma_\mu  \ell_l)+ \hat{C}^{ee}_{ijkl} (\overline{e_i} \gamma^\mu e_j) (\overline{e_k} \gamma_\mu  e_l) + \hat{C}^{\ell e}_{ ijkl}(\overline{\ell_i} \gamma^\mu \ell_j) (\overline{e_k} \gamma_\mu  e_l) \label{eq:SMEFTSIM1}\\&+\hat{C}_{ ijkl}^{\ell\ell, *}  (\overline{\ell_l} \gamma_\mu  \ell_k)(\overline{\ell_j} \gamma^\mu  \ell_i)+ \hat{C}^{ee, *}_{ijkl} (\overline{e_l} \gamma_\mu  e_k)(\overline{e_j} \gamma^\mu e_i)  + \hat{C}^{\ell e, *}_{ ijkl} (\overline{e_l} \gamma_\mu  e_k) (\overline{\ell_j} \gamma^\mu \ell_i)\nonumber,  \\
    &=2\hat{C}_{ ijkl}^{\ell\ell} (\overline{\ell_i} \gamma^\mu  \ell_j) (\overline{\ell_k} \gamma_\mu \ell_l)+ 2\hat{C}^{ee}_{ijkl} (\overline{e_i} \gamma^\mu e_j) (\overline{e_k} \gamma_\mu  e_l) + 2\hat{C}^{\ell e}_{ ijkl}(\overline{\ell_i} \gamma^\mu \ell_j) (\overline{e_k} \gamma_\mu  e_l), \label{eq:SMEFTSIM2}
\end{align}
where moving from~\cref{eq:SMEFTSIM1} to~\cref{eq:SMEFTSIM2} is the result of applying~\cref{eq:hermitian}. Contrasting~\cref{eq:SMEFTSIM2} with~\cref{eq:SMEFT} gives the following relations between our notation and that of SMEFTsim, 
\begin{align}
    \hat{C}_{ ijkl}^{\ell\ell} =  \frac{{C}_{ ijkl}^{\ell\ell}}{4}, \;\;\;
     \hat{C}_{ ijkl}^{ee} =  \frac{{C}_{ ijkl}^{ee}}{4}, \;\;\;
    \hat{C}_{ ijkl}^{\ell e} =  \frac{{C}_{ ijkl}^{\ell e}}{2}.
\end{align}
These expressions can be used to contrast our results to this alternate convention.

A further alternative is given by the Wilson Coefficient Exchange Format (WCxf) definition of the Warsaw basis~\cite{Aebischer:2017ugx}. In this convention, the Lagrangian is written as a sum over the independent operator set with complex-valued WCs, without an additional overall normalization factor or an explicit addition of the Hermitian-conjugate interactions. We do not use this convention here, but highlight this difference to facilitate comparison with results expressed in the WCxf convention elsewhere in the literature.

\section{Contrast with low-energy constraints}
\label{App:LEconstraints}
In this appendix we summarize the low-energy constraints on the four-lepton SMEFT operators relevant for our analysis. For the comparison of our results with the low-energy constraints  we assume that only one WC is non-zero at a time, allowing the bounds quoted in the literature to be directly translated into individual limits on the corresponding SMEFT coefficients. The constraints originate from a variety of low-energy precision measurements, each probing different flavor and chiral structures of the four--lepton interactions. In the following, we briefly summarize the observables used and quote the corresponding limits adopted in our analysis.

\subsection{Lepton Flavor Conserving Operators}
Several flavor--conserving four--lepton operators are constrained by the global electroweak fit of Ref.~\cite{Falkowski:2015krw}. The fit combines the $Z$ boson production and decay at LEP-1~\cite{ALEPH:2005ab}, two--lepton production in LEP-2~\cite{ALEPH:2013dgf} through the measurements of $e^+e^-\to\ell^+\ell^-$, low-energy neutrino--electron scattering from CHARM~\cite{CHARM:1988tlj} and CHARM-II~\cite{CHARM-II:1994dzw}, parity violation in low-energy M\o ller scattering~\cite{SLACE158:2005uay}, and the extraction of the Fermi constant $G_f$ from the flavor-conserving leptonic tau and muon decays. Different four--lepton operators are sensitive to different subsets of these observables. All relevant constraints with $68\%$ C.L. upper bounds are summarized in~\cref{tab:SmeftvLowE}.

The WC $C^{\ell\ell}_{1221}$ is the most constrained in this list, through its contribution to the extraction of $G_f$ from muon decay. Therefore, it modifies all other electroweak precision observables, which have been very accurately measured in LEP-1, as well as the precise $M_W$ measurements from LEP-2 and the Tevatron. 

The $C^{ee}_{1122}$ and $C^{\ell e}_{1122}$ WCs instead receive their dominant constraint from the $e^+e^-\to\mu^+\mu^-$ process at LEP-2, through measurements of total cross-section and forward-backward asymmetries. 
The coefficient $C^{\ell\ell}_{1122}$ is probed by a combination of the  $e^+e^-\to\mu^+\mu^-$ process at LEP-2, as well as the CHARM and CHARM-II measurements of neutrino--electron scattering ($\nu_\mu e^-\to\nu_\mu e^-$ and $\bar{\nu}_\mu e^-\to\bar{\nu}_\mu e^-$).

The coefficient $C^{\ell e}_{1221}$ is constrained by the Michel parameters of muon decay
$\mu^-\to e^-\nu_\mu\bar{\nu}_e$, measured at PSI~\cite{Danneberg:2005xv}, while $C^{\ell e}_{2332}$
is constrained by Michel-type form factors in leptonic tau decays: $\tau^-\to\mu^-\nu_\tau\bar{\nu}_\mu$, as compiled by the PDG~\cite{ParticleDataGroup:2014cgo}. The $C^{\ell\ell}_{2332}$ WC is also constrained by the leptonic tau decay $\tau\to e \mu^+ \mu^-$, which modifies the effective Fermi constant. 

The above constraints are mainly taken from Ref.~\cite{Falkowski:2015krw}. In addition to these, by using the observation of neutrino trident scattering  $\nu_\mu \gamma^* \to \nu_\mu \mu^+ \mu^-$ at CHARM-II~\cite{CHARM-II:1994dzw} one can derive upper bounds for the $C^{\ell\ell}_{2222}$ and $C^{\ell e}_{2222}$ WCs~\cite{Falkowski:2018dmy}, as listed in~\cref{tab:SmeftvLowE}. 

Finally, results from the NA64$\mu$ experiment at CERN yeild upper bounds for the $C^{\ell\ell}_{2233}$ and $C^{\ell e}_{3322}$ WCs, which are currently about $10^2~$TeV$^{-2}$~\cite{Crivelli:2025pjb}. The future NA62$\mu$ is expected to improve these upper bounds by about two orders of magnitude~\cite{Crivelli:2025pjb}. 


\subsection{Lepton flavor violating tau decays}
Lepton flavor violating (LFV) $\tau$ decays provide some of the most stringent probes of four--lepton operators involving second and third generation leptons. We derive these constraints following the EFT analysis of Ref.~\cite{Falkowski:2015krw}, which expresses the branching ratios of LFV $\tau$ decays in terms of the four-lepton SMEFT WCs. For the experimental upper bounds on the Branching fractions, they used the measurements of the Belle experiment~\cite{Hayasaka:2010np}: $Br(\tau^- \to \mu^+ \mu^- \mu^-)<2.1\times10^{-8}$ and $Br(\tau^- \to e^- \mu^+ \mu^-)<2.7\times10^{-8}$ at $90\%$ C.L., respectively. 
The computations are all performed at tree level. In this section, we present results in terms of dimensionless WCs denoted $\tilde c^X$, which are related to our notation defined in~\cref{sec:SMEFT} as 
\begin{align}
    C^X_{ijkl}\equiv\frac{\tilde c^X_{ijkl}}{v^2},
\end{align}
where $v\sim 246~$GeV is the electroweak vacuum expectation value. 

By using the above branching fractions Ref.~\cite{Falkowski:2015krw} find the following constraints. For $\tau^- \rightarrow \mu^- \mu^+ \mu^- $,
\begin{align}
\tilde c^{\rm{eff}}_{3222} 
< 6.9 \times 10^{-4}\,;
\end{align}
while for $\tau^- \rightarrow e^- \mu^+ \mu^- $
\begin{align}
\tilde c^{\rm{eff}}_{3122}
< &7.8 \times 10^{-4}\,,
\end{align}
with $\tilde c^{\rm{eff}}_{ijkl}$ defined as
\begin{align}
\label{eq:tautomumumuWEFT}
\tilde c^{\rm{eff}}_{3222}\equiv\Bigg\{ &2\Big| \tilde c^{\ell\ell}_{2223}+4 g_{L,SM}^{Ze}[\delta g_L^{Ze}]_{23} \Big|^2 + 2\Big| \tilde c^{ee}_{2223}+ 4 g_{R,SM}^{Ze}[\delta g_R^{Ze}]_{23} \Big|^2 
\nonumber\\+
& \Big| \tilde c^{\ell e}_{2223}+\tilde c^{\ell e}_{2322}+4 g_{R,SM}^{Ze}[\delta g_L^{Ze}]_{23}+4 g_{L,SM}^{Ze}[\delta g_R^{Ze}]_{23} \Big|^2 \Bigg\} ^{1/2} \,,
\end{align}
and
\begin{align}
\label{eq:tautoemumuWEFT}
\tilde c^{\rm{eff}}_{3122}=\Bigg\{ &\Big|\tilde c^{\ell\ell}_{1322}+\tilde c^{\ell\ell}_{1223}+4 g_{L,SM}^{Ze}[\delta g_L^{Ze}]_{13}\Big|^2+
\Big|\tilde c^{ee}_{1223}+4 g_{R,SM}^{Ze}[\delta g_{R}^{Ze}]_{13}\Big|^2  \nonumber\\
 &+\Big|\tilde c^{\ell e}_{2213}+\tilde c^{\ell e}_{2312}+4 g_{L,SM}^{Ze}[\delta g_R^{Ze}]_{13}\Big|^2+
\Big|\tilde c^{\ell e}_{1322}+\tilde c^{\ell e}_{1223}+4 g_{R,SM}^{Ze}[\delta g_L^{Ze}]_{13}\Big|^2\Bigg\}^{1/2}
\,.
\end{align}
The couplings $g_{L,SM}^{Ze}$ and $g_{R,SM}^{Ze}$ are the SM gauge couplings between the $Z$ gauge boson and fermions, and $\delta g_{L/R}^{Ze}$ are their corresponding vertex corrections. The vertex corrections are independent parameters.  

The upper bounds in these relations from Ref.~\cite{Falkowski:2015krw} are expressed as $90\%$ upper bounds, while for the sake of consistency with the other constraints we need the $68\%$ upper bounds. In order to achieve this we use the statistical treatment described in Section IV.D of Ref.~\cite{deGouvea:2022znk}. We assume that the best-fit values associated to all experimental results are zero, and we treat the upper bounds as quadratic $\chi^2$ functions $|\tilde c^{\rm{eff}}_{ijkl}|^2$, which are the relations in the left hand sides of~\cref{eq:tautomumumuWEFT,eq:tautoemumuWEFT}. Hence, we can write the effective $\chi^2$ function as :
\begin{equation}
\chi^2_{\rm{eff}}=\frac{(|\tilde c^{\rm{eff}}_{ijkl}|^2)^2}{\sigma^2_{\rm{eff}}}\,.
\end{equation}
Here the quoted $90\%$ upper bounds (right hand sides of of~\cref{eq:tautomumumuWEFT,eq:tautoemumuWEFT}) are related to the systematic uncertainty of the $\chi^2$ function as 
$ \sigma^2_{\rm{eff}}=\frac{(90\% ~\rm{Upper ~Bound})^2}{1.64^2}$. The $90\%$ C.L. corresponds to $1.64\sigma$. This approximately gives us a normal $\chi^2$ function, which we can use to place $68\%$ upper bounds on WCs, turning them on one at a time. From $\tau^- \rightarrow \mu^- \mu^+ \mu^- $ we find
\begin{align}
C^{\ell\ell}_{2223} , ~C^{ee}_{2223}&< 4.9\times10^{-3}~\mathrm{TeV}^{-2}, \nonumber\\
C^{\ell e}_{2223}, ~C^{\ell e}_{2322} &< 6.9\times10^{-3}~\mathrm{TeV}^{-2}\,.
\end{align}
Similarly, using the process $\tau^- \rightarrow e^- \mu^+ \mu^-$, we derive the bounds
\begin{align}
C^{\ell\ell}_{1322},~C^{\ell\ell}_{1223},~C^{ee}_{1223} ,~C^{\ell e}_{2213},~C^{\ell e}_{2312},~C^{\ell e}_{1322},~C^{\ell e}_{1223}&< 7.8\times10^{-3}~\mathrm{TeV}^{-2}\,.
\end{align}
All the relevant constraints are listed in~\cref{tab:SmeftvLowE}.

The Belle-II experiment recently updated the $90\%$ C.L. upper bounds for these processes as $Br(\tau^- \to \mu^+ \mu^-\mu^-)<1.9\times10^{-8}$~\cite{Belle-II:2024sce} and $Br(\tau^- \to e^- \mu^+ \mu^-)<2.4\times10^{-8}$~\cite{Belle-II:2025urb}. These updated upper bounds change the results of Ref.~\cite{Falkowski:2015krw} minimally, with an effect of less than $5\%$. However, the Belle-II experiment expects to improve the upper bounds on LFV $\tau$ decay by a factor of 70-80 times after gathering $50~$ab$^{-1}$ data sample~\cite{Belle-II:2018jsg}, which translates into approximately nine times improvement for the existing constraints on the LFV WCs.

\subsection{Lepton flavor violating constraints from loop corrections}
In this section, all the low-energy constraints described so far are sensitive to the SMEFT operators at tree level, so that the resulting bounds are unambiguous and directly comparable to the muon collider projections, which are similarly at tree level. For the flavor violating operators $C^{ee}_{1222}$, $C^{\ell e}_{1222}$, $C^{\ell e}_{2212}$ and $C^{\ell\ell}_{1222}$, however, the strongest low-energy constraints arise only once loop effects are also included. In this case, other low-energy observables can contribute, such as radiative muon decay $\mu \to e \gamma$ and the decay $\mu\to 3 e$. We extract all such relevant low energy constraints from Ref.~\cite{Asadi:2025dii}.

For example, the branching fraction of the radiative lepton decay $l_i \to l_f \gamma$ can be written in terms of the effective parameters $F^{if}_L$ and $F^{if}_R$~\cite{Asadi:2025dii}. Within tree level, these effective parameters can only be expressed as a linear combination of the dipole $C_{eW}$ and $C_{eB}$ WCs, which will not contribute to any of the four-lepton operators when turning on only one parameter at a time. However, when one-loop contributions to the radiative decay $l_i \to l_f \gamma$  are taken into account as well, then the $C^{\ell e}_{fjji}$ will also enter in the calculation of the effective parameters $F^{if}_L$ and $F^{if}_R$ (see e.g. Ref.~\cite{Crivellin:2013hpa} for full relations). As a result, the MEG-II experiment sets an upper bound of $Br(\mu\to e \gamma)<1.5\times 10^{-13}$~\cite{MEGII:2025gzr}, which translates to 
\begin{align}
    C^{\ell e}_{1222}<1.8\times 10^{-4}~\mathrm{TeV}^{-2}. 
\end{align}

On the other hand, when the leading-log running of the low-energy Wilson coefficients from the weak scale down to the lepton mass scale is taken into account through the one-loop anomalous dimensions, which can be written in terms of the low-energy coefficients obtained from the tree-level matching at $M_Z$~\cite{Asadi:2025dii}, the branching fraction of the $\ell_i\to 3\ell_f$ decay also depends on the $C^{\ell\ell}_{fijj}$ WC~\cite{Jenkins:2017dyc}. The SINDRUM experiment has found an upper bound of $Br(\mu\to 3 e)<10^{-12}$~\cite{SINDRUM:1987nra}, which translates to 
\begin{align}
    C^{\ell\ell}_{1222}<3.7\times 10^{-3}~\mathrm{TeV}^{-2}
\end{align} The Mu3e experiment~\cite{Mu3e:2020gyw} is expected to improve the upper bound on $Br(\mu\to 3 e)$ by four orders of magnitude, which means the constraints on these WCs will be two orders of magnitude improved. 

In a similar manner, through one-loop anomalous dimensions, from experiments sensitive to muon conversion to electrons when scattering off of atomic nuclei one can be sensitive to the four-lepton WCs $C^{ee}_{1222}$ and $C^{\ell e}_{2212}$. The SINDRUM-II experiment searches for $(\mu\to e)_{\rm{Au}}$ conversion in Gold, and results to the conversion rate of: $CR(\mu\to e)_{\rm{Au}}<7\times 10^{-13}$~\cite{SINDRUMII:2006dvw}. From this, one can find upper bounds of
\begin{align}
C^{ee}_{1222}<2.3\times 10^{-4}~\mathrm{TeV}^{-2}
\end{align}
and 
\begin{align}C^{\ell e}_{2212}<7.5\times 10^{-4}~\mathrm{TeV}^{-2}.
\end{align}
Using Aluminium targets, the Mu2e and COMET experiments~\cite{Mu2e:2014fns,COMET:2018auw} are expected to improve the reach on the conversion rate by more than four orders of magnitude with an aluminum nucleus. This translates into two orders of magnitude effective improvement on the WC constraints. 

\begin{table}[ht!]
\resizebox{\textwidth}{!}{\def\arraystretch{1.5}
    \centering
    \begin{tabular}{|c||c|c||c|c|}
    \hline
    \multicolumn{5}{|c|}{\textsc{Reach of a 10} TeV \textsc{Muon Collider at 10} ab$^{-1}$ \textsc{contrast with present ``Low Energy" constraints }}\\
    \hline
    \multirow{ 2}{*}{WC}& \multicolumn{2}{c||}{\textsc{Collider} (68 \{95\}\% C.L.)}& \multicolumn{2}{c|}{\textsc{Low Energy } (68\% C.L.)}
    \\
   \cline{2-5}
    & Units: TeV$^{-2}$ &Final state/s &  Units: TeV$^{-2}$ & Observable/s \\
    \hline
        $C^{ee}_{1122}$&  $1.8 \{3.6\}\times 10^{-5}$  &$e^+e^-$ & $8.6\times 10^{-2}$&LEP-II ($e^+e^-\to \mu^+\mu^-$)~\cite{Falkowski:2015krw}\\
        $C^{\ell e}_{1122}$&  $0.75 \{1.5\}\times 10^{-4}$ &$e^+e^-$ & $7.6\times 10^{-2}$& LEP-II ($e^+e^-\to \mu^+\mu^-$)~\cite{Falkowski:2015krw}\\
        $C^{\ell e}_{2211}$& $0.74 \{1.5\}\times 10^{-4}$&$e^+e^-$ &$7.7\times 10^{-2}$& CHARM-II ($\nu_\mu-e$ scattering)~\cite{Falkowski:2015krw}\\
        $C^{\ell e}_{1221}$& $1.8 \{2.5\}\times 10^{-4}$  &$ e^+e^-$ & $3.0\times 10^{-1}$&  PSI ($\mu^-\to e^-\nu_\mu\bar{\nu}_e$)~\cite{Falkowski:2015krw}\\
        $C^{\ell\ell}_{1122}$&  $3.3\{6.5\}\times 10^{-5}$ &$ e^+e^- $ & $7.8\times 10^{-2}$&LEP-II ($e^+e^-\to \mu^+\mu^-$) \&  CHARM-II ($\nu_\mu-e$ scattering)~\cite{Falkowski:2015krw}\\
        $C^{\ell\ell}_{1221}$&  $3.3 \{6.5\}\times 10^{-5}$ &$ e^+e^-$ &$2.8\times 10^{-2}$& LEP-2 ($W$-boson mass measurement)~\cite{Falkowski:2015krw} \\
         \hline 
        $C^{ee}_{2222}$ & $3.5\{6.9\}\times 10^{-5}$ &$\mu^+\mu^-$& & \\
        $C^{\ell e}_{2222}$&  $4.7\{9.3 \}\times 10^{-5}$ &$ \mu^+\mu^-$ &  $10.4$&  CHARM-II ($\nu_\mu$ Trident Scattering)~\cite{Falkowski:2017pss, Falkowski:2018dmy}\\
        $C^{\ell\ell}_{2222}$& $3.1\{6.0\}\times 10^{-5}$&$ \mu^+\mu^-$ & $3.46$& CHARM-II ($\nu_\mu$ Trident Scattering)~\cite{Falkowski:2017pss, Falkowski:2018dmy}\\
        \hline 
        $C^{ee}_{2233}$& $1.8\{3.6\}\times 10^{-5}$& $ \tau^+\tau^-$&& \\
        $C^{\ell e}_{3322}$&  $0.75\{1.5\}\times 10^{-4}$&$ \tau^+\tau^-$ &$1.0\times 10^2$&NA64$\mu$ (muon-nucleus scattering)~\cite{Crivelli:2025pjb}\\
        $C^{\ell e}_{2233}$&  $0.74\{1.5\}\times 10^{-4}$&$\tau^+\tau^-$ & & \\
        $C^{\ell e}_{2332}$&$1.8\{2.5\}\times 10^{-4}$&  $\tau^+\tau^-$& $5.6\times 10^0$& PDG ($\tau^-\to\mu^-\nu_\tau\bar{\nu}_\mu$ decay)~\cite{Falkowski:2015krw} \\
        $C^{\ell\ell}_{2233}$& $3.3\{6.5\}\times 10^{-5}$ &$\tau^+\tau^-$ & $1.0\times 10^2$& NA64$\mu$ (muon-nucleus scattering)~\cite{Crivelli:2025pjb} \\
        $C^{\ell\ell}_{2332}$& $3.3\{6.5\}\times 10^{-5}$ &$ \tau^+\tau^-$ & $6.1\times 10^{-1}$&PDG $(\tau^-\to\mu^-\nu_\tau\bar{\nu}_\mu$ decay)~\cite{Falkowski:2015krw} \\
        \hline 
        $C^{ee}_{1222}$ &$1.6\{ 2.5\}\times 10^{-5}$& $\mu^+e^-(\mu^-e^+)$&$2.3\times 10^{-4}$& SINDRUM-II $\Big((\mu\to e)_{\rm{Au}}\Big)$~\cite{SINDRUMII:2006dvw}\\
        $C^{\ell e}_{1222}$ &$1.6\{ 2.5\}\times 10^{-5}$&$ \mu^+e^-(\mu^-e^+)$ &$1.8\times 10^{-4}$& MEG-II ($\mu\to e \gamma$)~\cite{MEGII:2025gzr}\\
        $C^{\ell e}_{2212}$&$1.6\{ 2.5\}\times 10^{-5}$&$\mu^+e^-(\mu^-e^+)$&$7.5\times 10^{-4}$&SINDRUM-II $\Big((\mu\to e)_{\rm{Au}}\Big)$~\cite{SINDRUMII:2006dvw}\\
        $C^{\ell\ell}_{1222}$ &$1.6\{ 2.5\}\times 10^{-5}$&$ \mu^+e^-(\mu^-e^+)$&$3.7\times 10^{-3}$&SINDRUM $(\mu\to 3e)$~\cite{SINDRUM:1987nra}\\
        \hline       
        $C^{ee}_{2223}$ &$1.6\{ 2.5\}\times 10^{-5}$&$ \tau^+\mu^-(\mu^+\tau^-) $& $4.9\times 10^{-3}$& Belle ($\tau\to3\mu$), this work and~\cite{Falkowski:2015krw} \\
        $C^{\ell e}_{2322}$ &$1.6\{ 2.5\}\times 10^{-5}$&$ \tau^+\mu^-(\mu^+\tau^-) $& $6.9\times 10^{-3}$&Belle ($\tau\to3\mu$), this work and~\cite{Falkowski:2015krw}\\
        $C^{\ell e}_{2223}$ &$1.6\{2.5\}\times 10^{-5}$&$ \tau^+\mu^-(\mu^+\tau^-) $&$6.9\times 10^{-3}$&Belle ($\tau\to3\mu$), this work and~\cite{Falkowski:2015krw}\\
        $C^{\ell\ell}_{2223}$ &$1.6\{ 2.5\}\times 10^{-5}$&$\tau^+\mu^-(\mu^+\tau^-) $& $4.9\times 10^{-3}$&Belle ($\tau\to3\mu$), this work and~\cite{Falkowski:2015krw}\\
        \hline 
        $C^{ee}_{1322}$ &$1.6 \{2.5\}\times 10^{-5}$ &$ e^+\tau^-(e^-\tau^+)$&$7.8\times 10^{-3}$& Belle ($\tau^-\to e^- \mu^+ \mu^-$), this work and~\cite{Falkowski:2015krw}\\
        $C^{\ell e}_{1322}$ &$3.1\{5.1\}\times 10^{-5}$&$ e^+\tau^-(e^-\tau^+)$& $7.8\times 10^{-3}$&Belle ($\tau^-\to e^- \mu^+ \mu^-$), this work and~\cite{Falkowski:2015krw}\\
        $C^{\ell e}_{2213}$ &$3.1\{5.0\}\times 10^{-5}$&$ e^+\tau^-(e^-\tau^+)$& $7.8\times 10^{-3}$&Belle ($\tau^-\to e^- \mu^+ \mu^-$), this work and~\cite{Falkowski:2015krw}\\
        $C^{\ell e}_{2312}$ &$1.8\{2.9\}\times 10^{-5}$& $ e^+\tau^-(e^-\tau^+)$& $7.8\times 10^{-3}$&Belle ($\tau^-\to e^- \mu^+ \mu^-$), this work and~\cite{Falkowski:2015krw}\\
        $C^{\ell e}_{1223}$ &$1.8 \{2.9\}\times 10^{-5}$&$ e^+\tau^-(e^-\tau^+)$& $7.8\times 10^{-3}$& Belle ($\tau^-\to e^- \mu^+ \mu^-$), this work and~\cite{Falkowski:2015krw}\\
        $C^{\ell\ell}_{1322}$ &$3.1\{5.0\}\times 10^{-5}$& $e^+\tau^-(e^-\tau^+)$& $7.8\times 10^{-3}$&Belle ($\tau^-\to e^- \mu^+ \mu^-$), this work and~\cite{Falkowski:2015krw}\\
        $C^{\ell\ell}_{1223}$ &$3.1\{5.0\}\times 10^{-5}$&$ e^+\tau^-(e^-\tau^+)$& $7.8\times 10^{-3}$&Belle ($\tau^-\to e^- \mu^+ \mu^-$) , this work and~\cite{Falkowski:2015krw}\\
        \hline 
    \end{tabular}}
    \caption{Projected single-operator $68\%$ and $95\%$ C.L upper-bound from 10 TeV Muon Collider with 10$^{-1}$ integrated luminosity on four-lepton SMEFT Wilson coefficients. Constraints assume that the WCs are real, and represent two-sided bounds around zero. For reference, the PDF scale parameters used here are $(Q, q) = (\sqrt{{s}}/2, 1)$.  The low energy constraints listed here are also quoted assuming one operator at a time, and we quote the upper limit on their magnitude at $68\%$ C.L.
  }
    \label{tab:SmeftvLowE}
\end{table}

\newpage 

\bibliographystyle{utphys}
	\bibliography{ref}
\end{document}